\documentclass[11pt]{article}
\usepackage{amsmath,amsthm,natbib}
\usepackage{amssymb}
\usepackage{graphicx}
\usepackage{amsfonts}
\usepackage{xr}
\usepackage{epstopdf}
\usepackage{rotating}
\usepackage{caption}
\usepackage{placeins}
\usepackage[colorlinks, citecolor=blue]{hyperref}
\usepackage{setspace}

\newtheorem{theorem}{Theorem}

\newtheorem{corollary}[theorem]{Corollary}

\newtheorem{lemma}{Lemma}

\newtheorem{assumption}{Assumption}

\allowdisplaybreaks

\newcommand*\plim{\mathop{p\mkern2mu\mathrm{\mathchar"702D lim}}}
\newcommand*\txtIdioVolFM{\textit{IdioVol-FM}}
\newcommand*\swpSmall{\small}
\newcommand*\swpNormalsize{\normalsize}

\newtheorem*{def1}{Definition (Factor Model for Returns, R-FM)}
\newtheorem*{def2}{Definition (Idiosyncratic Volatility Factor Model, IdioVol-FM)}

\allowdisplaybreaks
\input{tcilatex}
\begin{document}

\title{Cross-sectional Dependence in Idiosyncratic Volatility%
}
\author{Ilze Kalnina\thanks{%
Corresponding author. Department of Economics, NC State University.
E-mail address: \texttt{ikalnin@ncsu.edu}.
}
\and Kokouvi Tewou\thanks{%
National Bank of Canada. E-mail address: \texttt{kokouvi.tewou@bnc.ca}.%
}
}
\date{%
\today }

\maketitle

\begin{abstract}

This paper
introduces an econometric framework for analyzing cross-sectional dependence in the idiosyncratic volatilities of assets using high frequency data.
We
first consider the estimation of standard measures of dependence in the idiosyncratic volatilities such as covariances and correlations.
Naive estimators of these measures are biased due to the use of the error-laden estimates of idiosyncratic volatilities.
We
provide bias-corrected estimators and the relevant asymptotic theory.
Next,
we introduce an idiosyncratic volatility factor model, in which we decompose the variation in idiosyncratic volatilities into two parts: the variation related to the systematic factors such as the market volatility, and the residual variation.
Again,  naive
estimators of the decomposition are biased, and we provide bias-corrected estimators.
We also
provide the asymptotic theory that allows us to test whether the residual (non-systematic) components of the idiosyncratic volatilities exhibit cross-sectional dependence.
We apply
 our methodology to the S\&P 100 index constituents,
 and document strong cross-sectional dependence in their idiosyncratic volatilities.
We consider two different sets of idiosyncratic volatility
factors, and find that neither can fully account for the cross-sectional dependence in
idiosyncratic volatilities. 
For each model, we map out the
network of dependencies in residual (non-systematic) idiosyncratic
volatilities across all stocks.



\bigskip

\end{abstract}

\sloppy

\raggedbottom

\newpage


\section{Introduction}

\label{sec:Intro}

In a panel of assets, returns are generally cross-sectionally dependent.
This dependence is usually modeled using the exposure of assets to some
common return factors, such as the Fama-French factors.
 In this Return Factor Model (R-FM), the total volatility of an asset return
can be decomposed into two parts: a component due to the exposure to the
common return factors (the systematic volatility), and a residual component
termed the Idiosyncratic Volatility (IdioVol). These two components of the
volatility of returns are the most popular measures of the systematic risk
and idiosyncratic risk of an asset.

Idiosyncratic Volatility is important in economics and finance for several
reasons. For example, when arbitrageurs exploit the mispricing of an
individual asset, they are exposed to the idiosyncratic risk of the asset
and not the systematic risk (see, e.g., \cite{CampbellLettauMalkielXu2001}).%
\footnote{%
An asset is said to be mispriced with respect to a given model if the
expected value of the return on the asset is not consistent with the model.}
Also, Idiosyncratic Volatility measures the exposure to the idiosyncratic
risk in imperfectly diversified portfolios. The cross-sectional dependence
in IdioVols is also important for option pricing, see \cite{gourier2016}. 
 The attention to IdioVols in empirical finance literature is exemplified by
two IdioVol puzzles, see \cite{CampbellLettauMalkielXu2001} and \cite%
{anghodrick06}. A recent observation is that the IdioVols seem
to be strongly correlated in the cross-section of stocks.\footnote{%
See, e.g., \cite{ConnorKorajczykLinton06}, \cite{duartekamara14}, \cite%
{herskovickellyCIV}, and \cite{ChristoffersenFournierJacobs}.} We propose
methods to formally study this empirical phenomenon with high-frequency
data, while fully accounting for the measurement errors in IdioVols.

This paper
provides an econometric framework for studying the cross-sectional
dependence in the Idiosyncratic Volatilities using high frequency data. The
analysis is based on a new general asymptotic theory that we develop for
estimators of quadratic covariations between nonlinear functions of spot
volatility matrices. We show that naive estimators, such as covariances and
correlations, are biased. The bias arises due to the use of error-laden
estimates of the spot volatility matrices. We provide the bias-corrected
estimators. We derive the asymptotic distribution of these estimators, and
propose consistent estimators of the asymptotic variances. We apply this new
asymptotic theory to construct tests of dependence between IdioVols and map
out the network of dependencies in IdioVols in a panel of assets.

To study Idiosyncratic Volatilities, we introduce the Idiosyncratic
Volatility Factor Model (IdioVol-FM).
Just like a Return Factor Model, R-FM, such as the Fama-French model,
decomposes returns into common and idiosyncratic returns, the IdioVol-FM
decomposes the IdioVols into systematic and residual (non-systematic)
components. The IdioVol factors may or may not be related to the return
factors. The IdioVol factors can include the volatility of the return
factors, or, more generally, (possibly non-linear) transformations of the
spot covariance matrices of any observable variables, such as the average
variance and average correlation factors of \cite{chenpetkova012}. We
propose bias-corrected estimators of the components of the IdioVol-FM model.

We provide the asymptotic theory for this model. For example, it allows us
to test whether the residual (non-systematic) components of the IdioVols
exhibit cross-sectional dependence. This allows us to identify the network
of dependencies in the residual IdioVols across stocks.

Reduced-form analysis of total and idiosyncratic volatilities can be useful
to inform the formulation of structural asset pricing models. For example,
\cite{herskovickellyCIV} document strong dependence in firm IdioVols, and
propose an incomplete markets asset pricing model, where IdioVol behavior is
explained by the idiosyncratic risk faced by households. When documenting the
cross-sectional dependence in IdioVol, \cite{herskovickellyCIV} estimate
several volatility factor models, for example, they regress IdioVols on
average firm volatilities, where the IdioVols are defined with respect to
the market return factor or the Fama-French factors. Our framework can be
used to estimate high-frequency regressions with these variables, on a fixed
time interval, while fully capturing the effect of the measurement error
from the preliminary estimation of both the dependent variable and the
factor.

Throughout the paper, we use factors that are specified by the researcher.
An example of our Return Factor Model is the so-called Fama-French factor
model, which has three observable factors, or the CAPM, which has one
observable factor (the market portfolio return). An example of our IdioVol
factors is the market volatility, which can be estimated from the market
index. Thus, our setup is different from settings such as PCA where factors
are identified from the cross-section of the assets studied. The treatment
of the latter case adds an additional layer of complexity to the model and
is beyond the scope of the current paper.

We apply our methodology 
to high-frequency data on the S\&P 100 index constituents. We study the
IdioVols with respect to two models for asset returns: the CAPM and the
three-factor Fama-French model.\footnote{%
The high frequency Fama-French factors are provided by \cite%
{yackalninaxiu-FF}.} In both cases, the average pairwise correlation between
the IdioVols is high (0.35). We verify that this dependence cannot be
explained by the missing return factors. This confirms the recent findings
of \cite{herskovickellyCIV} who use low frequency (daily and monthly) return
data.
We then consider the IdioVol-FM. We use two sets of IdioVol factors: the
market volatility alone and the market volatility together with volatilities
of nine industry ETFs. With the market volatility as the only IdioVol
factor, the average pairwise correlation between residual (non-systematic)
IdioVols is substantially lower (0.21) 
than between the total IdioVols. With the additional industry ETF
volatilities as IdioVol factors, average correlation between the residual
IdioVols decreases further (to 0.17). However, neither of the two sets of
the IdioVol factors can fully explain the cross-sectional dependence in the
IdioVols. For each model, we map out the network of dependencies in residual
IdioVols across all stocks.

This paper analyzes cross-sectional dependence in Idiosyncratic
Volatilities. This should \emph{not} be confused with the analysis of
cross-sectional dependence in total and idiosyncratic \emph{returns}. A
growing number of papers study the latter question using high frequency
data.
These date back to the analysis of realized covariances and their
transformations, see, e.g., \cite{barndorffnielsenshephard04} and \cite%
{ABDW2006}. A continuous-time factor model for asset returns with observable
return factors was first studied in \cite{myklandzhang2006}. Various return
factor models with observable factors have been studied by, among others,
\cite{BollerslevTodorovJOE2010}, \cite{FanFurgerXiu16}, \cite%
{litodorovtauchen17-adaptive, litodorovtauchen-jumpreg}, and \cite%
{yackalninaxiu-FF}. Emerging literature also studies the cross-sectional
dependence in returns using high-frequency data and latent return factors,
see 
\cite{AitXiuPCA,AitXiuPCA_HighDim} and \cite%
{Pelger2019_theory,Pelger2019_applied}. %
Importantly, the models in the above papers are silent on the
cross-sectional dependence structure in the IdioVols.

While this paper focuses on the study of cross-sectional dependence of
IdioVols, our new asymptotic theory can be used in various other
applications. For example, we can estimate dependence measures, in the form
of co-volatilities or the corresponding correlations, between the
time-varying asset betas.\footnote{%
Here, asset betas are the loadings of asset returns on return factors; these
are distinct from the asset volatility betas that we describe in the next
section.} While it is well-known that asset betas vary over time in
practice, there is no consensus as to what common factors drive this
variation, so accurate dependence measures of asset beta co-movement can be
helpful. Another example is the estimation of dependence measures between
total volatilities or systematic volatilities of asset returns. In addition,
we can estimate high-frequency regressions of one element of a spot
volatility matrix on other elements, such as regression of the asset
volatility on market volatility. Finally, we can estimate high-frequency
regressions of total asset volatility on average asset volatility, which
mirrors one more of the specifications considered in \cite{herskovickellyCIV}%
, in addition to the specifications described earlier.

Our inference theory is related to
several estimators in the existing literature. The closest are the
volatility of volatility estimator of \cite{vetter-vovo} and one of the
asymptotic bias estimators of \cite{jacodrosenbaum-sqrtn}. \cite{vetter-vovo} proposes an estimator of
volatility of volatility of the returns of one asset, and derives the
relevant theory for inference.\footnote{%
This estimator is also studied in \cite{yacjacod14} (Section 8.3) under
similar assumptions to \cite{vetter-vovo}. \cite{yacjacod14} cite 2011
working paper version of \cite{vetter-vovo}.} %
We extend the analysis to the multivariate case with nonlinear
transformations, return jumps, and volatility jumps. While \cite%
{jacodrosenbaum-sqrtn} focus on a different problem, one of the asymptotic
bias terms in their paper coincides with our quantity of interest in a
special case, see Section \ref{sec:estimation_general_functional} for
details. The setting in \cite{jacodrosenbaum-sqrtn} is multivariate and
robust to return and volatility jumps, but they only establish consistency
of the relevant estimator, and do not provide any asymptotic distribution
theory. In contrast, we derive the asymptotic distribution, as well as the
consistency of the estimator of the asymptotic variance. See also \cite%
{LiLiuZhang2012-VoV} who extend the results in \cite{vetter-vovo} to allow
for price jumps and market microstructure noise. They do not consider the
multivariate case, nonlinear transformations, or volatility jumps.
Finally,
\cite{ChongTodorov2024JOE} propose nonparametric estimators of the 
volatility of volatility and leverage effect using high-frequency data on short-dated options.

\cite{jacodrosenbaum13,jacodrosenbaum-sqrtn}, \cite%
{litodorovtauchen-dependencies} and \cite{LiLiuXiu2019-jackknife} estimate
integrated functionals of volatilities, which includes Idiosyncratic
Volatilities. The latter problem is simpler than the problem of the current
paper in the sense that $\sqrt{n}$-consistent estimation is possible, and
the estimators are consistent without a bias correction (see Section \ref%
{sec:estimation_general_functional} for details). In the literature on the
estimation of the leverage effect, preliminary estimation of volatility also
creates a bias, which also needs to be corrected to achieve consistency, see
\cite{aitfanli13}, \cite{Yacine-jump-lev}, \cite{kalninaxiu-lev} and \cite%
{WangMykland12}. 

One of the reasons why we can account for the measurement error from
preliminary estimation of volatilities is the fact that our framework only
uses one (in-fill) asymptotic approximation. It is interesting to contrast this approach with the
analysis of two-step estimators using joint in-fill and long-span
asymptotics, see, e.g., \cite{corrdidistaso06}, \cite{Todorov2009}, \cite%
{bandi-reno-2012}, \cite{kanaya-kristensen-2016}, and \cite{LiPatton2018}.
In these double asymptotic settings, the inference methods for the second
step typically do not depend on the first-step measurement error. This
provides a good approximation as long as the number of high-frequency
observations in every low-frequency period is large enough. A notable early
exception is \cite{BollerslevZhou2002} who use a simple parametric model for
the first-step measurement error.

The Realized Beta GARCH model of \cite{hansen-lunde-voev-2014} imposes a
structure on the cross-sectional dependence in IdioVols. This structure is
tightly linked with the Return Factor Model parameters, whereas our
stochastic volatility framework allows separate specification of the return
factors and the IdioVol factors.\footnote{%
In the Beta GARCH model, the IdioVol of a stock is a product of its own
(total) volatility, and one minus the square of the correlation between the
stock return and the market return.}

In the empirical section, we define a network of dependencies using
(functions of) quadratic covariations of IdioVols. This approach can be
compared with the network connectedness measures of \cite{dieboldyilmaz2014}%
.
The latter measures are based on forecast error variance decompositions from
vector autoregressions. They capture co-movements in forecast errors. In
contrast, we assume a general semimartingale setting, and our framework
captures realized co-movements in Idiosyncratic Volatilities, while
accounting for the measurement errors in these volatilities.

\indent The remainder of the paper is organized as follows. Section \ref%
{sec:model} introduces the model and the quantities of interest. Section \ref%
{sec:estimation} describes the identification and estimation. Section \ref%
{sec:AsyProperties} presents the asymptotic properties of our estimators.
Section \ref{sec:Empirical} uses high-frequency stock return data to study
the cross-sectional dependence in IdioVols using our framework. Section \ref%
{sec:MC} contains Monte Carlo simulations. The Online Supplementary Appendix
contains all proofs and additional figures.

\section{Model and Quantities of Interest}

\label{sec:model} We first describe a general Factor Model for the Returns
(R-FM), which allows us to define the Idiosyncratic Volatility. We then
introduce the Idiosyncratic Volatility Factor Model (IdioVol-FM). In this
framework, we proceed to define the cross-sectional measures of dependence
between the total IdioVols, as well as the residual IdioVols, which take
into account the dependence induced by the IdioVol factors.

Suppose we have (log) prices on $d_{S}$ assets such as stocks, $%
S_{t}=(S_{1,t},\ldots ,S_{d_{S},t})^{\top }$, and on $d_{F}$ observable
factors, $F_{t}=(F_{1,t},\ldots ,F_{d_{F},t})^{\top }$. We stack them into
the $d$-dimensional process $Y_{t}=(S_{1,t},\ldots
,S_{d_{S},t},F_{1,t},\ldots ,F_{d_{F},t})^{\top }$ where $d=d_{S}+d_{F}$.
The observable factors $F_{1},\ldots, F_{d_{F}}$ are used in the R-FM model
below. We assume that all observable variables jointly follow an It\^{o}
semimartingale, i.e., $Y_{t}$ follows
\begin{equation}
Y_{t}=Y_{0}+\int_{0}^{t}b_{s}ds+\int_{0}^{t}\sigma _{s}dW_{s}+J_{t}^{Y},
\label{eqn:Y_BSM}
\end{equation}%
where $W$ is a $d^{W}$-dimensional Brownian motion ($d^{W}\geq d$),
$C_{t}=\sigma _{t}\sigma _{t}^{\top }$ is the spot covariance process, and $%
J_{t}^{Y}$ denotes a finite variation jump process. The spot covariance
matrix process $C_{t}$
of $Y_{t}$ is a continuous It\^{o} semimartingale,\footnote{%
Note that assuming that $Y$ and $C$ are driven by the same $d^{W}$%
-dimensional Brownian motion $W$ is without loss of generality provided that
$d^{W}$ is large enough, see, e.g., equation (8.12) of \cite{yacjacod14}.}
\begin{equation}
C_{t}=C_{0}+\int_{0}^{t}\widetilde{b}_{s}ds+\int_{0}^{t}\widetilde{\sigma }%
_{s}dW_{s}+J_{t}^{\sigma }.  \label{eqn:C_BSM}
\end{equation}
We refer to the $\left( C_{t}\right) _{a,b}$ element of the matrix $C_{t}$
as $C_{ab,t}$. For convenience, we also use the alternative notation $%
C_{UV,t}$ to refer to the spot covariance between two elements $U$ and $V$
of $Y$, and $C_{U,t}$ to refer to $C_{UU,t}$.

We assume a standard continuous-time factor model for the asset returns.

\begin{def1}
\label{def:R-FM} For all $0\leq t\leq T$ and $j=1,\ldots ,d_{S}$,%
\footnote{\abovedisplayskip=1pt \belowdisplayskip=3pt Quadratic covariation
of two vector-valued It\^{o} semimartingales $X$ and $Y $, over the time
span $[0,T]$, is defined as
\begin{equation*}
\lbrack X,Y]_{T}=\plim_{M\rightarrow \infty
}\sum_{s=0}^{M-1}(X_{t_{s+1}}-X_{t_{s}})(Y_{t_{s+1}}-Y_{t_{s}})^{\top },
\end{equation*}%
for any $t_{0}<t_{1}<\ldots <t_{M}=T$ with $\sup_{s}\left\vert
t_{s+1}-t_{s}\right\vert \rightarrow 0$ as $M\rightarrow \infty $.
\par
\medskip \noindent Intuitively, quadratic covariation can be thought of as
the integrated covariance between the increments $dX_{t}$ and $dY_{t}$.}
\begin{equation}
\begin{split}
dS_{j,t}& =\beta _{j,t}^{\top }dF_{t}^{c}+\tilde{\beta}_{j,t}^{\top
}dF_{t}^{d}+dZ_{j,t}\text{ \ \ with} \\
\lbrack Z_{j},F]_{t}& =0.
\end{split}
\label{eqn:R-FM}
\end{equation}
\end{def1}

In the above, $dZ_{j,t}$ is the idiosyncratic return of stock $j$. The
superscripts $c$ and $d$ indicate the continuous and jump part of the
processes, so that $\beta_{j,t}$ and $\tilde{\beta}_{j,t}$ are the
continuous and jump factor loadings. For example, the $k$-th component of $%
\beta_{j,t}$ corresponds to the time-varying loading of the continuous part
of the return on stock $j$ to the continuous part of the return on the $k$%
-th factor. We set $\beta_t=(\beta_{1,t},\ldots,\beta_{d_S,t})^\top$ and $%
Z_t=(Z_{1,t},\ldots,Z_{d_S,t})^\top$.

We do not need the return factors $F_t$ to be the same across assets to
identify the model, but without loss of generality, we keep this structure
as it is standard in empirical finance. These return factors are assumed to
be observable, which is also standard. For example, in the empirical
application, we use two sets of return factors: the market portfolio and the
three Fama-French factors, which are constructed in \cite{yackalninaxiu-FF}.

A continuous-time factor model for returns with observable factors was
originally studied in \cite{myklandzhang2006} in the case of one factor and
in the absence of jumps. A burgeoning literature uses related models to
study the cross-sectional dependence of total and/or idiosyncratic returns.
However, this literature does not consider the cross-sectional dependence in
the IdioVols.

We define the idiosyncratic Volatility (IdioVol) to be the spot volatility
of $Z_{j,t}$ and denote it by $C_{Zj,t}$. 
Notice that R-FM in (\ref{eqn:R-FM}) implies that the factor loadings $\beta
_{t}$ as well as the IdioVols are functions of the total spot covariance
matrix $C_{t}$. In particular, the vector of factor loadings satisfies
\begin{equation}
\beta _{jt}=(C_{F,t})^{-1}C_{FSj,t},  \label{eqn:beta_ID}
\end{equation}%
for $j=1,\ldots ,d_{S}$, where $C_{F,t}$ 
denotes the spot covariance matrix of the factors $F$, which is the lower $%
d_{F}\times d_{F}$ sub-matrix of $C_{t}$; and $C_{FSj,t}$ denotes the
covariance of the factors and the $j^{th}$ stock, which is a vector
consisting of the last $d_{F}$ elements of the $j^{th}$ column of $C_{t}$.
The IdioVol of stock $j$ is then also a function of the total spot
covariance matrix $C_{t}$,
\begin{equation}
\underset{\text{IdioVol of stock j}}{\underbrace{C_{Zj,t}}}=\underset{\text{%
total volatility of stock j}}{\underbrace{C_{Yj,t}}}-~~~(C_{FSj,t})^{\top
}(C_{F,t})^{-1}C_{FSj,t}.  \label{eqn:IdioVol_ID}
\end{equation}%
By the It\^{o} lemma, (\ref{eqn:beta_ID}) and (\ref{eqn:IdioVol_ID}) imply
that factor loadings and IdioVols are also It\^{o} semimartingales with
characteristics that are functions of $C_{t}$.

We now introduce the Idiosyncratic Volatility Factor model (IdioVol-FM). In
IdioVol-FM, the cross-sectional dependence in the IdioVol shocks can be
potentially explained by certain IdioVol factors we denote as $\Pi _{t}$. A
simple example of IdioVol factor is the market volatility. Our model allows
IdioVol factors to be any given smooth functions of the matrix $C_{t}$; we
discuss  examples below.

\begin{def2}
For all $0\leq t\leq T$ and $j=1,\ldots ,d_{S}$, the Idiosyncratic
Volatility $C_{Zj}$ follows,
\begin{eqnarray}
dC_{Zj,t}\ \ &=&\gamma _{Zj}^{\top }d\Pi _{t}^{c}+\tilde{\gamma}_{Zj}^{\top
}d\Pi _{t}^{d}+dC_{Zj,t}^{resid}\hspace{3mm}\text{with}
\label{eqn:IdioVol-FM} \\
\lbrack C_{Zj}^{resid},\Pi ]_{t} &=&0,  \notag
\end{eqnarray}
where $\Pi _{t}=(\Pi _{1t},\ldots ,\Pi _{d_{\Pi }t})$ is a $\mathbb{R}%
^{d_{\Pi }}$-valued vector of IdioVol factors. IdioVol factors satisfy
\begin{equation}
\Pi _{kt}=\Pi _{k}(C_{t})  \label{eqn:pi}
\end{equation}%
with the function $\Pi _{k}(\cdot )$ being three times continuously
differentiable for $k=1,\ldots ,d_{\Pi }$.
\end{def2}

$\Pi \left( \cdot \right) $ is a smooth function of $C_{t}$. For example,
often $\Pi (C_{t})$ is $C_{F,t}$, i.e., $\Pi \left( \cdot \right) $ selects
the components of $C_{t}$ that correspond to the volatilities of the
observable factors $F_{t}$. 
More generally, $\Pi_{t} $ may also include the volatilities and covolatilities of other assets beyond $F_t$.
Even more generally, our theory permits a rather wide class of IdioVol factors,
since it includes general non-linear transforms of the spot covariance matrix process $C_{t}$. 
For example, IdioVol factors can be
linear combinations of the total volatilities of assets, see, e.g., the
average variance factor of \cite{chenpetkova012}. Another example is the
common IdioVol factor, or \textquotedblleft CIV\textquotedblright , which is
studied in \cite{herskovickellyCIV}. CIV is defined as the cross-sectional
average of the firm IdioVols from CAPM. 
The IdioVol factors can also be
the volatilities of any other observable processes.

We call the residual term $C_{Zj,t}^{resid}$ in the IdioVol-FM the residual
IdioVol of asset $j$.
Our assumptions imply that the components of the IdioVol-FM, $C_{Zj,t},\Pi
_{t}$ and $C_{Zj,t}^{resid}$, are It\^{o}
semimartingales. We remark that both the dependent variable and the
regressors in our IdioVol-FM are not directly observable and have to be
estimated, and our asymptotic theory takes that into account. As will see in
Section \ref{sec:estimation}, this preliminary estimation implies that the
naive estimators of all the dependence measures defined below are biased.
One of the contributions of this paper is to quantify this bias and provide
the bias-corrected estimators for all the quantities of interest.

Having specified our econometric framework, we now provide the definitions
of some natural measures of dependence of (the continuous parts of) the
(total) IdioVols and the residual IdioVols. We consider the estimation of these
measures in Section \ref{sec:estimation}.

Before studying the decomposition of the IdioVol-FM model,
one may be interested in  quantifying the dependence between
the (total) IdioVols of two stocks $j$ and $s$.
Quadratic covariation $%
[C_{Zj},C_{Zs}]_{T}^{c}$ is\ one natural measure of dependence between the
(continuous parts of) the IdioVols $C_{Zj}$ and $C_{Zs}$. Another natural
and scale invariant measure is the quadratic-covariation-based correlation
between the two IdioVol processes over a given time period $[0,T]$,
\begin{equation}
Corr\left( C_{Zj},C_{Zs}\right) =\frac{[C_{Zj},C_{Zs}]_{T}^{c}}{\sqrt{%
[C_{Zj},C_{Zj}]_{T}^{c}}\sqrt{[C_{Zs},C_{Zs}]_{T}^{c}}}.  \label{eqn:corrivs}
\end{equation}%
Correlation-based measure is more convenient for reporting the strength of
dependence, while the quadratic covariation $[C_{Zj},C_{Zs}]_{T}^{c}$
without normalization is more convenient for testing for the presence of
cross-sectional dependence in IdioVols. We consider such tests in Section %
\ref{sec:testing}.

Similarly, to measure the cross-sectional dependence between the residual
IdioVols of two stocks, after accounting for the effect of the IdioVol
factors, we use the quadratic-covariation-based correlation,
\begin{equation}
Corr\left( C_{Zj}^{resid},C_{Zs}^{resid}\right) =\frac{%
[C_{Zj}^{resid},C_{Zs}^{resid}]_{T}^{c}}{\sqrt{%
[C_{Zj}^{resid},C_{Zj}^{resid}]_{T}^{c}}\sqrt{%
[C_{Zs}^{resid},C_{Zs}^{resid}]_{T}^{c}}}.  \label{eqn:corrNSivs}
\end{equation}%
In Section \ref{sec:testing}, we use the quadratic covariation between the
two residual IdioVol processes $[C_{Zj}^{resid},C_{Zs}^{resid}]_{T}^{c}$
without normalization for testing purposes.

We want to capture how well the IdioVol factors explain the time variation
of IdioVols of the $j^{th}$ asset. For this purpose, we use the
quadratic-covariation based analog of the coefficient of determination. For $%
j=1,\ldots ,d_{S}$,
\begin{equation}
R_{Zj}^{2,\txtIdioVolFM}=\frac{\gamma _{Zj}^{\top }[\Pi ,\Pi ]_{T}^{c}\gamma
_{Zj}}{[C_{Zj},C_{Zj}]_{T}^{c}}.  \label{eqn:R2_univar}
\end{equation}

It is interesting to compare the correlation measure between IdioVols in
equation (\ref{eqn:corrivs}) with the correlation between the residual parts
of IdioVols in (\ref{eqn:corrNSivs}). We consider their difference,
\begin{equation}
Corr\left( C_{Zj},C_{Zs}\right) -Corr\left(
C_{Zj}^{resid},C_{Zs}^{resid}\right)
\end{equation}%
to see how much of the dependence between IdioVols can be attributed to the
IdioVol factors. In practice, if we compare assets that are known to have
positive covolatilities (typically, stocks have that property), another
useful measure of the common part in the overall covariation between
IdioVols is the following quantity,
\begin{equation}
Q_{Zj,Zs}^{\txtIdioVolFM}=\frac{\gamma _{Zj}^{\top }[\Pi ,\Pi
]_{T}^{c}\gamma _{Zs}}{[C_{Zj},C_{Zs}]_{T}^{c}}.  \label{eqn:R2_bivar}
\end{equation}%
This measure is bounded by 1 if the covariations between residual IdioVols
are nonnegative and smaller than the covariations between IdioVols, which is
what we find for every pair in our empirical application with high-frequency
observations on stock returns.

We remark that our framework can be compared with the following null
hypothesis studied in \cite{litodorovtauchen-dependencies}, $%
H_{0}:~C_{Zj,t}=a_{Zj}+\gamma _{Zj}^{\top }\Pi _{t},~0\leq t\leq T$. This $%
H_{0}$ implies that the IdioVol is a deterministic function of the factors,
which does not allow for an error term. In particular, this null hypothesis
implies $R_{Zj}^{2,\txtIdioVolFM}=1$. Our framework allows for testing
stochastic relationships, i.e., null hypotheses $H_{0}:\gamma _{Zj}^{\top
}=0 $ in the presence of an error term.


\section{Estimation}

\label{sec:estimation}

As we show below, the quantities of interest in Section \ref{sec:model} can
be expressed in terms of the continuous quadratic covariation between two
functions of the spot covariance matrix $C_{t}$,
\begin{equation}
\left[ H(C),G(C)\right] _{T}^{c}.  \label{eqn:HC_GC}
\end{equation}%
Section \ref{sec:estimation_general_functional} proposes estimators of this
general functional, and Section \ref{sec:estimation_special_case} explains
how to use these formulas to obtain estimators of the quantities of interest
in Section \ref{sec:model}.

\subsection{Estimation of a General Functional}

\label{sec:estimation_general_functional}

%

This section proposes estimators of the continuous quadratic covariation
between two functions of the spot covariance matrix $[H(C),G(C)]_{T}^{c}$,
where $H$ and $G$ are given real-valued smooth functions. Recall that $C_{t}$
is the spot covariance matrix of the observable variables, see equations (%
\ref{eqn:Y_BSM})-(\ref{eqn:C_BSM}).

Suppose we have discrete observations on $Y_{t}$ over an interval $[0,T]$.
Denote by $\Delta _{n}$ the distance between observations. It is well known
that we can estimate the spot covariance matrix $C_{t}$ at time $(i-1)\Delta
_{n}$ with a local truncated realized volatility estimator,
\begin{equation}
\widehat{C}_{i\Delta _{n}}=\frac{1}{k_{n}\Delta _{n}}\sum_{m=0}^{k_{n}-1}%
\left( \Delta _{i+m}^{n}Y\right) \left( \Delta _{i+m}^{n}Y\right) ^{\top
}1_{\{\Vert \Delta _{i+m}^{n}Y\Vert \leq u_{n}\}},  \label{eqn:spotC_hat}
\end{equation}%
where $\Delta _{i}^{n}Y=Y_{i\Delta _{n}}-Y_{(i-1)\Delta _{n}}$ and where $%
k_{n}$ is the number of observations in a local window.\footnote{%
It is also possible to define more flexible kernel-based estimators as in
\cite{kristensen10}.} We refer to the $\left( \widehat{C}_{i\Delta
_{n}}\right) _{a,b}$ element of the matrix $\widehat{C}_{i\Delta _{n}}$ as $%
\widehat{C}_{ab,i\Delta _{n}}$. 

If $C_{i\Delta _{n}}$ was observed and in the absence of volatility jumps,
we could estimate $[H(C),G(C)]_{T}$ by the realized covariance between $%
G(C_{i\Delta _{n}})$ and $H(C_{i\Delta _{n}})$, which is the sample analog
of the definition of $[H(C),G(C)]_{T}$. However, we do not observe $%
C_{i\Delta _{n}}$. If we replace it with $\widehat{C}_{i\Delta _{n}}$ in (%
\ref{eqn:spotC_hat}), we obtain the plug-in estimator
\begin{equation}
\widehat{\lbrack H(C),G(C)]}_{T}^{\mathit{Naive}}=\frac{1}{k_{n}}%
\sum_{i=1}^{[T/\Delta _{n}]-2k_{n}+1}\left( H(\widehat{C}_{(i+k_{n})\Delta
_{n}})-H(\widehat{C}_{i\Delta _{n}})\right) \left( G(\widehat{C}%
_{(i+k_{n})\Delta _{n}})-G(\widehat{C}_{i\Delta _{n}})\right) .
\label{eqn:estNAIVE}
\end{equation}%
However, it turns out that due to the measurement errors in $\widehat{C}%
_{i\Delta _{n}}$, this estimator is inconsistent.

We propose two estimators for the general quantity $[H(C),G(C)]_{T}^{c}$.
Our first estimator is a bias-corrected sample analog of the definition of
quadratic covariation between two It\^{o} processes,
\begin{align}
\widehat{\left[ H(C),G(C)\right] _{T}^{c}}^{AN}& =\frac{3}{2k_{n}}%
\sum_{i=k_{n}+1}^{[T/\Delta _{n}]-3k_{n}+1}\Bigg(\left( H(\widehat{C}%
_{(i+k_{n})\Delta _{n}})-H(\widehat{C}_{i\Delta _{n}})\right) \left( G(%
\widehat{C}_{(i+k_{n})\Delta _{n}})-G(\widehat{C}_{i\Delta _{n}})\right)
\notag \\
& -\frac{2}{k_{n}}\sum_{g,h,a,b=1}^{d}(\partial _{gh}H\partial _{ab}G)(%
\widehat{C}_{i\Delta _{n}})\left( \widehat{C}_{ga,i\Delta _{n}}\widehat{C}%
_{hb,i\Delta _{n}}+\widehat{C}_{gb,i\Delta _{n}}\widehat{C}_{ha,i\Delta
_{n}}\right) \Bigg)1_{\left\{ A_{i}\cap A_{i+k_{n}}\right\} },
\label{eqn:estAN}
\end{align}%
where the indicator function should only be applied if we are concerned
about volatility jumps, and thus we want to truncate them.\ In the above, we
denote by $A_{i}$ the event of not detecting a volatility jump in the
interval $\left( i\Delta _{n},\left( i+k_{n}\right) \Delta _{n}\right] $,
defined as $A_{i}\equiv \{||\widehat{C}_{\left( i+k_{n}\right) \Delta _{n}}-%
\widehat{C}_{\left( i-k_{n}\right) \Delta _{n}}||<u_{n}^{\prime }\}$, where $%
u_{n}^{\prime }$ is some threshold.

Our second estimator is based on the following equality, which follows by
the It\^{o} lemma,
\begin{equation}
\lbrack H(C),G(C)]_{T}^{c}=\sum_{g,h,a,b=1}^{d}\int_{0}^{T}\left( \partial
_{gh}H\partial _{ab}G\right) (C_{t})\overline{C}_{t}^{gh,ab}dt,
\label{eqn:Ito_on_HCGC}
\end{equation}%
where $\overline{C}_{t}^{gh,ab}$ denotes the continuous covariation between
the volatility processes $C_{gh,t}$ and $C_{ab,t}$. The quantity is thus a
non-linear functional of the spot covariance and spot volatility of
volatility matrices. Our second estimator is a bias-corrected version of the
sample counterpart of the \textquotedblleft linearized\textquotedblright\
expression in (\ref{eqn:Ito_on_HCGC}),\footnote{%
The computation time for any of our two estimators is increasing with the
number of stocks and factors $d$. In practice, we compute all the quantities
of interest for pairs of stocks, so $d_{S}=2$ and thus $d=d_{F}+2$.}
\begin{eqnarray}
&&\widehat{\left[ H(C),G(C)\right] _{T}^{c}}^{LIN}  \notag \\
&=&\frac{3}{2k_{n}}\sum_{g,h,a,b=1}^{d}\sum_{i=k_{n}+1}^{[T/\Delta
_{n}]-3k_{n}+1}(\partial _{gh}H\partial _{ab}G)(\widehat{C}_{i\Delta
_{n}})\left( (\widehat{C}_{gh,(i+k_{n})\Delta _{n}}-\widehat{C}_{gh,i\Delta
_{n}})(\widehat{C}_{ab,(i+k_{n})\Delta _{n}}-\widehat{C}_{ab,i\Delta
_{n}})\right.  \notag \\
&&\left. -\frac{2}{k_{n}}(\widehat{C}_{ga,i\Delta _{n}}\widehat{C}%
_{hb,i\Delta _{n}}+\widehat{C}_{gb,i\Delta _{n}}\widehat{C}_{ha,i\Delta
_{n}})\right) 1_{\left\{ A_{i}\cap A_{i+k_{n}}\right\} }.  \label{eqn:estLIN}
\end{eqnarray}

We now provide the intuition for the bias terms. Suppose volatility is
continuous. If we had observations on $C_{i\Delta _{n}}$, the estimators of $%
[H(C),G(C)]_{T}$ would not need any bias-correction terms. It is useful to
think of $\widehat{C}_{i\Delta _{n}}$ as an estimator of integrated
volatility matrix, $\widehat{C}_{i\Delta _{n}}=\frac{1}{k_{n}\Delta _{n}}%
\int_{i\Delta _{n}}^{\left( i+k_{n}\right) \Delta _{n}}C_{s}ds+U_{i\Delta
_{n}}$, where $U_{i\Delta _{n}}$ is the estimation error.
The first part of the bias-correction in (\ref{eqn:estAN}) and (\ref%
{eqn:estLIN}) is an additive term%
\begin{equation}
-\frac{3}{k_{n}^{2}}\sum_{i=k_{n}+1}^{[T/\Delta _{n}]-3k_{n}+1}\Bigg(%
\sum_{g,h,a,b=1}^{d}(\partial _{gh}H\partial _{ab}G)(\widehat{C}_{i\Delta
_{n}})\Big(\widehat{C}_{ga,i\Delta _{n}}\widehat{C}_{hb,i\Delta _{n}}+%
\widehat{C}_{gb,i\Delta _{n}}\widehat{C}_{ha,i\Delta _{n}}\Big)\Bigg).
\label{eqn:bias_correcton_term}
\end{equation}%
This term arises because of the estimation error $U_{i\Delta _{n}}$.
Intuitively, estimation of, e.g., variance of functionals of $C_{i\Delta
_{n}}$ by variance of functionals of $\widehat{C}_{i\Delta _{n}}$
overestimates it due to the additional variability of $U_{i\Delta _{n}}$. In
particular, one can show that the additive bias-correction term in (\ref%
{eqn:bias_correcton_term}) is, up to a scale factor, an estimator of the
asymptotic covariance between the estimators of $\int_{0}^{T}H(C_{t})dt$ and
$\int_{0}^{T}G(C_{t})dt$.

The second part of the bias-correction in (\ref{eqn:estAN}) and (\ref%
{eqn:estLIN}) is the multiplicative correction factor $3/2$. This correction
factor is needed because of a smoothing bias that arises due to the
replacement of $C_{i\Delta _{n}}$ by $\frac{1}{\Delta _{n}}\int_{i\Delta
_{n}}^{\left( i+k_{n}\right) \Delta _{n}}C_{s}ds$. To gain some intuition,
consider the special case of $d=1$ and $H\left( \cdot \right) =G\left( \cdot
\right) =\cdot \ $. 
Suppose we had observations on $\frac{1}{\Delta _{n}}\int_{i\Delta
_{n}}^{\left( i+k_{n}\right) \Delta _{n}}C_{s}ds$. The $i^{th}$ summand in
the naive estimator of
$\left[ C,C\right] _{T}$ would be
\begin{equation}
\left( \int_{\left( i+k_{n}\right) \Delta _{n}}^{\left( i+2k_{n}\right)
\Delta _{n}}C_{s}ds-\int_{i\Delta _{n}}^{\left( i+k_{n}\right) \Delta
_{n}}C_{s}ds\right) ^{2}=\left( \int_{i\Delta _{n}}^{\left( i+k_{n}\right)
\Delta _{n}}\left( C_{s+\Delta _{n}k_{n}}-C_{s}\right) ds\right) ^{2},
\label{eqn:bias_intuition_additive}
\end{equation}%
divided by $\Delta _{n}^{2}k_{n}^{3}$. Consider the weights that the
integral $\int_{i\Delta _{n}}^{\left( i+k_{n}\right) \Delta _{n}}\left(
C_{s+\Delta _{n}k_{n}}-C_{s}\right) ds$ puts on $\Delta _{n}$\mbox{-}%
increments 
of the volatility $C_{t}$:
these weights are triangular, i.e.,
$\left( \Delta _{n}k_{n}-\left\vert \Delta _{n}k_{n}+i\Delta
_{n}-s\right\vert \right) I\left\{ s\in \left[ i\Delta _{n},\left(
i+2k_{n}\right) \Delta _{n}\right] \right\} $. 
One can show that the squared integral in (\ref{eqn:bias_intuition_additive}%
) is proportional to the integral of the squared triangular weights, $\frac{1%
}{\left( \Delta _{n}k_{n}\right) ^{3}}\int_{i\Delta _{n}}^{\left(
2k_{n}+i\right) \Delta _{n}}\left( \Delta _{n}k_{n}-\left\vert \Delta
_{n}k_{n}+i\Delta _{n}-s\right\vert \right) ^{2}ds$. The latter integral
equals $\frac{2}{3}$, hence the estimator needs a multiplicative correction
factor $\frac{3}{2}$.%

When $H(\cdot )=G(\cdot )$, the estimand is nonnegative, $\left[ H(C),G(C)%
\right] _{T}^{c}\geq 0$, so our estimators are nonnegative in large samples.
However, due to the presence of an additive bias-correction, our estimators
are not guaranteed to be nonnegative in finite samples. We remark that \cite%
{vetter-vovo} constructs a univariate volatility of volatility estimator
that is guaranteed to be nonnegative, at the cost of a slower rate of
convergence.%

Our two estimators, AN in equation (\ref{eqn:estAN}) and LIN in (\ref%
{eqn:estLIN}), are identical when $H$ and $G$ are linear, for example, when
estimating the covariation between two volatility processes. In the
univariate case $d=1$, when $H(\cdot )=G(\cdot )=\cdot $\ , and when one
assumes no price or volatility jumps and omits the price and volatility jump
truncation, both of our estimators coincide with the volatility of
volatility estimator of \cite{vetter-vovo}. 

While \cite{jacodrosenbaum-sqrtn} focus on a different problem, one of the
asymptotic bias terms in their paper is of the form $\left[ H(C),H(C)\right]
_{T}^{c}$. In the special case $H(\cdot )=G(\cdot )$, aside from a scale
factor, the end-effects, and the form of the volatility jump truncation, our
LIN estimator in equation (\ref{eqn:estLIN}) coincides with their estimator.
Our approach to volatility jumps differs as we truncate these jump from
below, while \cite{jacodrosenbaum-sqrtn} truncate from above, and we use a
simpler form of truncation that in finite samples is robust to consecutive
volatility jumps. \cite{jacodrosenbaum-sqrtn} only establish consistency of
the relevant estimator, and do not provide any asymptotic distribution
theory. In contrast, we derive the asymptotic distribution of the estimators
of $\left[ H(C),G(C)\right] _{T}^{c}$, and provide a consistent estimator of
the asymptotic variance.%

\subsection{Estimation in R-FM and IdioVol-FM models}

\label{sec:estimation_special_case}

In this section, we explain how to use the formulas in equations (\ref{eqn:estAN}%
) and (\ref{eqn:estLIN}) to obtain estimators for the objects of interest in
Section \ref{sec:model}, see equations (\ref{eqn:IdioVol-FM})--(\ref%
{eqn:R2_bivar}). In particular, each of these objects of interest,
\begin{equation}
\begin{split}
& [C_{Zj},C_{Zs}]_{T}^{c},~Corr\left( C_{Zj},C_{Zs}\right) ,~\gamma
_{Z_{j}},~[C_{Zj}^{resid},C_{Zs}^{resid}]_{T}^{c}, \\
& ~Corr\left( C_{Zj}^{resid},C_{Zs}^{resid}\right) ,~Q_{Zj,Zs}^{\txtIdioVolFM%
}\hspace{0.5mm}\text{,}\hspace{2mm}\text{and}\hspace{2mm}R_{Zj}^{2,%
\txtIdioVolFM},
\end{split}
\label{eqn:list_of_estimands}
\end{equation}%
for $j,s=1,\ldots ,d_{S}$, can be written as
\begin{equation}
\varphi \left( \left[ H_{1}(C),G_{1}(C)\right] _{T}^{c},\ldots ,\left[
H_{\kappa }(C),G_{\kappa }(C)\right] _{T}^{c}\right) ,
\label{eqn:general_functional}
\end{equation}%
for some smooth, real-valued functions $\varphi $, $H_{r}$, $G_{r}$, $%
r=1,\ldots ,\kappa $. Each element in (\ref{eqn:general_functional}) is of
the form $[H_{r}(C),G_{r}(C)]_{T}^{c}$, i.e., it is the continuous part of a
quadratic covariation between functions of $C_{t}$, and hence can be
estimated using the estimators proposed in Section \ref%
{sec:estimation_general_functional}.

Consider the first quantity in equation (\ref{eqn:list_of_estimands}),
which is the continuous part of the quadratic covariation between $j^{th}$
and $s^{th}$ IdioVol, $[C_{Zj},C_{Zs}]_{T}^{c}$. 
By equation (\ref{eqn:IdioVol_ID}), $C_{Z\ell }=C_{Y\ell
,t}-(C_{FS\ell ,t})^{\top }(C_{F,t})^{-1}C_{FS\ell ,t}$, and the quantity is
of the form $[C_{Zj},C_{Zs}]_{T}^{c}=\left[ H\left( C_{t}\right) ,G\left(
C_{t}\right) \right] _{T}^{c}$, where%
\begin{eqnarray*}
H\left( C_{t}\right)  &=&C_{Yj,t}-(C_{FSj,t})^{\top }(C_{F,t})^{-1}C_{FSj,t}
\\
G\left( C_{t}\right)  &=&C_{Ys,t}-(C_{FSs,t})^{\top }(C_{F,t})^{-1}C_{FSs,t}.
\end{eqnarray*}%
As per equation (\ref{eqn:corrivs}), $Corr\left( C_{Zj},C_{Zs}\right) $ is also of
the form of equation~(\ref{eqn:general_functional}).

Next, note that IdioVol-FM implies
\begin{eqnarray}
\gamma _{Zj} &=&\left( [\Pi ,\Pi ]_{T}^{c}\right) ^{-1}[\Pi ,C_{Zj}]_{T}^{c},%
\text{\ \ \ and}  \label{eqn:ID_gamma} \\
{}[C_{Zj}^{resid},C_{Zs}^{resid}]^c_{T} &=&[C_{Zj},C_{Zs}]_{T}^{c}-\gamma
_{Zj}^{\top }[\Pi ,\Pi ]_{T}^{c}\gamma _{Zs}  \label{eqn:ID_QC_C_resid}
\end{eqnarray}
for $j,s=1,\ldots ,d_{S}$. Recall that $C_{Zj,t}$, $C_{Zs,t}$, and every
element of $\Pi _{t}$ are given real-valued functions of $C_{t}$. 
For example, if volatility factors are the volatilities of
return factors $F_t$, we have $\Pi \left( C_{t}\right) =C_{F,t}$, so $\Pi
\left( \cdot \right) $ selects the last $d_{F}$ diagonal elements from $%
C_{t}$ (recall that $F_{t}$ are the last $d_{F}$ elements of vector $Y_t$).
 Thus, the
right-hand-sides of (\ref{eqn:ID_gamma}) and (\ref{eqn:ID_QC_C_resid}) have
the form of equation (\ref{eqn:general_functional}) for a finite number of
quantities of the form $[H_{r}(C),G_{r}(C)]_{T}^{c}$.

Finally, the remaining quantities in equation (\ref{eqn:list_of_estimands}), $Corr\left(C^{resid}_{Zj},C^{resid}_{Zs} \right)$, $Q^{\txtIdioVolFM%
}_{Zj,Zs}$ and $R^{2,\txtIdioVolFM}_{Zj}$, are smooth functions of $%
[C_{Zj}^{resid}, C_{Zj}^{resid}]^c_T$, $[C_{Zj}, C_{Zs}]^c_T$, $\gamma_{Zj}$%
, and $[\Pi,\Pi]^c_T$, each of which is of the form of equation (\ref%
{eqn:general_functional}), and hence are themselves of the form of equation (%
\ref{eqn:general_functional}).

\section{Asymptotic Properties}

\label{sec:AsyProperties}

In this section, we first present the full list of assumptions for our
asymptotic results. We then obtain the joint asymptotic distribution between
the general functionals $[H_r(C),G_r(C)]^c_T$ for $r=1,\ldots,\kappa$
introduced in Section \ref{sec:estimation_general_functional}. We also
develop estimators for the asymptotic variance-covariance matrix. The
asymptotic distributions of the estimators of $Corr\left(C_{Zi},C_{Zj}%
\right) $ and other quantities of interest in Section \ref{sec:model} follow
by the Delta method (see Section \ref{sec:estimation_special_case} for
details). Finally, to illustrate the application of the general theory, we
describe three statistical tests about the IdioVols, which we later
implement in the empirical and Monte Carlo analysis.

\subsection{Assumptions}

\label{sec:Assumptions}

Recall that the $d$-dimensional process $Y_t$ represents the (log) prices of
stocks, $S_t$, and factors $F_t$.

\begin{assumption}
\label{ass:itoprice} Suppose $Y$ is an It\^{o} semimartingale on a filtered
space $(\Omega ,\mathcal{F},(\mathcal{F}_{t})_{t\geq 0},\mathbb{P})$,
\begin{equation}
Y_{t}=Y_{0}+\int_{0}^{t}b_{s}ds+\int_{0}^{t}\sigma
_{s}dW_{s}+\int_{0}^{t}\int_{E}\delta (s,z)\mu (ds,dz),
\end{equation}%
where $W$ is a $d^{W}$-dimensional Brownian motion ($d^{W}\geq d$) and $\mu $
is a Poisson random measure on $\mathbb{R}_{+}\times E$, with $E$ an
auxiliary Polish space with intensity measure $\nu (dt,dz)=dt\otimes \lambda
(dz)$ for some $\sigma $-finite measure $\lambda $ on $E$. The process $%
b_{t} $ is $\mathbb{R}^{d}$-valued optional, $\sigma _{t}$ is $\mathbb{R}%
^{d}\times \mathbb{R}^{d^{W}}$-valued, and $\delta =\delta (w,t,z)$ is a
predictable $\mathbb{R}^{d}$ -valued function on $\Omega \times \mathbb{R}%
_{+}\times E$. Moreover, $\Vert \delta (w,t\wedge {\tau _{m}(w)},z)\Vert
\wedge 1\leq \Gamma _{m}(z)$, for all (w,t,z), where ($\tau _{m}$) is a
localizing sequence of stopping times and, for some $r\in \lbrack 0,1/2)$,
the function $\Gamma _{m}$ on $E$ satisfies $\int_{E}\Gamma
_{m}(z)^{r}\lambda (dz)<\infty $. The spot volatility matrix of $Y$ is then
defined as $C_{t}=\sigma _{t}\sigma _{t}^{\top }$. We assume that $C_{t}$ is
an It\^{o} semimartingale,\footnote{%
Note that $\widetilde{\sigma }_{s}=(\widetilde{\sigma }_{s}^{gh,m})$ is $%
(d\times d\times d^{W})$-dimensional and $\widetilde{\sigma }_{s}dW_{s}$ is $%
(d\times d)$-dimensional with $(\widetilde{\sigma }_{s}dW_{s})^{gh}=%
\sum_{m=1}^{d^{W}}\widetilde{\sigma }_{s}^{gh,m}dW_{s}^{m}$.}
\begin{equation}
C_{t}=C_{0}+\int_{0}^{t}\widetilde{b}_{s}ds+\int_{0}^{t}\widetilde{\sigma }%
_{s}dW_{s}+J_{t}^{\sigma },  \label{eqn:LPP}
\end{equation}%
where $\widetilde{b}$ is $\mathbb{R}^{d}\times \mathbb{R}^{d}$-valued
optional, and $J_{t}^{\sigma }$ is a finite activity jump process. $C_{t}$
takes values in the space $\mathcal{M}_{d}$ consisting of $d\times d$
positive definite matrices. For a sequence of convex compact subsets $(%
\mathcal{K}_{m})_{m\geq 1}$ of $\mathcal{M}_{d}$, $C_{t}\in \mathcal{K}_{m}$
for all $t\leq \tau _{m}$.
\end{assumption}

With the above notation, the elements of the spot volatility of volatility
matrix and spot covariation of the continuous martingale parts of $X$ and $c$
are defined as follows,
\begin{equation}
\overline{C}_{t}^{gh,ab}=\sum_{m=1}^{d^{W}}\widetilde{\sigma }_{t}^{gh,m}%
\widetilde{\sigma }_{t}^{ab,m},\hspace{2mm}\overline{C}_{t}^{\prime
g,ab}=\sum_{m=1}^{d^{W}}\sigma _{t}^{gm}\widetilde{\sigma }_{t}^{ab,m}.
\label{eqn:SVP}
\end{equation}%
We assume the following for the process $\widetilde{\sigma }_{t}$:

\begin{assumption}
\label{ass:volvol} $\widetilde{\sigma }_{t}$ is a continuous It\^{o}
semimartingale with its characteristics satisfying the same requirements as
that of $C_{t}-J_{t}^{\sigma }$.
\end{assumption}

\indent Assumption \ref{ass:itoprice} is very general and nests most of the
multivariate continuous-time models used in economics and finance. It allows
for potential stochastic volatility and jumps in returns. Assumption \ref%
{ass:volvol} is required to obtain the asymptotic distribution of estimators
of the quadratic covariation between functionals of the spot covariance
matrix $C_t$. It is not needed to prove consistency. This assumption also
appears in \cite{WangMykland12}, \cite{vetter-vovo}, and \cite%
{kalninaxiu-lev}.%

\subsection{Asymptotic Distribution}

We have seen in Section \ref{sec:estimation} that all quantities of interest
in (\ref{eqn:list_of_estimands}) are functions of multiple objects of the
form $[H(C),G(C)]^c_T$. Therefore, if we can obtain a multivariate
asymptotic distribution for a vector with elements of the form $%
[H(C),G(C)]^c_T$, the asymptotic distributions for all our estimators follow
by the Delta method. The current section presents this asymptotic
distribution.

Let $H_{1},G_{1},\ldots ,H_{\kappa },G_{\kappa }$ be given smooth
real-valued functions. We are interested in the asymptotic behavior of
vectors
\begin{equation}
\begin{aligned}
&\Big(\widehat{[H_1(C),G_1(C)]^c_T}^{AN},\ldots,\widehat{[H_\kappa(C),G_%
\kappa(C)]^c_T}^{AN}\Big)^\top \text{ and } \\ &
\Big(\widehat{[H_1(C),G_1(C)]^c_T}^{LIN},\ldots,\widehat{[H_\kappa(C),G_%
\kappa(C)]^c_T}^{LIN}\Big)^\top. \end{aligned}
\end{equation}
The following theorem summarizes the joint asymptotic behavior of the
estimators.

\begin{theorem}
\label{thm:clt} Let $\widehat{[H_{r}(C),G_{r}(C)]_{T}^{c}}$ denote either $%
\widehat{[H_{r}(C),G_{r}(C)]_{T}^{c}}^{AN}$ or $\widehat{%
[H_{r}(C),G_{r}(C)]_{T}^{c}}^{LIN}$ defined in equations (\ref{eqn:estAN})
and (\ref{eqn:estLIN}), where $H_{r}$ and $G_{r}$ are three times
differentiable real-valued functions, for $r=1,\ldots ,\kappa $. 
Suppose Assumptions \ref{ass:itoprice} and \ref{ass:volvol} hold. 
Fix $k_{n}=\theta \Delta _{n}^{-1/2}$ for some $\theta \in (0,\infty )$. 
Set $u_{n}\asymp \Delta _{n}^{\varpi }$ 
with 
$\frac{2\varpi^{\prime }+9}{4\left( 5-r\right) }
<
 \varpi
  <\frac{1}{2}$, and $%
u_{n}^{\prime }\asymp \Delta _{n}^{\varpi ^{\prime }}$ 
with
$0 < \varpi ^{\prime }<\min \left( \frac{1}{2}-r,\frac{1}{8}\right) $%
. Then, as $%
\Delta _{n}\rightarrow 0$,
\begin{equation}
\Delta _{n}^{-1/4}\left(
\begin{array}{c}
\widehat{\lbrack H_{1}(C),G_{1}(C)]_{T}^{c}}-[H_{1}(C),G_{1}(C)]_{T}^{c} \\
\ldots \\
\widehat{\lbrack H_{\kappa }(C),G_{\kappa }(C)]_{T}^{c}}-[H_{\kappa
}(C),G_{\kappa }(C)]_{T}^{c}%
\end{array}%
\right) \overset{\text{L-s}}{\longrightarrow }MN(0,\Sigma _{T}).
\end{equation}
Let $\Sigma _{T}^{r,s}$ be the $\left( \Sigma _{T}\right) _{r,s}$ element of
the $\kappa \times \kappa $ matrix $\Sigma _{T}$. We have
\begin{align*}
& \Sigma _{T}^{r,s}=\Sigma _{T}^{r,s,(1)}+\Sigma _{T}^{r,s,(2)}+\Sigma
_{T}^{r,s,(3)}, \\
& \Sigma _{T}^{r,s,(1)}=\frac{6}{\theta ^{3}}\sum_{g,h,a,b=1}^{d}%
\sum_{j,k,l,m=1}^{d}\int_{0}^{T}\big(\partial _{gh}H_{r}\partial
_{ab}G_{r}\partial _{jk}H_{s}\partial _{lm}G_{s}(C_{s})\big)\Big[%
C_{t}(gh,jk)C_{t}(ab,lm) \\
& \hspace{30mm}+C_{t}(ab,jk)C_{t}(gh,lm)\Big]dt, \\
& \Sigma _{T}^{r,s,(2)}=\frac{151\theta }{140}\sum_{g,h,a,b=1}^{d}%
\sum_{j,k,l,m=1}^{d}\int_{0}^{T}\big(\partial _{gh}H_{r}\partial
_{ab}G_{r}\partial _{jk}H_{s}\partial _{lm}G_{s}(C_{t})\big)\Big[\overline{C}%
_{t}^{gh,jk}\overline{C}_{t}^{ab,lm} \\
& \hspace{30mm}+\overline{C}_{t}^{ab,jk}\overline{C}_{t}^{gh,lm}\Big]dt, \\
& \Sigma _{T}^{r,s,(3)}=\frac{3}{2\theta }\sum_{g,h,a,b=1}^{d}%
\sum_{j,k,l,m=1}^{d}\int_{0}^{T}\big(\partial _{gh}H_{r}\partial
_{ab}G_{r}\partial _{jk}H_{s}\partial _{lm}G_{s}(C_{t})\big)\Big[C_{t}(gh,jk)%
\overline{C}_{t}^{ab,lm} \\
& \hspace{30mm}+C_{t}(ab,lm)\overline{C}_{t}^{gh,jk}+C_{t}(gh,lm)\overline{C}%
_{t}^{ab,jk}+C_{t}(ab,jk)\overline{C}_{t}^{gh,lm}\Big]dt,
\end{align*}%
with
\begin{equation*}
C_{t}(gh,jk)=C_{gj,t}C_{hk,t}+C_{gk,t}C_{hj,t}.
\end{equation*}
\end{theorem}

The convergence in Theorem \ref{thm:clt} is stable in law (denoted $L$-$s$,
see for example \cite{AldousEagle78} and \cite{jacodprotter2012}). The limit
is mixed gaussian and the precision of the estimators depends on the paths
of the spot covariance and the volatility of volatility process. The rate of
convergence $\Delta_n^{-1/4}$ has been shown to be the optimal for
volatility of volatility estimation (in the absence of volatility jumps).

The asymptotic variance of the estimators depends on the tuning parameter $%
\theta$ whose choice may be crucial for the reliability of the inference. We
document the sensitivity of the inference theory to the choice of the
parameter $\theta$ in a Monte Carlo experiment (see Section \ref{sec:MC}).

\subsection{Estimation of the Asymptotic Covariance Matrix}

To provide a consistent estimator for the element $\Sigma _{T}^{r,s}$ of the
asymptotic covariance matrix in Theorem \ref{thm:clt}, we introduce the
following quantities:
\begin{align*}
\widehat{\Omega }_{T}^{r,s,(1)}& =\Delta
_{n}\sum_{g,h,a,b=1}^{d}\sum_{j,k,l,m=1}^{d}\sum_{i=k_{n}+1}^{[T/\Delta
_{n}]-5k_{n}+1}\left( \partial _{gh}H_{r}\partial _{ab}G_{r}\partial
_{jk}H_{s}\partial _{lm}G_{s}(\widehat{C}_{i\Delta _{n}})\right) \\
& \times \left[ \widetilde{C}_{i\Delta _{n}}(gh,jk)\widetilde{C}_{i\Delta
_{n}}(ab,lm)+\widetilde{C}_{i\Delta _{n}}(ab,jk)\widetilde{C}_{i\Delta
_{n}}(gh,lm)\right] 1_{\cap _{j=0}^{3}A_{i+jk_{n}}}, \\
\widehat{\Omega }_{T}^{r,s,(2)}&
=\sum_{g,h,a,b=1}^{d}\sum_{j,k,l,m=1}^{d}\sum_{i=k_{n}+1}^{[T/\Delta
_{n}]-5k_{n}+1}\left( \partial _{gh}H_{r}\partial _{ab}G_{r}\partial
_{jk}H_{s}\partial _{lm}G_{s}(\widehat{C}_{i\Delta _{n}})\right) \left[
\frac{1}{2}\widehat{\lambda }_{i}^{n,gh}\widehat{\lambda }_{i}^{n,jk}%
\widehat{\lambda }_{i+2k_{n}}^{n,ab}\widehat{\lambda }_{i+2k_{n}}^{n,lm}%
\right. \\
& +\frac{1}{2}\widehat{\lambda }_{i}^{n,ab}\widehat{\lambda }_{i}^{n,lm}%
\widehat{\lambda }_{i+2k_{n}}^{n,gh}\widehat{\lambda }_{i+2k_{n}}^{n,jk}+%
\frac{1}{2}\widehat{\lambda }_{i}^{n,ab}\widehat{\lambda }_{i}^{n,jk}%
\widehat{\lambda }_{i+2k_{n}}^{n,gh}\widehat{\lambda }_{i+2k_{n}}^{n,lm}+%
\left. \frac{1}{2}\widehat{\lambda }_{i}^{n,gh}\widehat{\lambda }_{i}^{n,lm}%
\widehat{\lambda }_{i+2k_{n}}^{n,ab}\widehat{\lambda }_{i+2k_{n}}^{n,jk}%
\right] 1_{\cap _{j=0}^{3}A_{i+jk_{n}}}, \\
\widehat{\Omega }_{T}^{r,s,(3)}& =\frac{3}{2k_{n}}\sum_{g,h,a,b=1}^{d}%
\sum_{j,k,l,m=1}^{d}\sum_{i=k_{n}+1}^{[T/\Delta _{n}]-5k_{n}+1}\left(
\partial _{gh}H_{r}\partial _{ab}G_{r}\partial _{jk}H_{s}\partial _{lm}G_{s}(%
\widehat{C}_{i\Delta _{n}})\right) \\
& \left[ \widetilde{C}_{i\Delta _{n}}(gh,jk)\widehat{\lambda }_{i}^{n,ab}%
\widehat{\lambda }_{i}^{n,lm}+\widetilde{C}_{i\Delta _{n}}(ab,lm)\widehat{%
\lambda }_{i}^{n,gh}\widehat{\lambda }_{i}^{n,jk}\right. \\
& \left. +\widetilde{C}_{i\Delta _{n}}(gh,lm)\widehat{\lambda }_{i}^{n,ab}%
\widehat{\lambda }_{i}^{n,jk}+(\widetilde{C}_{i\Delta _{n}}(ab,jk)\widehat{%
\lambda }_{i}^{n,gh}\widehat{\lambda }_{i}^{n,lm}\right] 1_{\cap
_{j=0}^{3}A_{i+jk_{n}}}.
\end{align*}%
with 
$\widehat{\lambda }_{i}^{n,jk}=\widehat{C}_{i+k_{n}}^{n,jk}-\widehat{C}%
_{i}^{n,jk}$, 
$\widetilde{C}_{i\Delta _{n}}(gh,jk)=
\widehat{C}_{gj,i\Delta_{n}}\widehat{C}_{hk,i\Delta _{n}}
+
\widehat{C}_{gk,i\Delta _{n}}\widehat{C}_{hj,i\Delta _{n}}$, 
and  
$A_{i}=\{||\widehat{C}_{\left( i+k_{n}\right) \Delta _{n}}-\widehat{C}_{\left( i-k_{n}\right) \Delta _{n}}||<u_{n}^{\prime }\}  $.

The following result holds,

\begin{theorem}
\label{thm:avar} Suppose the assumptions of Theorem \ref{thm:clt} hold.
Then, as $\Delta _{n}\rightarrow 0$,
\begin{align}
& \frac{6}{\theta ^{3}}\widehat{\Omega }_{T}^{r,s,(1)}\overset{\mathbb{P}}{%
\longrightarrow }\Sigma _{T}^{r,s,(1)},  \label{eqn:avar1} \\
& \frac{3}{2\theta }[\widehat{\Omega }_{T}^{r,s,(3)}-\frac{6}{\theta }%
\widehat{\Omega }_{T}^{r,s,(1)}]\overset{\mathbb{P}}{\longrightarrow }\Sigma
_{T}^{r,s,(3)},\text{ and}  \label{eqn:avar2} \\
& \frac{151\theta }{140}\frac{9}{4\theta ^{2}}[\widehat{\Omega }%
_{T}^{r,s,(2)}+\frac{4}{\theta ^{2}}\widehat{\Omega }_{T}^{r,s,(1)}-\frac{4}{%
3}\widehat{\Omega }_{T}^{r,s,(3)}]\overset{\mathbb{P}}{\longrightarrow }%
\Sigma _{T}^{r,s,(2)}.  \label{eqn:avar3}
\end{align}
\end{theorem}

The estimated matrix $\widehat{\Sigma}_T$ is symmetric but is not guaranteed
to be positive semi-definite. By Theorem \ref{thm:clt}, $\widehat{\Sigma}_T$
is positive semi-definite in large samples.
An interesting question is the estimation of the asymptotic variance using
subsampling or bootstrap methods, see \cite{kalnina11-sub,kalnina23-multisub}%
, and we leave it for future research.

\textbf{Remark 1:} The rate of convergence in equation (\ref{eqn:avar1}) can
be shown to be $\Delta_n^{-1/2}$, and the rate of convergence in (\ref%
{eqn:avar2}) and (\ref{eqn:avar3}) can be shown to be $\Delta_n^{-1/4}$.

\textbf{Remark 2:} In the one-dimensional case ($d=1$), much simpler
estimators of $\Sigma_T^{r,s,(2)}$ can be constructed using the quantities $%
\widehat{\lambda}_i^{n,jk}\widehat{\lambda}_i^{n,lm}\widehat{\lambda}%
_{i+k_n}^{n,gh}\widehat{\lambda}_{i+k_n}^{n,xy}$ or $\widehat{\lambda}%
_i^{n,jk}\widehat{\lambda}_i^{n,lm}\widehat{\lambda}_{i}^{n,gh}\widehat{%
\lambda}_{i}^{n,xy}$ as in \cite{vetter-vovo}. However, in the
multidimensional case, the latter quantities do not identify separately the
quantity $\overline{C_t}^{jk,lm}\overline{C_t}^{gh,xy}$ since the
combination $\overline{C_t}^{jk,lm}\overline{C_t}^{gh,xy}+\overline{C_t}%
^{jk,gh}\overline{C_t}^{lm,xy}+\overline{C_t}^{jk,xy}\overline{C_t}^{gh,lm}$
shows up in a non-trivial way in the limit of the estimator.

\begin{corollary}
\label{corr:r1}
Let $\widehat{[H_{r}(C),G_{r}(C)]_{T}^{c}}$ denote either $\widehat{%
[H_{r}(C),G_{r}(C)]_{T}^{c}}^{AN}$ or $\widehat{[H_{r}(C),G_{r}(C)]_{T}^{c}}%
^{LIN}$ defined in equations (\ref{eqn:estAN}) and (\ref{eqn:estLIN}).
Suppose the assumptions of Theorem \ref{thm:clt} hold. Then, as $\Delta
_{n}\rightarrow 0$,
\begin{equation}
\Delta _{n}^{-1/4}\ \widehat{\Sigma }_{T}^{-1/2}\left(
\begin{array}{l}
\widehat{\lbrack H_{1}(C),G_{1}(C)]_{T}^{c}}-[H_{1}(C),G_{1}(C)]_{T}^{c} \\
\hspace{25mm}\vdots \\
\widehat{\lbrack H_{\kappa }(C),G_{\kappa }(C)]_{T}^{c}}-[H_{\kappa
}(C),G_{\kappa }(C)]_{T}^{c}%
\end{array}%
\right) \overset{L}{\longrightarrow }N(0,I_{\kappa }).
\end{equation}

\end{corollary}

In the above, we use $L$ to denote the convergence in distribution and $%
I_{\kappa}$ the identity matrix of order $\kappa$. Corollary \ref{corr:r1}
states the standardized asymptotic distribution, which follows directly from
the properties of the stable-in-law convergence. Similarly, by the Delta
method, standardized asymptotic distribution can also be derived for the
estimators of the quantities in (\ref{eqn:list_of_estimands}). These
standardized distributions allow the construction of confidence intervals
for all the latent quantities of the form $[H_r(C),G_r(C)]^c_T$ and, more
generally, functions of these quantities.

\subsection{Tests}

\label{sec:testing}

As an illustration of application of the general theory, we provide three
tests about the dependence of Idiosyncratic Volatility. Our framework allows
to test general hypotheses about the joint dynamics of any subset of the
available stocks. The three examples below are stated for one pair of
stocks, and correspond to the tests we implement in the empirical and Monte
Carlo studies.

First, one can test for the absence of dependence between the continuous
components of the IdioVols of the returns on assets $j$ and $s$,
\begin{equation}
H_{0}^{1}:[C_{Zj},C_{Zs}]_{T}^{c}=0.  \label{eqn:H0_no_IV_dependence}
\end{equation}%
Under $H_{0}^{1}$, $\Delta _{n}^{-1/4}\widehat{[C_{Zj},C_{Zs}]_{T}^{c}}%
\widehat{V}^{-1/2}\overset{L}{\rightarrow }N\left( 0,1\right) $, so we can
use a t-test.

Second, we can test the hypothesis that none of the IdioVol factors $\Pi $
explaining the dynamics of IdioVol shocks of stock $j$,
\begin{equation}
H_{0}^{2}:[C_{Zj},\Pi ]_{T}^{c}=0  \label{eqn:H0_IdioVol_factor_irrelevant}
\end{equation}%
Under this null hypothesis, the vector of IdioVol factor loadings equals
zero, $\gamma _{Z_{j}}=0$. Under $H_{0}^{2}$,
\begin{equation}
\Delta _{n}^{-1/4}\left( \widehat{[C_{Zj},\Pi ]_{T}^{c}}\right) ^{\top
}\left( \widehat{V}\right) ^{-1}\widehat{[C_{Zj},\Pi ]}_{T}^{c}\overset{L}{%
\rightarrow }\mathcal{X}_{d_{\Pi },1-\alpha }^{2},
\label{eqn:test_IdioVolfactor}
\end{equation}%
so we can use a Wald test. One can of course also construct a t-test for
irrelevance of any one particular IdioVol factor. The final example is a
test for absence of dependence between the residual IdioVols of stock $j$
and $s$,
\begin{equation}
H_{0}^{3}:[C_{Zj}^{resid},C_{Zs}^{resid}]_{T}^{c}=0.
\label{eqn:H0:no_resid-IV_dependence}
\end{equation}
Under $H_{0}^{1}$, $\Delta _{n}^{-1/4}\widehat{%
[C_{Zj}^{resid},C_{Zs}^{resid}]_{T}^{c}}\widehat{V}^{-1/2}\overset{L}{%
\rightarrow }N\left( 0,1\right) $, so we can use a t-test.

Each of the above estimators
\begin{equation*}
\widehat{\lbrack C_{Zj},C_{Zs}]_{T}^{c}},\widehat{[C_{Zj},\Pi ]_{T}^{c}},%
\text{ and }\widehat{[C_{Zj}^{resid},C_{Zs}^{resid}]_{T}^{c}}
\end{equation*}%
can be obtained by choosing appropriate pair(s) of transformations $H$ and $%
G $ in the general estimator $\widehat{[H(C),G(C)]_{T}^{c}}$, see Section %
\ref{sec:estimation} for details. Any of the two types of the latter
estimator can be used,
\begin{equation*}
\widehat{\lbrack H(C),G(C)]_{T}^{c}}^{AN}\text{ \ or \ }\widehat{%
[H(C),G(C)]_{T}^{c}}^{LIN}.
\end{equation*}
For the first two tests, the expression for the true asymptotic variance, $V$%
, is obtained using Theorem \ref{thm:clt} and its estimation follows from
Theorem \ref{thm:avar}. The asymptotic variance in the third test is
obtained by applying the Delta method to the joint convergence result in
Theorem \ref{thm:clt}. The expression for the estimator of the asymptotic
variance, $\widehat{V}$, follows from Theorem \ref{thm:avar}.
Under R-FM and the assumptions of Theorem \ref{thm:clt}, Corollary \ref%
{corr:r1} implies that the asymptotic size of the two types of tests for the
null hypotheses $H_{0}^{1}$ and $H_{0}^{2}$ is $\alpha $, and their power
approaches 1. The same properties apply for the tests of the null hypotheses
$H_{0}^{3}$ with our R-FM and IdioVol-FM representations.

Theoretically, it is possible to test for absence of dependence in the
IdioVols at each point in time. In this case the null hypothesis is $%
H_{0}^{1\prime }:[C_{Zj},C_{Zs}]_{t}^{c}=0\hspace{3mm}\text{for all}\hspace{%
3mm}0\leq t\leq T$, which is, in theory, stronger than our $H_{0}^{1\prime }$%
. In particular, Theorem \ref{thm:clt} can be used to set up
Kolmogorov-Smirnov type of tests for $H_{0}^{\prime 1}$ in the same spirit
as \cite{vetter-vovo}. However, we do not pursue this direction in the
current paper for two reasons. First, the testing procedure would be more
involved. Second, empirical evidence suggests nonnegative dependence between
IdioVols, which means that in practice, it is not too restrictive to assume $%
[C_{Zj},C_{Zs}]_{t}^{c}\geq 0~\forall t$, under which $H_{0}^{1}$ and $%
H_{0}^{1\prime }$ are equivalent.

\section{Empirical Analysis}

\label{sec:Empirical}

We apply our methods to study the cross-sectional dependence in IdioVols
using high frequency data. One of our main findings is that stocks' IdioVols
co-move strongly with the market volatility. This is a quite surprising
finding. It is of course well known that the total volatility of stocks
moves with the market volatility. However, we stress that we find that the
strong effect is still present when considering the IdioVols.

We use transaction prices from NYSE TAQ database for S\&P 100 index
constituents from 2003 to 2012. Starting with the union of constituents over
this period, we select only those stocks for which complete data is
available; this results in a full sample of 104 stocks. After excluding the
non-trading days, our sample contains 2517 days. We also use the
high-frequency data on nine industry Exchange-Traded Funds, ETFs (Consumer
Discretionary, Consumer Staples, Energy, Financial, Health Care, Industrial,
Materials, Technology, and Utilities), and the high-frequency size and value
Fama-French factors, see \cite{yackalninaxiu-FF}. To aid visualization, we
report additional results for a subset of 30 stocks. We obtain the subset of
30 stocks by selecting at least two stocks from each of the nine GICS
sectors, together with the most liquid stocks; see Table \ref{tbl:stocks}
for details. For each day, we consider data from the regular exchange
opening hours from time stamped between 9:30 a.m. until 4 p.m.

We clean the data following the procedure suggested by \cite%
{barndorffnielsenhansenlundeshephard08}, remove the overnight returns and
then sample at 5 minutes. This sparse sampling has been widely used in the
literature because the effect of the microstructure noise and potential
asynchronicity of the data is less important at this frequency, see also
\cite{LiuPattonSheppard15}. The return jump truncation threshold is the same
as in simulations, see Section \ref{sec:MC}. The number of observations in
the local window is taken as in Theorem \ref{thm:clt} to be $k_{n}=\theta
\Delta _{n}^{-1/2}$. We take $\theta =2.5$ and $\Delta _{n}=1/252/(6.5\times
12)$, i.e., $\Delta _{n}$ is 5 minutes (with one year being a unit of time),
which corresponds to the local window of approximately one week. The
threshold for volatility jumps is based on the individual asset volatility
changing by more than 10 percentage points. The optimal selection of this
tuning parameter is a complex issue that falls outside the scope of this
paper. We find that both
types of estimators, AN and LIN, produce very similar results and report
only the AN estimator for brevity.

To obtain the Idiosyncratic Volatilities, the preliminary step is to
estimate the Return Factor Model (R-FM) for each stock. Figures \ref%
{fig:R2_a} and \ref{fig:R2_b} contain plots of the time series of the
estimated $R_{Yj}^{2}$ of the R-FM for the subset of 30 stocks.\footnote{%
For the $j^{th}$ stock, our analog of the coefficient of determination in
the R-FM is $R_{Yj}^{2}=1-\frac{\int_{0}^{T}C_{Zj,t}dt}{%
\int_{0}^{T}C_{Yj,t}dt}$. 
We estimate $R_{Yj}^{2}$ using the general method of \cite{jacodrosenbaum13}%
. The resulting estimator of $R_{Yj}^{2}$ requires a choice of a block size
for the spot volatility estimation; we choose two hours in practice (the
number of observations in a block, say $l_{n}$, has to satisfy $%
l_{n}^{2}\Delta _{n}\rightarrow 0$ and $l_{n}^{3}\Delta _{n}\rightarrow
\infty $, so it is of smaller order than the number of observations $k_{n}$
in our estimators of Section \ref{sec:estimation}).} Each plot contains
monthly $R_{Yj}^{2}$ from two Return Factor Models, CAPM and the Fama-French
regression with market, size, and value factors.
Figures \ref{fig:R2_a} and \ref{fig:R2_b} show that these time series of all
stocks follow approximately the same trend with a considerable increase in
the contribution around the crisis year 2008. Higher $R_{Yj}^{2}$ indicates
that the systematic risk is relatively more important, which is typical
during crises. $R_{Yj}^{2}$ is consistently higher in the Fama-French
regression model compared to the CAPM regression model, albeit not by much.
We proceed to investigate the dynamic properties of the panel of
Idiosyncratic Volatilities.

\begin{figure}[!h]
\centering
\includegraphics[trim = 0mm 0mm 0mm
5mm,scale=0.4]{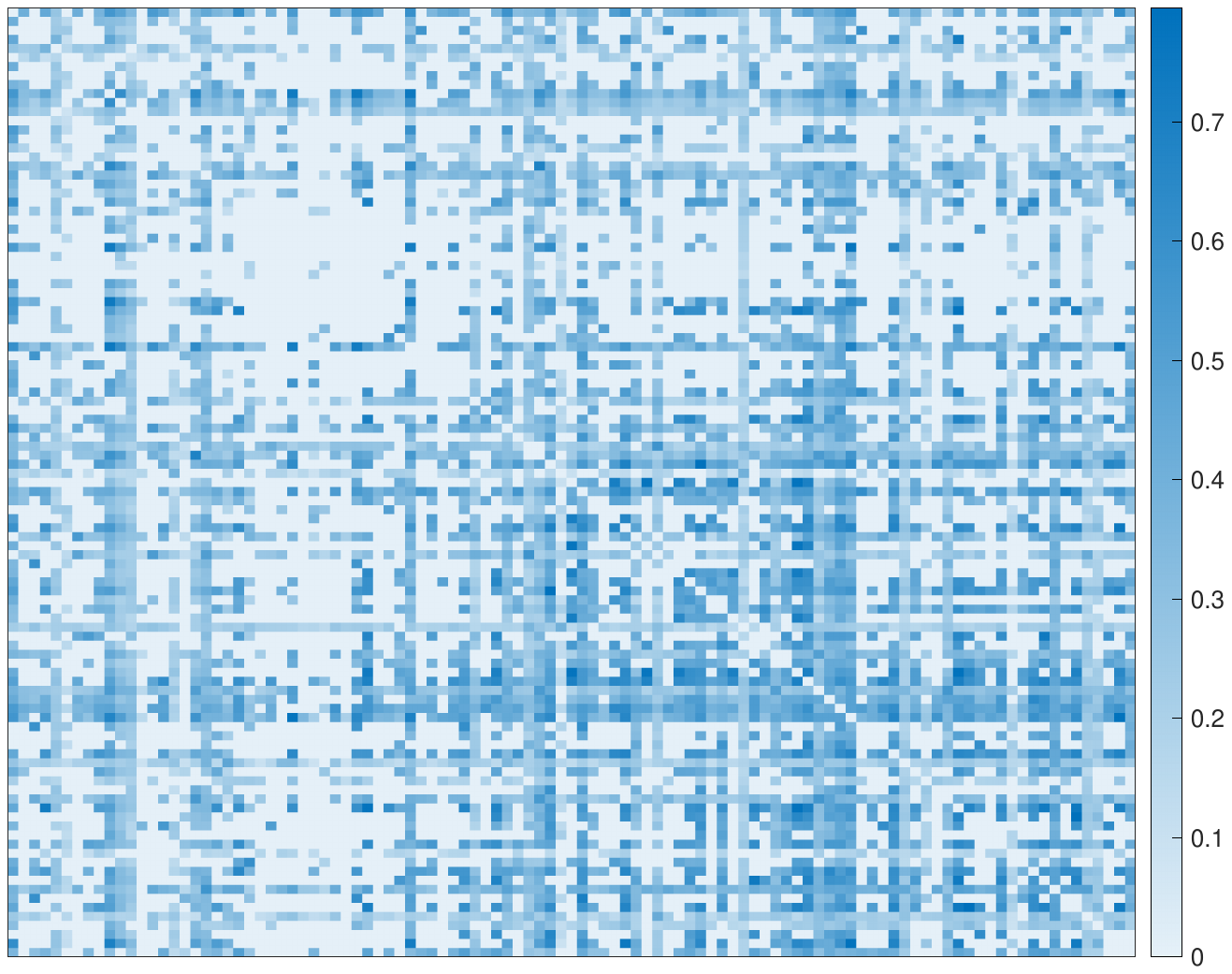}
\caption{The heatmap of dependencies in total IdioVols. 104 stocks. For
every pair, we test the null hypothesis of no dependence in IdioVols. If the
null is rejected, the heatmap color is proportional to the estimated value
of $Corr\left(C_{Zi},C_{Zj}\right)$, the quadratic-covariation based
correlation between the IdioVols, defined in equation (\protect\ref%
{eqn:corrivs}). Zero value is assigned to pairs where the null is not
rejected as well as the diagonal elements.}
\label{fig:heatmap_F00}
\end{figure}

\begin{figure}[!h]
\centering
\includegraphics[trim = 15mm 10mm 5mm
20mm,scale=0.5]{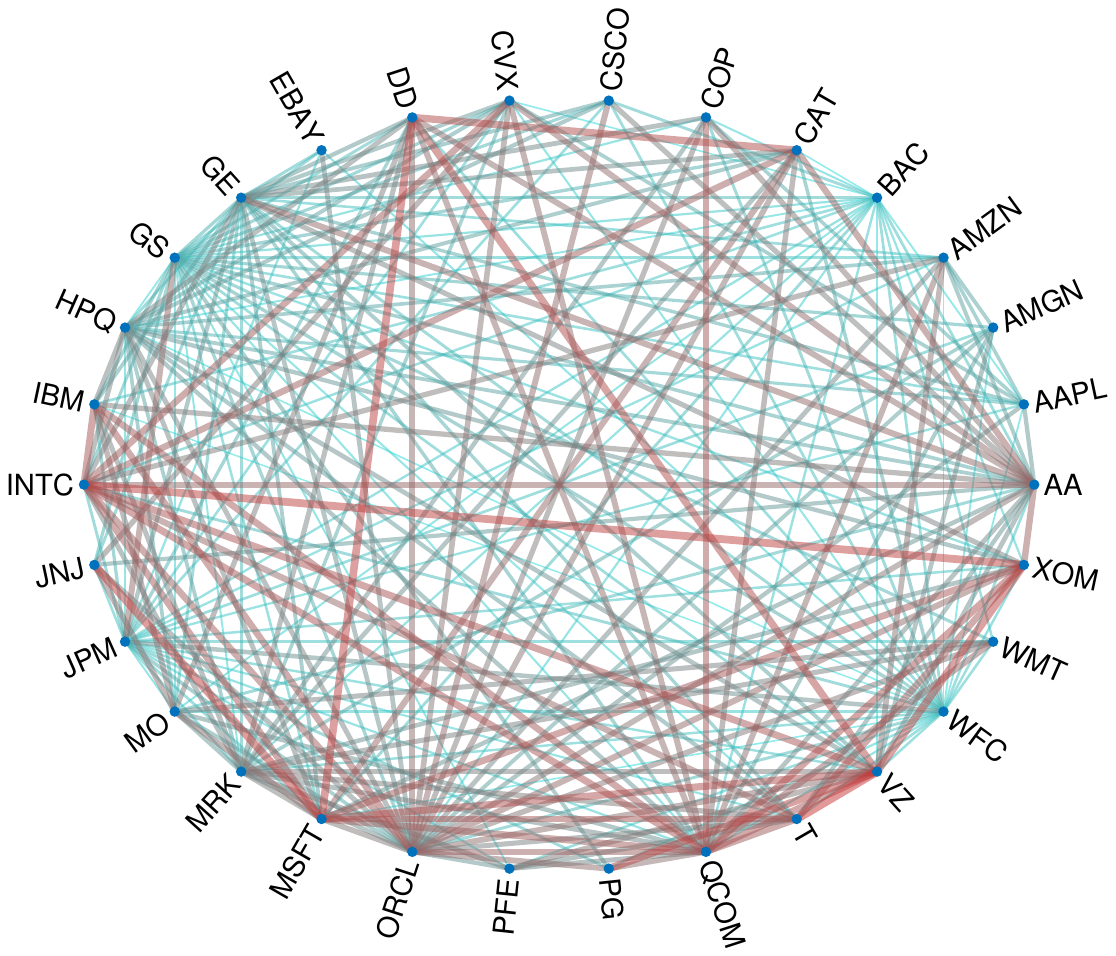}
\caption{The network of dependencies in total IdioVols. 30 stocks. The color
and thickness of each line is proportional to the estimated value of $%
Corr\left(C_{Zi},C_{Zj}\right)$, the quadratic-covariation based correlation
between the IdioVols, defined in equation (\protect\ref{eqn:corrivs}) (red
and thick lines indicate high correlation). We simultaneously test 435 null
hypotheses of no correlation, and the lines are only plotted when the null
is rejected.}
\label{fig:network_F00}
\end{figure}

We first investigate the dependence in the (total) Idiosyncratic
Volatilities. Our panel has 5356 
pairs of stocks. For each pair of stocks, we compute the correlation between
the IdioVols, $Corr\left( C_{Zi},C_{Zj}\right) $, see Section \ref%
{sec:estimation_special_case} for the implementation details. All pairwise
correlations are positive in our sample, and their average is $0.35$. Figure %
\ref{fig:heatmap_F00} contains a heatmap of this dependency measure in the
IdioVols. We simultaneously test 5356 hypotheses of no correlation, and
Figure \ref{fig:heatmap_F00} assigns non-zero correlations only for those
pairs of assets, for which the null is rejected; the diagonal contains
zeros, too. We account for multiple testing by controlling the false
discovery rate at $5\%$. Overall, Figure \ref{fig:heatmap_F00} shows that
the cross-sectional dependence between the IdioVols is very strong.\ To aid
visualization, Figure \ref{fig:network_F00} maps the network of dependencies
in the IdioVols for the subset of 30 stocks. Similarly to Figure \ref%
{fig:heatmap_F00}, in Figure \ref{fig:network_F00}, we simultaneously test
435 hypotheses of no correlation,
and Figure \ref{fig:network_F00} connects only the assets, for which the
null is rejected. Unsurprisingly, the cross-sectional dependence between the
IdioVols is also very strong among this subset of stocks.

Could missing factors in the R-FM provide an explanation? Omitted return
factors in the R-FM are captured by the idiosyncratic returns, and can
therefore induce correlation between the estimated IdioVols, provided these
missing return factors have non-negligible volatility of volatility. To
investigate this possibility, we consider the correlations between
idiosyncratic returns, $Corr(Z_{i},Z_{j})$.\footnote{%
Our measure of correlation between the idiosyncratic returns $dZ_{i}$ and $%
dZ_{j}$ is
\begin{equation}
Corr(Z_{i},Z_{j})=\frac{\int_{0}^{T}C_{ZiZj,t}dt}{\sqrt{%
\int_{0}^{T}C_{Zi,t}dt}\sqrt{\int_{0}^{T}C_{Zj,t}dt}},\hspace{3mm}%
i,j=1,\ldots ,d_{S},  \label{eqn:corrZ12}
\end{equation}%
where $C_{ZiZj,t}$ is the spot covariation between
$Z_{i}$ and $Z_{j}$.
Similarly to $R_{Yj}^{2}$, we estimate $Corr(Z_{i},Z_{j})$ using the
estimator of \cite{jacodrosenbaum13}.} Table \ref{tab:idioretVscorr}
presents a summary of how estimates of $Corr(Z_{i},Z_{j})$ are related to
the estimates of correlation in IdioVols, \textit{Corr}$\left(
C_{Zi},C_{Zj}\right) $. In particular, different rows in Table \ref%
{tab:idioretVscorr} display average values of $\widehat{Corr}\left(
C_{Zi},C_{Zj}\right) $ among those pairs, for which $|\widehat{Corr}%
(Z_{i},Z_{j})|$ is below some threshold.
We observe that even among pairs with virtually
uncorrelated idiosyncratic returns, the correlations among IdioVols are
still high. This conclusion holds both for the idiosyncratic returns and
volatilities defined with respect to CAPM, as well as the R-FM with three
Fama-French factors. Moreover, we observe that IdioVol correlations, $%
\widehat{Corr}\left( C_{Zi},C_{Zj}\right) $, are similar compared among
pairs that have high or low idiosyncratic return correlations, $\widehat{Corr%
}\left( C_{Zi},C_{Zj}\right) $. These results suggest that missing return
factors cannot explain dependence in IdioVols for all considered stocks.
This finding is in line with the empirical analysis of \cite%
{herskovickellyCIV} with daily and monthly returns.

\begin{figure}[!h]
\centering
\includegraphics[trim = 0mm 0mm 0mm
5mm,scale=0.4]{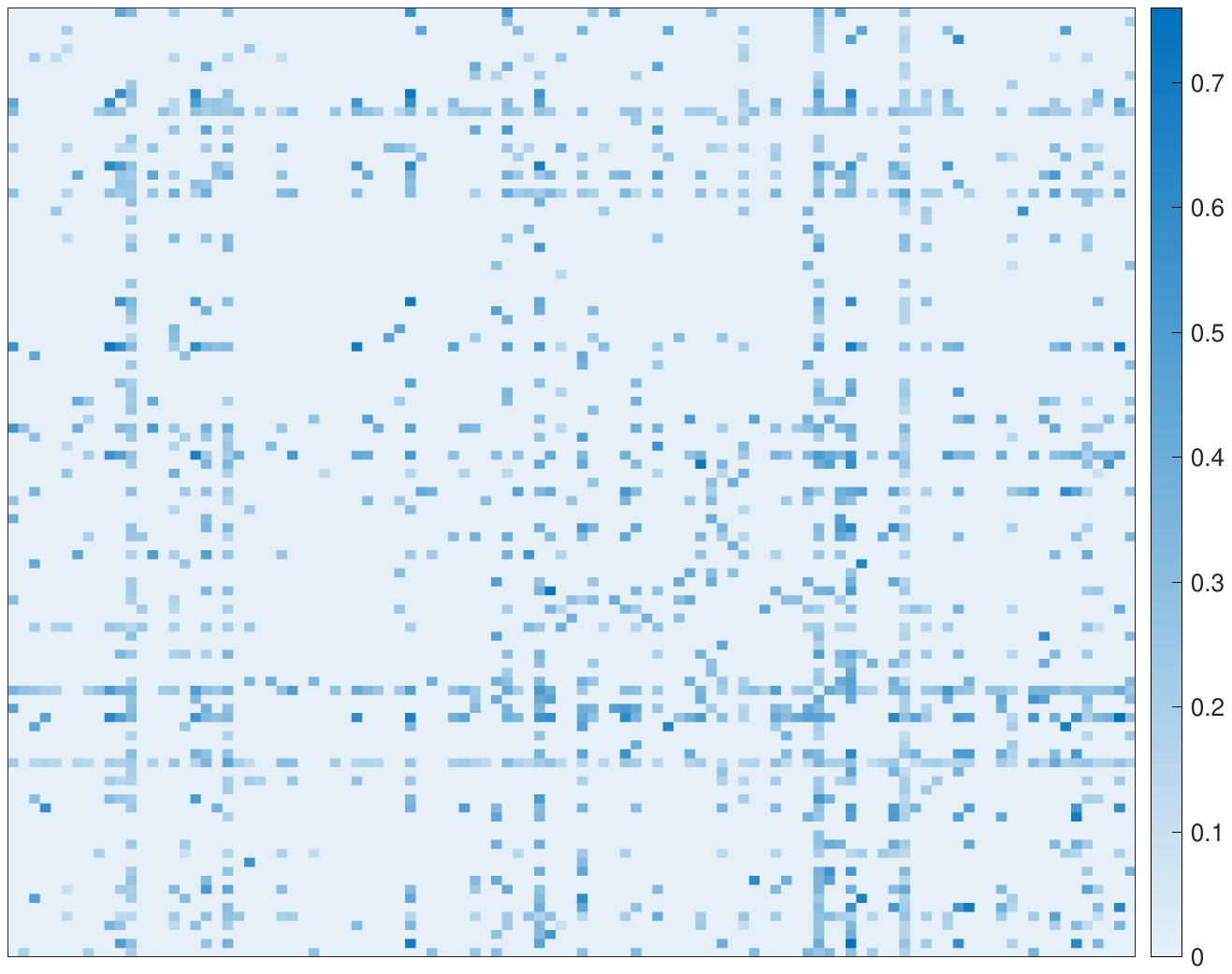}
\caption{The heatmap of dependencies in residual IdioVols after accounting
for a single IdioVol factor: the market variance. 104 stocks. For every
pair, we test the null hypothesis of no dependence in residual IdioVols. If
the null is rejected, the heatmap color is proportional to the estimated
value of $Corr\left(C^{resid}_{Zi},C^{resid}_{Zj}\right)$, the
quadratic-covariation based correlation between the residual IdioVols,
defined in equation (\protect\ref{eqn:corrNSivs}). Zero value is assigned to
pairs where the null is not rejected as well as the diagonal elements.}
\label{fig:heatmap_F01}
\end{figure}

\begin{figure}[!h]
\centering
\includegraphics[trim = 15mm 25mm 5mm
25mm,scale=0.47]{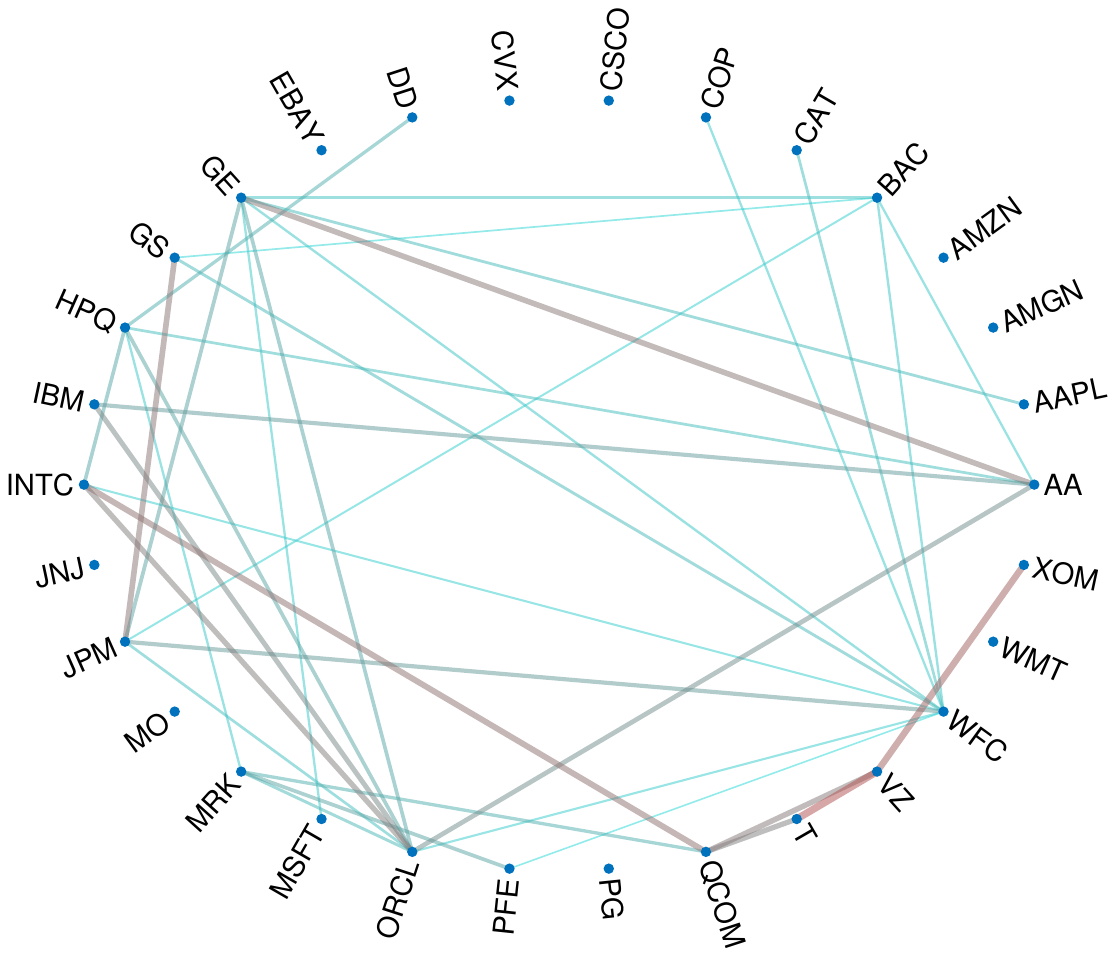}
\caption{The network of dependencies in residual IdioVols after accounting
for a single IdioVol factor: the market variance. 30 stocks. The color and
thickness of each line is proportional to the estimated value of $%
Corr\left(C^{resid}_{Zi},C^{resid}_{Zj}\right)$, the quadratic-covariation
based correlation between the residual IdioVols, defined in equation (%
\protect\ref{eqn:corrNSivs}) (red and thick lines indicate high
correlation). We simultaneously test 435 null hypotheses of no correlation,
and the lines are only plotted when the null is rejected.}
\label{fig:network_F01}
\end{figure}

To understand the source of the strong cross-sectional dependence in the
IdioVols, we consider the Idiosyncratic Volatility Factor Model (IdioVol-FM)
of Section \ref{sec:model}. We first use the market volatility as the only
IdioVol factor ($d_{\Pi}=1$).\footnote{%
We also considered the volatility of size and value Fama-French factors.
However, both these factors turned out to have very low volatility of
volatility and therefore did not significantly change the results.} 
Panel (a) of Table \ref{tbl:Empirical_gammas_R2} 
reports the estimates of the IdioVol loading ($%
\widehat{\gamma }_{Zi}$) and the $R^{2}$ of the IdioVol-FM ($R_{Zi}^{2,%
\txtIdioVolFM}$, see equation (\ref{eqn:R2_univar})). 
Panel (a) uses two different definitions of IdioVol, one
defined with respect to CAPM, and a second IdioVol defined with respect to
Fama-French three factor model. For virtually every stock, the estimated
IdioVol factor loading is positive, suggesting that the Idiosyncratic
Volatility co-moves with the market volatility. We have also calculated the
relevant t-statistics, showing that for virtually every stock, IdioVol
loading $\widehat{\gamma }_{Zi}$ is highly statistically significant. Next,
Figures \ref{fig:heatmap_F01} and \ref{fig:network_F01} show dependencies
among residual IdioVols after accounting for the market volatility as the sole
IdioVol factor. The average pairwise correlations between the residual
IdioVols, $\widehat{Corr}(C_{Zi}^{resid},C_{Zj}^{resid})$, across all pairs
of stocks, 
decrease to $0.21$. However, the market volatility cannot explain all
cross-sectional dependence in residual IdioVols, as evidenced by the
remaining links in both Figure \ref{fig:heatmap_F01} and \ref%
{fig:network_F01}.

\begin{figure}[!h]
\centering
\includegraphics[trim = 0mm 0mm 0mm
5mm,scale=0.4]{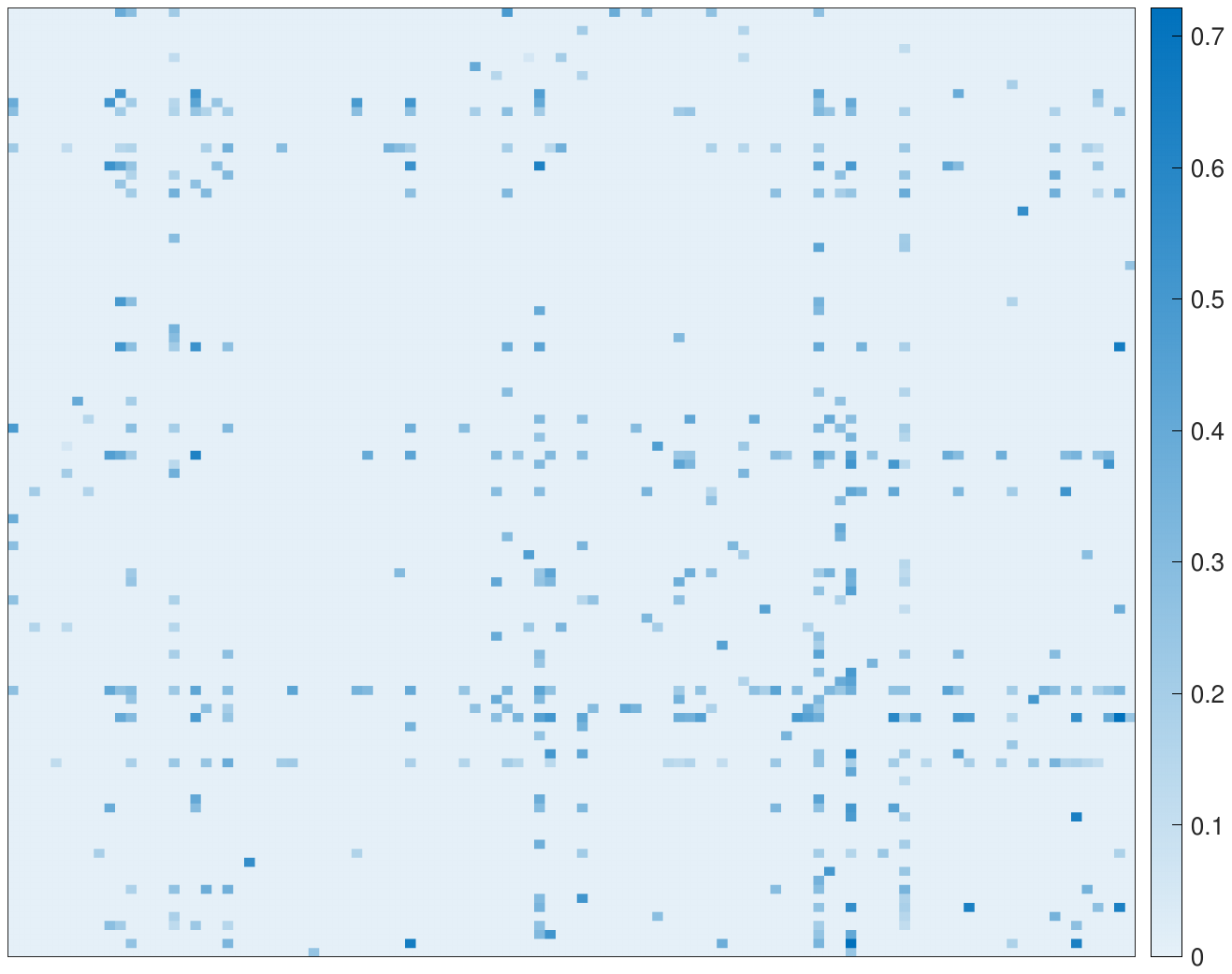}
\caption{The heatmap of dependencies in residual IdioVols after accounting
for ten IdioVol factors: the market variance and the variances of nine
industry ETFs. 104 stocks. For every pair, we test the null hypothesis of no
dependence in residual IdioVols. If the null is rejected, the heatmap color
is proportional to the estimated value of $Corr%
\left(C^{resid}_{Zi},C^{resid}_{Zj}\right)$, the quadratic-covariation based
correlation between the residual IdioVols, defined in equation (\protect\ref%
{eqn:corrNSivs}). Zero value is assigned to pairs where the null is not
rejected as well as the diagonal elements.}
\label{fig:heatmap_F10}
\end{figure}

\begin{figure}[!h]
\centering
\includegraphics[trim = 15mm 25mm 5mm
25mm,scale=0.47]{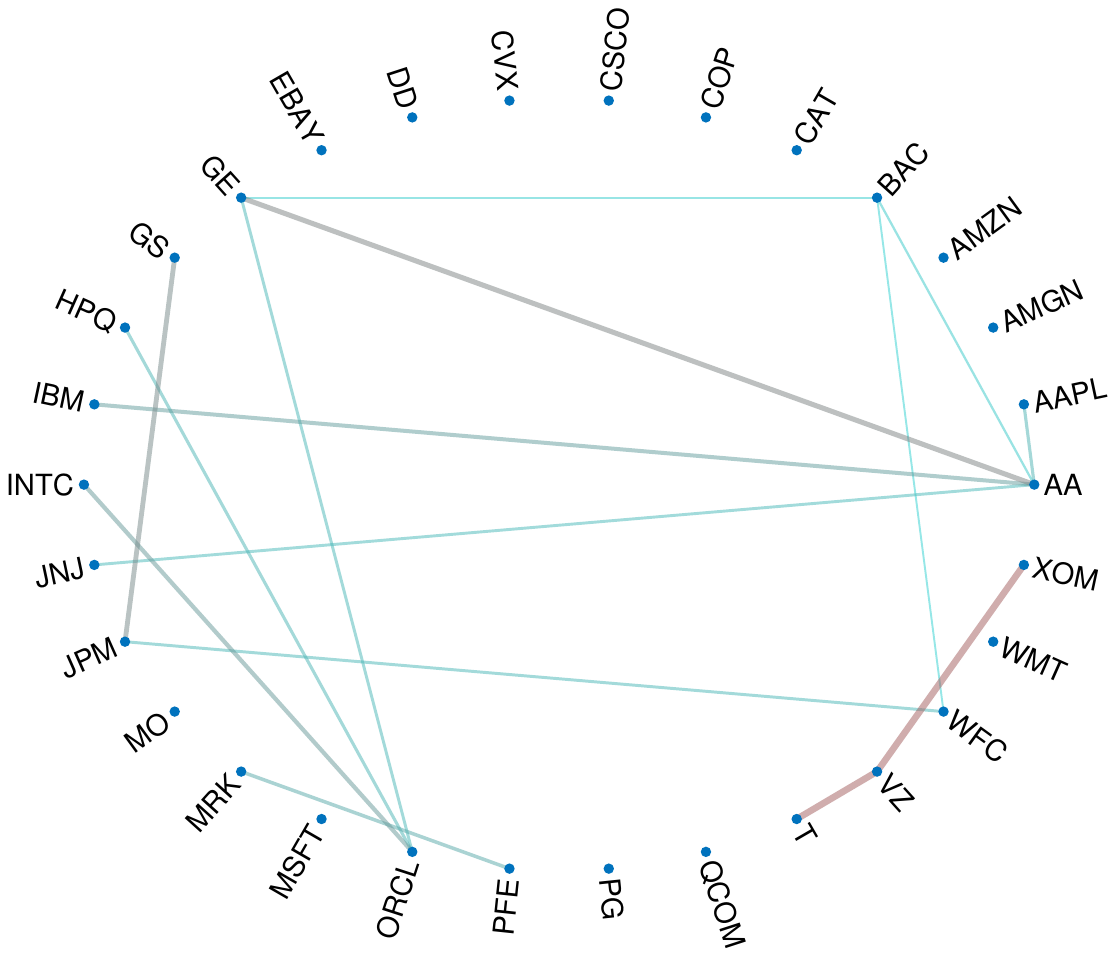}
\caption{The network of dependencies in residual IdioVols after accounting
for ten IdioVol factors: the market variance and the variances of nine
industry ETFs. 30 stocks. The color and thickness of each line is
proportional to the estimated value of $Corr%
\left(C^{resid}_{Zi},C^{resid}_{Zj}\right)$, the quadratic-covariation based
correlation between the residual IdioVols, defined in equation (\protect\ref%
{eqn:corrNSivs}) (red and thick lines indicate high correlation). We
simultaneously test 435 null hypotheses of no correlation, and the lines are
only plotted when the null is rejected.}
\label{fig:network_F10}
\end{figure}

Finally, we consider an IdioVol-FM with ten IdioVol factors, $d_{\Pi}=10$,
 market volatility and the volatilities of nine industry ETFs. 
 We use CAPM IdioVols.
Panel (b) of Table \ref{tbl:Empirical_gammas_R2} 
reports the corresponding $R_{Zi}^{2,\txtIdioVolFM}$,
which is considerably higher than in the one-factor case, $d_{\Pi}=1$.
Figures \ref%
{fig:heatmap_F10} and \ref{fig:network_F10} show the implications for the
cross-section of this ten-factor IdioVol-FM,
 for 104 and \ 30 stocks, respectively. The average pairwise
correlations between the residual IdioVols, $\widehat{Corr}%
(C_{Zi}^{resid},C_{Zj}^{resid})$, decrease further to $0.17$. However,
significant dependence between the residual IdioVols remains, as evidenced
by the remaining links in both Figures \ref{fig:heatmap_F10} and \ref%
{fig:network_F10}. Our results suggest that there is room for considering
the construction of additional IdioVol factors based on economic theory, for
example, along the lines of the heterogeneous agents model of \cite%
{herskovickellyCIV}.

For comparison, we also calculate the naive estimators, see equation (\ref%
{eqn:estNAIVE}). Of course, since the naive estimators are inconsistent, we
do not have valid confidence intervals to accompany them. We focus on the
one-factor IdioVol-FM. In our data set, the absolute values of the
differences between\ the naive and the bias-corrected estimators range,
across all pairs of stocks, between $0$ and $0.045$ for $Corr\left(
C_{Zi},C_{Zj}\right) $, between $0$ and $0.051$ for $%
Corr(C_{Zi}^{resid},C_{Zj}^{resid})$, and between $0.06$ and $0.13$ for $%
R_{Zj}^{2,\txtIdioVolFM}$. However, the relative errors can be large, for
example, for $R_{Zj}^{2,\txtIdioVolFM}$, it is $42\%$ on average. We find that in the
instances where the differences are small, the multiplicative bias, i.e.,
the factor $2/3$, dominates the additive bias both in the numerator and the
denominator, so that the multiplicative bias approximately cancels out for
these estimands.


\bigskip

\begin{table}[thb]
\begin{center}
\begin{tabular}{lll}
\hline\hline
Sector & Stock & Ticker \\ \hline
Financials & Bank of America Corp & BAC \\
& Goldman Sachs Group Inc & GS \\
& JPMorgan Chase \& Co & JPM \\
& Wells Fargo \& Co & WFC \\ \hline
Energy & ConocoPhillips & COP \\
& Chevron Corp & CVX \\
& Exxon Mobil Corp & XOM \\ \hline
Consumer Staples & Altria Group Inc & MO \\
& Procter \& Gamble Co & PG \\
& Walmart Inc & WMT \\ \hline
Industrials & Caterpillar Inc & CAT \\
& GE Aerospace & GE \\ \hline
Information Technology & Apple Inc & AAPL \\
& Cisco Systems Inc & CSCO \\
& HP Inc & HPQ \\
& Intl Business Machines Corp & IBM \\
& Intel Corp & INTC \\
& Microsoft Corp & MSFT \\
& Oracle Corp & ORCL \\
& Qualcomm Inc & QCOM \\ \hline
Health Care & Amgen Inc & AMGN \\
& Johnson \& Johnson & JNJ \\
& Merck \& Co & MRK \\
& Pfizer Inc & PFE \\ \hline
Consumer Discretionary & Amazon.com Inc & AMZN \\
& Ebay Inc & EBAY \\ \hline
Materials & Alcoa Corp & AA \\
& Dupont de Nemours Inc & DD \\ \hline
Communication Services & AT\&T Inc & T \\
& Verizon Communications Inc & VZ \\ \hline\hline
\end{tabular}%
\end{center}
\caption{The set of 30 stocks used in the maps of network of dependencies in
(residual) IdioVols in Figures \protect\ref{fig:network_F00}, \protect\ref%
{fig:network_F01}, and \protect\ref{fig:network_F10}.}
\label{tbl:stocks}
\end{table}

\begin{sidewaystable}[thb]
\begin{center}
\begin{tabular}{lcccccccc}
\hline\hline
& &  \multicolumn{3}{c}{\text{CAPM}} & & \multicolumn{3}{c}{\text{FF3 Model}}\\
$|\widehat{\text{Corr}}(Z_i,Z_j)|$ & &  Pairs & Avg\,$|\widehat{\text{Corr}}(Z_i,Z_j)|$  &
Avg\,$\widehat{\text{Corr}}\left(C_{Zi},C_{Zj} \right)$ 
& & Pairs & Avg\,$|\widehat{\text{Corr}}(Z_i,Z_j)|$ &
Avg\,$\widehat{\text{Corr}}\left(C_{Zi},C_{Zj} \right)$ 
 \\
\cline{1-1} \cline{3-5} \cline{7-9}
    ~~~~$<0.6$ &    & 5356 & 0.045 & 0.347 &   & 5356 & 0.045 & 0.347\\
    ~~~~$<0.5$ &    & 5354 & 0.045 & 0.347 &   & 5353 & 0.045 & 0.347\\
    ~~~~$<0.4$ &    & 5334 & 0.044 & 0.346 &   & 5335 & 0.044 & 0.346\\
    ~~~~$<0.3$ &    & 5300 & 0.042 & 0.344 &   & 5300 & 0.042 & 0.345\\
    ~~~~$<0.2$ &    & 5236 & 0.039 & 0.343 &   & 5236 & 0.039 & 0.344\\
    ~~~~$<0.1$ &    & 4925 & 0.033 & 0.338 &   & 4928 & 0.033 & 0.339\\
    ~~~~$<0.075$ &    & 4642 & 0.030 & 0.333 &   & 4647 & 0.030 & 0.333\\
    ~~~~$<0.050$ &    & 3873 & 0.024 & 0.320 &   & 3895 & 0.024 & 0.320\\
    ~~~~$<0.025$ &    & 2049 & 0.013 & 0.296 &   & 2044 & 0.013 & 0.296\\
    ~~~~$<0.010$ &    & 757 & 0.005 & 0.293 &   & 748 & 0.005 & 0.293\\
    ~~~~$<0.005$ &    & 374 & 0.003 & 0.297 &   & 373 & 0.002 & 0.296\\
\hline\hline
\end{tabular}
\caption{Each row in this table describes the subset of pairs of stocks with $|\widehat{\text{Corr}(Z_i,Z_j)}|$ below a threshold in column one.
The table considers two R-FMs: the left panel defines the IdioVol with respect to CAPM, and the right panel defines the IdioVol with respect to the three-factor Fama-French model. In both cases, the market volatility is the only IdioVol factor. Each panel reports three quantities for the given subset of pairs: the number of pairs, average absolute pairwise correlation in idiosyncratic returns, and average pairwise correlation between IdioVols.
}\label{tab:idioretVscorr}
\end{center}
\end{sidewaystable}

\newcommand*\tabRSqIVFM{${R}_{Z}^{2,\text{\scalebox{0.8}[1.0]{\textit{IVFM}}}}$}

\clearpage
\begin{table}[!thb]
  \thispagestyle{empty}%
  \swpSmall
  \centering
  \renewcommand{\arraystretch}{0.9}
  \setlength{\tabcolsep}{5pt}
\begin{tabular}{|c|cc|cc|c|c|c|cc|cc|c|}
\cline{1-6}\cline{1-6}\cline{8-13}
& \multicolumn{4}{c|}{(a)} & (b) &  &  & \multicolumn{4}{c|}{(a)} & (b) \\
\cline{2-6}\cline{9-13}
\rule{0pt}{2.1ex} & \multicolumn{2}{c|}{CAPM} & \multicolumn{2}{c|}{FF3 Model} & CAPM &  &  &
\multicolumn{2}{c|}{CAPM} & \multicolumn{2}{c|}{FF3 Model} & CAPM \\
& \multicolumn{2}{|c}{$d_{\Pi }=1$} & \multicolumn{2}{|c|}{$d_{\Pi }=1$} & $%
d_{\Pi }=10$ &  &  & \multicolumn{2}{|c}{$d_{\Pi }=1$} &
\multicolumn{2}{|c|}{$d_{\Pi }=1$} & $d_{\Pi }=10$ \\
Stock & $\widehat{\gamma }_{Z_{\rule{0pt}{1ex}}}$ & \tabRSqIVFM
& $\widehat{\gamma }_{Z}$ & \tabRSqIVFM & \tabRSqIVFM &  & Stock & $\widehat{\gamma}_{Z}
$ & \tabRSqIVFM & $\widehat{\gamma }_{Z}$ & \tabRSqIVFM & \tabRSqIVFM \\ \cline{1-6}\cline{8-13}
 \rule{0pt}{2.3ex} AA & 0.57 & 0.12 & 0.56 & 0.12 & 0.25 &  & HNZ & 0.36 & 0.52 & 0.36 & 0.52 & 0.72\\ 
  AAPL & 0.30 & 0.07 & 0.30 & 0.07 & 0.19 &  & HON & 0.32 & 0.21 & 0.31 & 0.21 & 0.41\\ 
  ABT & 0.23 & 0.20 & 0.23 & 0.20 & 0.35 &  & HPQ & 0.44 & 0.15 & 0.44 & 0.15 & 0.29\\ 
  AEP & 0.37 & 0.29 & 0.36 & 0.29 & 0.44 &  & HSH & 0.19 & 0.11 & 0.19 & 0.12 & 0.18\\ 
  AES & 0.49 & 0.07 & 0.49 & 0.07 & 0.19 &  & IBM & 0.35 & 0.44 & 0.35 & 0.45 & 0.55\\ 
  AIG & 0.37 & 0.02 & 0.36 & 0.02 & 0.11 &  & INTC & 0.37 & 0.25 & 0.37 & 0.25 & 0.42\\ 
  ALL & 0.29 & 0.07 & 0.29 & 0.07 & 0.20 &  & IP & 0.34 & 0.07 & 0.33 & 0.07 & 0.22\\ 
  AMGN & 0.29 & 0.19 & 0.29 & 0.18 & 0.28 &  & JNJ & 0.37 & 0.62 & 0.37 & 0.62 & 0.69\\ 
  AMZN & 0.56 & 0.19 & 0.55 & 0.19 & 0.31 &  & JPM & 0.34 & 0.06 & 0.34 & 0.06 & 0.24\\ 
  APA & 0.33 & 0.14 & 0.32 & 0.14 & 0.36 &  & KO & 0.32 & 0.55 & 0.31 & 0.54 & 0.62\\ 
  APC & 0.27 & 0.05 & 0.26 & 0.04 & 0.17 &  & LLY & 0.40 & 0.46 & 0.39 & 0.46 & 0.55\\ 
  ATI & 0.35 & 0.03 & 0.35 & 0.03 & 0.14 &  & LMT & 0.40 & 0.28 & 0.39 & 0.28 & 0.39\\ 
  AVP & 0.25 & 0.04 & 0.24 & 0.03 & 0.18 &  & LOW & 0.47 & 0.28 & 0.45 & 0.27 & 0.41\\ 
  AXP & 0.40 & 0.09 & 0.39 & 0.09 & 0.29 &  & MCD & 0.30 & 0.20 & 0.30 & 0.20 & 0.28\\ 
  BA & 0.37 & 0.30 & 0.36 & 0.29 & 0.37 &  & MDLZ & 0.28 & 0.26 & 0.27 & 0.26 & 0.31\\ 
  BAC & 0.42 & 0.03 & 0.42 & 0.04 & 0.10 &  & MDT & 0.50 & 0.55 & 0.50 & 0.56 & 0.60\\ 
  BAX & 0.22 & 0.05 & 0.22 & 0.05 & 0.17 &  & MET & 0.25 & 0.05 & 0.25 & 0.05 & 0.14\\ 
  BHI & 0.25 & 0.06 & 0.25 & 0.06 & 0.21 &  & MMM & 0.25 & 0.30 & 0.24 & 0.29 & 0.43\\ 
  BK & 0.55 & 0.09 & 0.54 & 0.09 & 0.30 &  & MO & 0.43 & 0.36 & 0.43 & 0.36 & 0.39\\ 
  BMY & 0.30 & 0.25 & 0.30 & 0.25 & 0.30 &  & MON & 0.34 & 0.05 & 0.33 & 0.05 & 0.15\\ 
  C & 0.26 & 0.02 & 0.26 & 0.02 & 0.18 &  & MRK & 0.40 & 0.20 & 0.39 & 0.20 & 0.30\\ 
  CAT & 0.53 & 0.32 & 0.53 & 0.33 & 0.40 &  & MSFT & 0.51 & 0.59 & 0.50 & 0.60 & 0.68\\ 
  CI & 0.46 & 0.08 & 0.45 & 0.08 & 0.18 &  & NKE & 0.53 & 0.45 & 0.53 & 0.45 & 0.52\\ 
  CL & 0.19 & 0.29 & 0.19 & 0.30 & 0.37 &  & NOV & 0.30 & 0.04 & 0.29 & 0.04 & 0.18\\ 
  CMCSA & 0.37 & 0.20 & 0.37 & 0.20 & 0.27 &  & NSC & 0.41 & 0.17 & 0.41 & 0.16 & 0.36\\ 
  COF & 0.56 & 0.08 & 0.56 & 0.08 & 0.21 &  & ORCL & 0.36 & 0.25 & 0.36 & 0.25 & 0.40\\ 
  COP & 0.35 & 0.18 & 0.35 & 0.18 & 0.40 &  & OXY & 0.35 & 0.11 & 0.34 & 0.10 & 0.31\\ 
  COST & 0.32 & 0.30 & 0.32 & 0.30 & 0.35 &  & PEP & 0.27 & 0.44 & 0.27 & 0.45 & 0.57\\ 
  CPB & 0.17 & 0.09 & 0.17 & 0.09 & 0.26 &  & PFE & 0.30 & 0.20 & 0.30 & 0.21 & 0.24\\ 
  CSC & 0.32 & 0.08 & 0.32 & 0.08 & 0.10 &  & PG & 0.27 & 0.58 & 0.27 & 0.58 & 0.65\\ 
  CSCO & 0.39 & 0.27 & 0.39 & 0.27 & 0.44 &  & QCOM & 0.45 & 0.24 & 0.45 & 0.24 & 0.36\\ 
  CVS & 0.33 & 0.15 & 0.33 & 0.15 & 0.24 &  & RF & 0.50 & 0.03 & 0.50 & 0.03 & 0.14\\ 
  CVX & 0.29 & 0.23 & 0.28 & 0.23 & 0.47 &  & ROK & 0.54 & 0.22 & 0.53 & 0.22 & 0.27\\ 
  DD & 0.46 & 0.46 & 0.45 & 0.46 & 0.54 &  & S & 0.39 & 0.02 & 0.38 & 0.02 & 0.11\\ 
  DELL & 0.32 & 0.15 & 0.32 & 0.15 & 0.26 &  & SBUX & 0.49 & 0.24 & 0.48 & 0.24 & 0.32\\ 
  DIS & 0.42 & 0.36 & 0.42 & 0.36 & 0.50 &  & SO & 0.36 & 0.66 & 0.35 & 0.66 & 0.72\\ 
  DOW & 0.47 & 0.16 & 0.47 & 0.16 & 0.21 &  & T & 0.53 & 0.30 & 0.53 & 0.30 & 0.47\\ 
  DVN & 0.31 & 0.09 & 0.31 & 0.09 & 0.30 &  & TGT & 0.62 & 0.31 & 0.62 & 0.31 & 0.41\\ 
  EBAY & 0.50 & 0.26 & 0.50 & 0.26 & 0.45 &  & TWX & 0.53 & 0.41 & 0.52 & 0.41 & 0.47\\ 
  EMC & 0.47 & 0.20 & 0.47 & 0.20 & 0.37 &  & TXN & 0.42 & 0.30 & 0.42 & 0.30 & 0.46\\ 
  EMR & 0.26 & 0.11 & 0.26 & 0.11 & 0.19 &  & UIS & 0.30 & 0.01 & 0.29 & 0.01 & 0.04\\ 
  ETR & 0.34 & 0.32 & 0.34 & 0.32 & 0.47 &  & UNH & 0.63 & 0.23 & 0.64 & 0.24 & 0.28\\ 
  EXC & 0.44 & 0.26 & 0.42 & 0.25 & 0.40 &  & UNP & 0.56 & 0.29 & 0.55 & 0.29 & 0.43\\ 
  F & 0.52 & 0.05 & 0.51 & 0.05 & 0.15 &  & UPS & 0.33 & 0.49 & 0.33 & 0.49 & 0.56\\ 
  FDX & 0.34 & 0.30 & 0.34 & 0.30 & 0.41 &  & USB & 0.60 & 0.18 & 0.60 & 0.18 & 0.37\\ 
  GD & 0.45 & 0.43 & 0.44 & 0.44 & 0.55 &  & UTX & 0.38 & 0.33 & 0.38 & 0.33 & 0.49\\ 
  GE & 0.38 & 0.09 & 0.38 & 0.09 & 0.31 &  & VZ & 0.41 & 0.38 & 0.40 & 0.38 & 0.51\\ 
  GILD & 0.37 & 0.15 & 0.38 & 0.16 & 0.26 &  & WFC & 0.33 & 0.05 & 0.32 & 0.05 & 0.21\\ 
  GS & 0.43 & 0.12 & 0.43 & 0.12 & 0.31 &  & WMB & 0.35 & 0.03 & 0.35 & 0.03 & 0.10\\ 
  HAL & 0.29 & 0.05 & 0.29 & 0.05 & 0.21 &  & WMT & 0.28 & 0.47 & 0.28 & 0.48 & 0.56\\ 
  HD & 0.36 & 0.22 & 0.36 & 0.22 & 0.47 &  & XOM & 0.35 & 0.24 & 0.35 & 0.24 & 0.35\\ 
  HIG & 0.27 & 0.03 & 0.26 & 0.03 & 0.09 &  & XRX & 0.52 & 0.18 & 0.52 & 0.18 & 0.24\\ 
   \cline{1-6}\cline{8-13}
  \end{tabular}%
  \caption{\small Panel (a) presents estimates of the IdioVol factor loading
  ($\widehat{\gamma}_Z $, see eq. (\protect\ref{eqn:IdioVol-FM})) 
  and the contribution of the market volatility 
  to the variation in the IdioVols 
  (\tabRSqIVFM$=\widehat{R}^{2, \txtIdioVolFM}_Z$, 
  see eq. (\protect\ref{eqn:R2_univar})) 
  in the one-factor IdioVol-FM, $d_{\Pi }=1$. 
  Panel (a) considers two R-FMs, CAPM or the three-factor Fama-French model (FF3).
  Panel (b) presents \tabRSqIVFM in the ten-factor IdioVol-FM, $d_{\Pi }=10$, with CAPM IdioVols. }
\label{tbl:Empirical_gammas_R2}
\end{table}

\FloatBarrier

\section{Monte Carlo}

\label{sec:MC} This section investigates the finite sample properties of our
estimators and tests. The data generating process (DGP) is similar to that
of \cite{litodorovtauchen-dependencies-WP2013} and is constructed as
follows. Denote by $Y_{1}$ and $Y_{2}$ the log-prices of two individual
stocks, and by $X$ the log-price of the market portfolio. Recall that the
superscript $c$ indicates the continuous part of a process. We assume
\begin{align*}
dX_{t}=dX_{t}^c+dJ_{3,t}, \hspace{3mm} dX_{t}^c=\sqrt{C_{X,t}}dW_t,
\end{align*}
and, for $j=1,2$,
\begin{align*}
dY_{j,t}=\beta_tdX_{t}^c+d\widetilde{Y}_{j,t}^c+dJ_{j,t}, \hspace{3mm} d%
\widetilde{Y}_{j,t}^c=\sqrt{C_{Zj,t}}d\widetilde{W}_{j,t}.
\end{align*}
In the above, $C_{X}$ is the spot volatility of the market portfolio, $%
\widetilde{W}_1$ and $\widetilde{W}_2$ are Brownian motions with $\text{Corr}%
(d\widetilde{W}_{1,t},d\widetilde{W}_{2,t})=0.4$, and $W$ is an independent
Brownian motion; $J_1, J_2$, and $J_3$ are independent compound Poisson
processes with intensity equal to 2 jumps per year and jump size
distribution $N(0,0.02^2)$. The beta process is time-varying and is
specified as $\beta_t=0.5+0.1\hspace{1mm}\text{sin}(100t). $

We next specify the volatility processes. As our building blocks, we first
generate four processes $f_{1},\ldots ,f_{4}$ as mutually independent
Cox-Ingersoll-Ross processes,%
\begin{align*}
df_{1,t}& =5(0.09-f_{1,t})dt+0.35\sqrt{f_{1,t}}\Big(-0.8dW_{t}+\sqrt{%
1-0.8^{2}}dB_{1,t}\Big), \\
df_{j,t}& =5(0.09-f_{j,t})dt+0.35\sqrt{f_{j,t}}dB_{j,t}~~,\text{ for }%
j=2,3,4,
\end{align*}%
where $B_{1},\ldots ,B_{4}$ are independent standard\ Brownian Motions,
which are also independent from the Brownian Motions of the return Factor
Model.\footnote{%
The Feller property is satisfied implying the positiveness of the processes $%
(f_{j,t})_{1\leq j\leq 4}$.} We use the first process $f_{1}$ as the market
volatility, i.e., $C_{X,t}=f_{1,t}$. We use the other three processes $%
f_{2},f_{3}$, and $f_{4}$ to construct two 
different specifications for the IdioVol processes $C_{Z1,t}$ and $C_{Z2,t}$%
, see Table \ref{tbl:spec} for details. The common Brownian Motion $W_{t}$
in the market portfolio price process $X_{t}$ and its volatility process $%
C_{X,t}=f_{1,t}$ generates a leverage effect for the market portfolio. The
value of the leverage effect is $-0.8$, which is standard in the literature,
see \cite{kalninaxiu-lev}, \cite{aitfanli13} and \cite{Yacine-jump-lev}.%
\footnote{%
Notice that by It\^{o} Lemma, each of these three models can be expressed in
terms of equation (\ref{eqn:Y_BSM}) for the vector $\left(
X_{t},Y_{1,t},Y_{2,t}\right) ^{\prime }$ and equation (\ref{eqn:C_BSM}) for
the volatility matrix of this vector.}

\begin{table}[htp]
\begin{center}
\begin{tabular}{ccc}
\hline\hline
& \rule{0pt}{13pt} $C_{Z1,t}$ & $C_{Z2,t}$ \\ \hline
Model 1 & \rule{0pt}{13pt} $0.1+1.5f_{2,t}$ & $0.1+1.5f_{3,t}$ \\
Model 2 & $0.1+0.45C_{X,t}+f_{2,t}+0.4f_{4,t}$ & $%
0.1+0.35C_{X,t}+0.3f_{3,t}+0.6f_{4,t}$ \\ \hline\hline
\end{tabular}%
\end{center}
\par
{}
\caption{Different specifications for the Idiosyncratic Volatility processes
$C_{Z1,t}$ and $C_{Z2,t}$. }
\label{tbl:spec}
\end{table}

We set the time span $T$ equal to 1,260 or 2,520 days, which correspond
approximately to 5 and 10 business years. These values are standard in the
nonparametric leverage effect estimation literature (see \cite{aitfanli13}
and \cite{kalninaxiu-lev}), where the rate of convergence is also $\Delta
^{-1/4}$. Each day consists of 6.5 trading hours. We consider two different
values for the sampling frequency, $\Delta _{n}=$ 1 minute and $\Delta _{n}=$
5 minutes.
We follow \cite{litodorovtauchen-dependencies} and set the jump truncation
threshold $u_{n}$ in day $t$ at $3\widehat{\sigma }_{t}\Delta _{n}^{0.49}$,
where $\widehat{\sigma }_{t}$ is the squared root of the annualized bipower
variation of \cite{barndorffnielsenshephard04}. We choose four different
values for the width of the subsamples, which corresponds to $\theta
=1.5,2,2.5$ and $3$ (recall that the number of observations in a window is $%
k_{n}=\theta /\sqrt{\Delta _{n}}$). We use 10,000 Monte Carlo replications
in all the experiments.

We first investigate the finite sample properties of the estimators (using
Model 3). We consider the following estimands:

\begin{itemize}
\item the IdioVol factor loading of the first stock, $\gamma _{Z1}$,

\item the contribution of the market volatility to the variation of the
IdioVol of the first stock $R_{Z1}^{2,\txtIdioVolFM}$,

\item the correlation between the Idiosyncratic Volatilities of stocks 1 and
2, $Corr\left( C_{Z1},C_{Z2}\right) $,

\item the correlation between the residual Idiosyncratic Volatilities, $%
Corr\left( C_{Z1}^{resid},C_{Z2}^{resid}\right) $.
\end{itemize}

In Table \ref{tbl:const1}, we report the median bias, the interquartile
range (IQR), and the RMSE of the two type of the bias-corrected estimators
as well as the naive estimator for each estimand using 5 minutes data over
10 years. In Tables \ref{tbl:const1}-\ref{tbl:r4}, in order to simplify the
interpretation of the results, we fix the volatility paths $C_{X,t}$ and $%
(f_{j,t})_{0\leq j \leq 4}$ across simulations.
 
Consider first the comparison of the AN and LIN estimators. One does not
consistently over-perform the other in terms of the bias or the IQR.
Interestingly, in terms of the RMSE, the LIN estimator outperforms the AN
estimator in every scenario considered. The naive estimators are
substantially biased. The comparison of the bias-corrected estimators and
the naive estimators reveals the usual bias-variance trade-off, as the
bias-corrected estimators have smaller bias but larger IQR than the naive
estimator. In terms of RMSE, the bias-corrected estimators generally
outperform the naive estimator: RMSE is significantly lower when estimating $%
\gamma_{Z1}$, $R^{2,\txtIdioVolFM}_{Z1}$, or $Corr\left(C_{Z1},C_{Z2}
\right) $, while the results for $Corr\left(C^{resid}_{Z1},C^{resid}_{Z2}
\right)$ are mixed.

It is also informative to see how these results change when we increase the
sampling frequency. In Table \ref{tbl:const2}, we report the results with $%
\Delta_n=1\text{ minute}$ in the same setting. The qualitative conclusions
of Table \ref{tbl:const1} remain true in Table \ref{tbl:const2}. Compared to
Table \ref{tbl:const1}, the bias and IQR are smaller. However, the magnitude
of the decrease of the IQR is small.

Finally, Table \ref{tbl:r4} contains results from same experiment using data
sampled at one minute over 5 years. Despite using more than twice as many
observations than in the first experiment, the precision is not as good. In
other words, increasing the time span is more effective for precision gain
than increasing the sampling frequency.
The qualitative conclusions generally remain the same as in Table \ref%
{tbl:const1}.

\begin{sidewaystable}

\begin{tabular*}{1.0\textwidth}{l|@{\extracolsep{\fill}}cccccccccccc}
\hline\hline
\rule{0pt}{15pt}\hspace{0.1in}\text{{}} & \multicolumn{4}{c}{LIN} &
\multicolumn{4}{c}{AN} & \multicolumn{4}{c}{Naive} \\
$\widehat{\theta }$ & 1.5 & 2 & 2.5 & 3 & 1.5 & 2 & 2.5 & 3 & 1.5 & 2 & 2.5
& 3 \\ \hline
&  &  &  &  &  &  &  &  &  &  &  &  \\
& \multicolumn{12}{c}{\textbf{Median Bias}} \\
\rule{0pt}{15pt}$\widehat{\gamma }_{Z1}$ & -0.007 & -0.004 & -0.005 & -0.011
& -0.032 & -0.027 & -0.025 & -0.028 & -0.257 & -0.230 & -0.209 & -0.177 \\
\rule{0pt}{15pt}$\widehat{R}_{Z1}^{2,\txtIdioVolFM}$ & -0.153 & -0.138
& -0.127 & -0.115 & -0.146 & -0.132 & -0.121 & -0.110 & -0.484 & -0.465 &
-0.448 & -0.417 \\
\rule{0pt}{15pt}$\widehat{Corr}\left( C_{Z1},C_{Z2}\right) $ & -0.129 &
-0.104 & -0.086 & -0.059 & -0.147 & -0.118 & -0.100 & -0.070 & -0.342 &
-0.334 & -0.325 & -0.307 \\
\rule{0pt}{15pt}$\widehat{Corr}\left( C_{Z1}^{resid},C_{Z2}^{resid}\right) $
& -0.089 & -0.064 & -0.045 & -0.018 & -0.109 & -0.082 & -0.061 & -0.029 &
-0.245 & -0.239 & -0.232 & -0.218 \\
& \multicolumn{12}{c}{\textbf{\rule{0pt}{15pt}IQR}} \\
\rule{0pt}{15pt}$\widehat{\gamma }_{Z1}$ & 0.173 & 0.157 & 0.141 & 0.118 &
0.173 & 0.154 & 0.140 & 0.118 & 0.079 & 0.078 & 0.078 & 0.076 \\
\rule{0pt}{15pt}$\widehat{R}_{Z1}^{2,\txtIdioVolFM}$ & 0.180 & 0.166 &
0.154 & 0.133 & 0.201 & 0.185 & 0.170 & 0.141 & 0.040 & 0.042 & 0.044 & 0.046
\\
\rule{0pt}{15pt}$\widehat{Corr}\left( C_{Z1},C_{Z2}\right) $ & 0.279 & 0.257
& 0.238 & 0.211 & 0.321 & 0.289 & 0.266 & 0.229 & 0.039 & 0.041 & 0.043 &
0.048 \\
\rule{0pt}{15pt}$\widehat{Corr}\left( C_{Z1}^{resid},C_{Z2}^{resid}\right) $
& 0.330 & 0.304 & 0.280 & 0.249 & 0.381 & 0.344 & 0.311 & 0.273 & 0.040 &
0.042 & 0.044 & 0.049 \\
& \multicolumn{12}{c}{\textbf{\rule{0pt}{15pt}RMSE}} \\
\rule{0pt}{15pt}$\widehat{\gamma }_{Z1}$ & 0.130 & 0.116 & 0.105 & 0.090 &
0.132 & 0.118 & 0.108 & 0.093 & 0.263 & 0.238 & 0.217 & 0.185 \\
\rule{0pt}{15pt}$\widehat{R}_{Z1}^{2,\txtIdioVolFM}$ & 0.206 & 0.185 &
0.170 & 0.150 & 0.242 & 0.192 & 0.174 & 0.152 & 0.484 & 0.466 & 0.449 & 0.418
\\
\rule{0pt}{15pt}$\widehat{Corr}\left( C_{Z1},C_{Z2}\right) $ & 0.257 & 0.226
& 0.203 & 0.169 & 0.309 & 0.260 & 0.229 & 0.187 & 0.343 & 0.335 & 0.327 &
0.309 \\
\rule{0pt}{15pt}$\widehat{Corr}\left( C_{Z1}^{resid},C_{Z2}^{resid}\right) $
& 0.300 & 0.261 & 0.235 & 0.199 & 0.394 & 0.309 & 0.266 & 0.213 & 0.247 &
0.241 & 0.234 & 0.221 \\ \\  \hline\hline
\end{tabular*}

{\caption{Finite sample properties of our estimators using 10 years of data sampled at 5 minutes. The true values are $\gamma_{Z1}=0.450$, 
$R^{2,\txtIdioVolFM}_{Z1}=0.342$, $Corr \left( C_{Z1},C_{Z2} \right)=0.523$, 
$Corr\left(C^{resid}_{Z1},C^{resid}_{Z2} \right)=0.424$. Model 2. 
 }\label{tbl:const1}}

\end{sidewaystable} 

\begin{sidewaystable}
\begin{tabular*}{1.0\textwidth}{l|@{\extracolsep{\fill}}cccccccccccc}
\hline\hline
\rule{0pt}{15pt}\hspace{0.1in}\text{{}} & \multicolumn{4}{c}{LIN} &
\multicolumn{4}{c}{AN} & \multicolumn{4}{c}{Naive} \\
$\widehat{\theta }$ & 1.5 & 2 & 2.5 & 3 & 1.5 & 2 & 2.5 & 3 & 1.5 & 2 & 2.5
& 3 \\ \hline
&  &  &  &  &  &  &  &  &  &  &  &  \\
& \multicolumn{12}{c}{\textbf{Median Bias}} \\
\rule{0pt}{15pt}$\widehat{\gamma }_{Z1}$ & -0.034 & -0.029 & -0.022 & -0.013
& -0.052 & -0.044 & -0.036 & -0.025 & -0.304 & -0.295 & -0.275 & -0.267 \\
\rule{0pt}{15pt}$\widehat{R}_{Z1}^{2,\txtIdioVolFM}$ & -0.140 & -0.123
& -0.109 & -0.086 & -0.135 & -0.117 & -0.103 & -0.080 & -0.496 & -0.492 &
-0.477 & -0.473 \\
\rule{0pt}{15pt}$\widehat{Corr}\left( C_{Z1},C_{Z2}\right) $ & -0.146 &
-0.128 & -0.114 & -0.091 & -0.163 & -0.143 & -0.129 & -0.104 & -0.327 &
-0.323 & -0.321 & -0.317 \\
\rule{0pt}{15pt}$\widehat{Corr}\left( C_{Z1}^{resid},C_{Z2}^{resid}\right) $
& -0.118 & -0.105 & -0.095 & -0.076 & -0.138 & -0.123 & -0.111 & -0.092 &
-0.220 & -0.216 & -0.216 & -0.212 \\
& \multicolumn{12}{c}{\textbf{\rule{0pt}{15pt}IQR}} \\
\rule{0pt}{15pt}$\widehat{\gamma }_{Z1}$ & 0.147 & 0.132 & 0.118 & 0.100 &
0.146 & 0.131 & 0.117 & 0.099 & 0.062 & 0.062 & 0.063 & 0.063 \\
\rule{0pt}{15pt}$\widehat{R}_{Z1}^{2,\txtIdioVolFM}$ & 0.165 & 0.148 &
0.137 & 0.119 & 0.176 & 0.158 & 0.145 & 0.125 & 0.032 & 0.032 & 0.034 & 0.034
\\
\rule{0pt}{15pt}$\widehat{Corr}\left( C_{Z1},C_{Z2}\right) $ & 0.260 & 0.232
& 0.209 & 0.175 & 0.287 & 0.249 & 0.224 & 0.188 & 0.032 & 0.032 & 0.033 &
0.033 \\
\rule{0pt}{15pt}$\widehat{Corr}\left( C_{Z1}^{resid},C_{Z2}^{resid}\right) $
& 0.312 & 0.280 & 0.254 & 0.211 & 0.341 & 0.303 & 0.273 & 0.225 & 0.032 &
0.032 & 0.033 & 0.033 \\
& \multicolumn{12}{c}{\textbf{\rule{0pt}{15pt}RMSE}} \\
\rule{0pt}{15pt}$\widehat{\gamma }_{Z1}$ & 0.115 & 0.102 & 0.091 & 0.076 &
0.121 & 0.106 & 0.095 & 0.078 & 0.307 & 0.299 & 0.279 & 0.271 \\
\rule{0pt}{15pt}$\widehat{R}_{Z1}^{2,\txtIdioVolFM}$ & 0.192 & 0.165 &
0.147 & 0.121 & 0.198 & 0.168 & 0.148 & 0.121 & 0.496 & 0.493 & 0.478 & 0.474
\\
\rule{0pt}{15pt}$\widehat{Corr}\left( C_{Z1},C_{Z2}\right) $ & 0.251 & 0.220
& 0.196 & 0.162 & 0.283 & 0.243 & 0.215 & 0.177 & 0.328 & 0.324 & 0.322 &
0.318 \\
\rule{0pt}{15pt}$\widehat{Corr}\left( C_{Z1}^{resid},C_{Z2}^{resid}\right) $
& 0.291 & 0.249 & 0.221 & 0.182 & 0.760 & 0.279 & 0.245 & 0.199 & 0.221 &
0.218 & 0.218 & 0.214 \\
  \\ \hline\hline
\end{tabular*}

{\caption{Finite sample properties of our estimators using 10 years of data sampled at 1 minute. The true values are $\gamma_{Z1}=0.450$, 
$R^{2,\txtIdioVolFM}_{Z1}=0.336$, $Corr \left( C_{Z1},C_{Z2} \right)=0.514$, $Corr\left(C^{resid}_{Z1},C^{resid}_{Z2} \right)=0.408$. Model 2.}\label{tbl:const2}}
\end{sidewaystable} 


\begin{sidewaystable}

\begin{tabular*}{1.0\textwidth}{l|@{\extracolsep{\fill}}cccccccccccc}
\hline\hline
\rule{0pt}{15pt}\hspace{0.1in}\text{{}} & \multicolumn{4}{c}{LIN} &
\multicolumn{4}{c}{AN} & \multicolumn{4}{c}{Naive} \\
$\widehat{\theta }$ & 1.5 & 2 & 2.5 & 3 & 1.5 & 2 & 2.5 & 3 & 1.5 & 2 & 2.5
& 3 \\ \hline
&  &  &  &  &  &  &  &  &  &  &  &  \\
& \multicolumn{12}{c}{\textbf{Median Bias}} \\
\rule{0pt}{15pt}$\widehat{\gamma }_{Z1}$ & -0.075 & -0.072 & -0.068 & -0.061
& -0.096 & -0.089 & -0.083 & -0.075 & -0.323 & -0.315 & -0.299 & -0.291 \\
\rule{0pt}{15pt}$\widehat{R}_{Z1}^{2,\txtIdioVolFM}$ & -0.183 & -0.169
& -0.155 & -0.139 & -0.183 & -0.169 & -0.156 & -0.137 & -0.500 & -0.496 &
-0.484 & -0.480 \\
\rule{0pt}{15pt}$\widehat{Corr}\left( C_{Z1},C_{Z2}\right) $ & -0.187 &
-0.169 & -0.161 & -0.145 & -0.214 & -0.194 & -0.185 & -0.166 & -0.321 &
-0.316 & -0.317 & -0.313 \\
\rule{0pt}{15pt}$\widehat{Corr}\left( C_{Z1}^{resid},C_{Z2}^{resid}\right) $
& -0.144 & -0.128 & -0.125 & -0.116 & -0.167 & -0.155 & -0.146 & -0.139 &
-0.209 & -0.205 & -0.207 & -0.202 \\
& \multicolumn{12}{c}{\textbf{\rule{0pt}{15pt}IQR}} \\
\rule{0pt}{15pt}$\widehat{\gamma }_{Z1}$ & 0.229 & 0.205 & 0.184 & 0.154 &
0.225 & 0.202 & 0.184 & 0.154 & 0.092 & 0.092 & 0.093 & 0.093 \\
\rule{0pt}{15pt}$\widehat{R}_{Z1}^{2,\txtIdioVolFM}$ & 0.246 & 0.223 &
0.206 & 0.177 & 0.265 & 0.238 & 0.218 & 0.187 & 0.047 & 0.047 & 0.049 & 0.049
\\
\rule{0pt}{15pt}$\widehat{Corr}\left( C_{Z1},C_{Z2}\right) $ & 0.407 & 0.357
& 0.325 & 0.281 & 0.453 & 0.394 & 0.354 & 0.299 & 0.047 & 0.046 & 0.049 &
0.048 \\
\rule{0pt}{15pt}$\widehat{Corr}\left( C_{Z1}^{resid},C_{Z2}^{resid}\right) $
& 0.475 & 0.419 & 0.387 & 0.324 & 0.529 & 0.462 & 0.420 & 0.352 & 0.047 &
0.047 & 0.049 & 0.049 \\
& \multicolumn{12}{c}{\textbf{\rule{0pt}{15pt}RMSE}} \\
\rule{0pt}{15pt}$\widehat{\gamma }_{Z1}$ & 0.184 & 0.165 & 0.150 & 0.127 &
0.192 & 0.172 & 0.156 & 0.134 & 0.330 & 0.321 & 0.307 & 0.298 \\
\rule{0pt}{15pt}$\widehat{R}_{Z1}^{2,\txtIdioVolFM}$ & 0.330 & 0.240 &
0.218 & 0.188 & 0.420 & 0.246 & 0.225 & 0.192 & 0.501 & 0.497 & 0.486 & 0.482
\\
\rule{0pt}{15pt}$\widehat{Corr}\left( C_{Z1},C_{Z2}\right) $ & 0.409 & 0.342
& 0.307 & 0.260 & 0.500 & 0.388 & 0.345 & 0.285 & 0.322 & 0.318 & 0.319 &
0.314 \\
\rule{0pt}{15pt}$\widehat{Corr}\left( C_{Z1}^{resid},C_{Z2}^{resid}\right) $
& 0.510 & 0.399 & 0.355 & 0.287 & 0.813 & 0.481 & 0.417 & 0.323 & 0.212 &
0.207 & 0.209 & 0.205 \\
  \\ \hline\hline
\end{tabular*}

{\caption{Finite sample properties of our estimators using 5 years of data sampled at 1 minute. The true values are $\gamma_{Z1}=0.450$,
$R^{2,\txtIdioVolFM}_{Z1}=0.35$,
 $Corr \left( C_{Z1},C_{Z2} \right)=0.517$, $Corr \left( C^{resid}_{Z1},C^{resid}_{Z2} \right)=0.417$. Model 2.}\label{tbl:r4}}
\end{sidewaystable} 

Next, we study the 
empirical rejection probabilities of the three statistical tests as outlined
in Section \ref{sec:testing}. The first null hypothesis is the absence of
dependence between the IdioVols, $H^1_0 : [C_{Z1},C_{Z2}]_T=0$. The second
null hypothesis we test is the absence of dependence between the IdioVol of
the first stock and the market volatility, $H_0^2 : [C_{Z1},C_{X}]_T=0$. The
third null hypothesis is the absence of dependence in the two residual
IdioVols, $H_0^3 : [C_{Z1}^{resid},C_{Z2}^{resid}]_T=0$.

Table \ref{tbl:MC_size_M1} presents the empirical rejection probabilities of
the t-tests corresponding to the null hypotheses $H_{0}^{1},$ $H_{0}^{2}$,
and $H_{0}^{3}$ in the above, in Model 1. In Model 1, these null hypotheses
are true, so numbers in Table \ref{tbl:MC_size_M1} represent empirical size.
We present the results for two sampling frequencies ($\Delta _{n}=1$ minute
and $\Delta _{n}=5$ minutes) and the two type of estimators (AN and LIN). We
see that the empirical rejection probabilities are reasonably close to the
nominal size of the test.
Neither type of estimator (AN or LIN) seems to dominate the other.
Consistent with the asymptotic theory, the empirical rejection probabilities
of the three tests become closer to the nominal size of the test
when frequency is higher.

Table \ref{tbl:MC_size_M2} presents the empirical rejection probabilities of
the t-tests for the same null hypotheses in Model 2. In this model, all
three null hypotheses are false, so the numbers in the table represent
power. The magnitude of dependence between the residual IdioVols, $%
[C_{Z1}^{resid},C_{Z2}^{resid}]_{T}$, is of course smaller than the
magnitude of the dependence between total IdioVols, $[C_{Z1},C_{Z2}]_{T}$,
so the power in Panel C is lower than in Panel A. However, in most of the
cases the power is still nontrivial, especially for larger block sizes $%
\theta $, and clearly increasing with higher frequency.

\bigskip


\begin{table}[thb]
\begin{center}
\begin{tabular}{cccccccccccccc}
\hline\hline
& \multicolumn{6}{c}{\rule{0pt}{12pt}$\Delta _{n}=5\text{ minutes}$} &  & 
\multicolumn{6}{c}{$\Delta _{n}=1\text{ minute}$} \\ \cline{2-7}\cline{9-14}
& \multicolumn{2}{c}{\rule{0pt}{12pt}$\theta =1.5$} & \multicolumn{2}{c}{$%
\theta =2.0$} & \multicolumn{2}{c}{$\theta =2.5$} &  & \multicolumn{2}{c}{$%
\theta =1.5$} & \multicolumn{2}{c}{$\theta =2.0$} & \multicolumn{2}{c}{$%
\theta =2.5$} \\ 
& \text{AN} & \text{LIN} & \text{AN} & \text{LIN} & \text{AN} & \text{LIN} & 
& \text{AN} & \text{LIN} & \text{AN} & \text{LIN} & \text{AN} & \text{LIN}
\\ \cline{2-7}\cline{9-14}
& \multicolumn{13}{c}{\rule{0pt}{19pt}\textbf{Panel A : } $%
H_{0}^{1}:[C_{Z1},C_{Z2}]_{T}=0$} \\ 
$\alpha =10\%\rule{0pt}{15pt}$ & 9.8 & 12.1 & 10.8 & 12.6 & 11.1 & 12.6 &  & 
10.8 & 11.4 & 11.3 & 11.1 & 10.7 & 11.2 \\ 
$\alpha =5\%$ & 5.5 & 5.2 & 5.4 & 6.1 & 6.0 & 6.9 &  & 6.6 & 6.3 & 5.7 & 5.3
& 5.3 & 5.1 \\ 
$\alpha =1\%$ & 1.0 & 1.5 & 0.9 & 1.7 & 0.5 & 1.1 &  & 1.6 & 1.3 & 1.2 & 1.1
& 0.9 & 0.6 \\ 
& \multicolumn{13}{c}{\rule{0pt}{19pt}\textbf{Panel B : } $%
H_{0}^{2}:[C_{Z1},C_{X}]_{T}=0$} \\ 
$\alpha =10\%\rule{0pt}{15pt}$ & 10.2 & 10.3 & 10.4 & 10.9 & 9.9 & 10.0 &  & 
9.7 & 8.9 & 9.2 & 9.0 & 10.4 & 10.4 \\ 
$\alpha =5\%$ & 4.6 & 4.5 & 4.5 & 4.6 & 4.8 & 5.1 &  & 5.1 & 4.5 & 4.8 & 5.4
& 5.4 & 5.3 \\ 
$\alpha =1\%$ & 0.8 & 0.5 & 1.1 & 0.9 & 1.1 & 1.3 &  & 1.1 & 1.3 & 1.1 & 1.2
& 0.9 & 1.1 \\ 
& \multicolumn{13}{c}{\rule{0pt}{19pt}\textbf{Panel C : }$%
H_{0}^{3}:[C_{Z1}^{resid},C_{Z2}^{resid}]_{T}=0$} \\ 
$\alpha =10\%\rule{0pt}{15pt}$ & 10.0 & 11.7 & 10.8 & 12.7 & 11.5 & 12.6 & 
& 11.0 & 11.2 & 11.2 & 10.7 & 10.7 & 11.7 \\ 
$\alpha =5\%$ & 5.9 & 4.9 & 5.7 & 6.1 & 5.9 & 7.3 &  & 6.4 & 6.4 & 5.4 & 5.2
& 4.9 & 4.9 \\ 
$\alpha =1\%$ & 1.0 & 1.5 & 0.9 & 1.5 & 0.6 & 1.0 &  & 1.8 & 1.4 & 1.3 & 1.2
& 0.9 & 0.6 \\ 
&  &  &  &  &  &  &  &  &  &  &  &  &  \\ \hline\hline
\end{tabular}
\end{center}
\caption{
The size of the t-tests. Model 1. $T=10$ years. $\alpha$ denotes the nominal size of the test. }
\label{tbl:MC_size_M1}
\end{table}

\begin{table}[thb]
\begin{center}
\begin{tabular}{cccccccccccccc}
\hline\hline
& \multicolumn{6}{c}{\rule{0pt}{12pt}$\Delta _{n}=5\text{ minutes}$} &  & 
\multicolumn{6}{c}{$\Delta _{n}=1\text{ minute}$} \\ \cline{2-7}\cline{9-14}
& \multicolumn{2}{c}{\rule{0pt}{12pt}$\theta =1.5$} & \multicolumn{2}{c}{$%
\theta =2.0$} & \multicolumn{2}{c}{$\theta =2.5$} &  & \multicolumn{2}{c}{$%
\theta =1.5$} & \multicolumn{2}{c}{$\theta =2.0$} & \multicolumn{2}{c}{$%
\theta =2.5$} \\ 
& \text{AN} & \text{LIN} & \text{AN} & \text{LIN} & \text{AN} & \text{LIN} & 
& \text{AN} & \text{LIN} & \text{AN} & \text{LIN} & \text{AN} & \text{LIN}
\\ \cline{2-7}\cline{9-14}
& \multicolumn{13}{c}{\rule{0pt}{19pt}\textbf{Panel A : } $%
H_{0}^{1}:[C_{Z1},C_{Z2}]_{T}=0$} \\ 
$\alpha =10\%\rule{0pt}{15pt}$ & 20.3 & 31.5 & 36.8 & 47.0 & 54.2 & 64.8 & 
& 32.5 & 39.8 & 64.6 & 69.6 & 88.0 & 91.0 \\ 
$\alpha =5\%$ & 11.9 & 21.4 & 25.4 & 36.5 & 41.0 & 54.0 &  & 21.8 & 28.1 & 
49.5 & 57.2 & 79.2 & 84.4 \\ 
$\alpha =1\%$ & 3.0 & 7.0 & 8.2 & 16.9 & 20.1 & 28.6 &  & 9.9 & 13.2 & 27.6
& 32.2 & 54.8 & 62.2 \\ 
& \multicolumn{13}{c}{\rule{0pt}{19pt}\textbf{Panel B : } $%
H_{0}^{2}:[C_{Z1},C_{X}]_{T}=0$} \\ 
$\alpha =10\%\rule{0pt}{15pt}$ & 60.2 & 67.6 & 83.0 & 87.9 & 93.9 & 96.3 & 
& 91.8 & 93.6 & 99.6 & 99.6 & 100.0 & 100.0 \\ 
$\alpha =5\%$ & 45.8 & 57.2 & 72.9 & 79.0 & 88.5 & 91.9 &  & 86.6 & 89.5 & 
98.4 & 98.8 & 100.0 & 100.0 \\ 
$\alpha =1\%$ & 23.4 & 31.6 & 50.8 & 58.6 & 70.6 & 76.4 &  & 68.5 & 72.5 & 
94.0 & 95.2 & 99.2 & 99.3 \\ 
& \multicolumn{13}{c}{\rule{0pt}{19pt}\textbf{Panel C : }$%
H_{0}^{3}:[C_{Z1}^{resid},C_{Z2}^{resid}]_{T}=0$} \\ 
$\alpha =10\%\rule{0pt}{15pt}$ & 14.2 & 19.9 & 22.6 & 29.5 & 30.9 & 38.6 & 
& 19.6 & 22.3 & 33.5 & 36.5 & 52.9 & 58.4 \\ 
$\alpha =5\%$ & 7.4 & 12.6 & 14.1 & 20.5 & 21.6 & 29.2 &  & 12.1 & 14.8 & 
22.4 & 26.6 & 39.8 & 44.6 \\ 
$\alpha =1\%$ & 1.5 & 3.3 & 4.8 & 6.9 & 8.4 & 12.1 &  & 3.2 & 5.2 & 10.0 & 
12.1 & 19.5 & 22.9 \\ 
&  &  &  &  &  &  &  &  &  &  &  &  &  \\ \hline\hline
\end{tabular}
\end{center}
\caption{The power of the t-tests. Model 2. 
$T=10$ years. $\alpha$ denotes the nominal size of the test. }
\label{tbl:MC_size_M2}
\end{table}

\newpage

\section{Conclusion}

We introduce an econometric framework for analysis of cross-sectional
dependence in the IdioVols of assets using high frequency data. First, we
provide bias-corrected estimators of standard measures of dependence between
IdioVols, as well as the associated asymptotic theory. Second, we study an
IdioVol Factor Model, in which we decompose the variation in IdioVols into
two parts: the variation related to the systematic factors such as the
market volatility, and the residual variation.
We provide the asymptotic theory that allows us to test, for example,
whether the residual (non-systematic) components of the IdioVols exhibit
cross-sectional dependence.

To provide the bias-corrected estimators and inference results, we develop a
new asymptotic theory for general estimators of quadratic covariation of
vector-valued (possibly) nonlinear transformations of the spot covariance
matrices. This theoretical contribution is of its own interest, and can be
applied in other contexts. For example, our results can be used to conduct
inference for the cross-sectional dependence in asset betas.

We apply our methodology to the S\&P100 index components, and document
strong cross-sectional dependence in their Idiosyncratic Volatilities. We
consider two different sets of idiosyncratic volatility factors, and find
that neither can fully account for the cross-sectional dependence in
idiosyncratic volatilities. For each model, we map out the network of
dependencies in residual (non-systematic) Idiosyncratic Volatilities across
all stocks.

\section{Acknowledgements}

We are grateful to co-editors Torben Andersen and Serena Ng, two associate
editors, and four anonymous referees for numerous helpful suggestions. We
benefited from discussions with Marine Carrasco, Yoosoon Chang, Valentina
Corradi, Russell Davidson, Jean-Marie Dufour, Prosper Dovonon, Kirill
Evdokimov, S\'ilvia Gon\c calves, Peter Hansen, Jean Jacod, Dennis
Kristensen, Joon Park, Benoit Perron, and Dacheng Xiu. We thank seminar
participants at University of Amsterdam, Bank of Canada, Concordia, HEC
Montreal, Indiana, LSE, McGill, NC State, Pennsylvania, Surrey, Toulouse,
UCL, Warwick, Western Ontario, as well as participants of various
conferences, for helpful comments and suggestions.
Ilze Kalnina is grateful to UCL and CeMMaP for their hospitality and
support. 
She is also grateful to the Economics Department, the Gregory C. Chow
Econometrics Research Program, and the Bendheim Center for Finance at
Princeton University for their hospitality. 

\newpage 

\singlespacing
\bibliographystyle{econometrica}
\bibliography{KKbib}
\FloatBarrier

\clearpage
\begin{appendix}%
{\noindent \textbf{\LARGE Appendix}}



\setcounter{theorem}{0} \setcounter{section}{0} \setcounter{subsection}{0} %
\setcounter{equation}{0} \renewcommand
{\thesection}{\Alph{section}} \renewcommand\thesubsection{\Alph{section}.%
\arabic{subsection}} \renewcommand{\theequation}{\Alph{section}.%
\arabic{equation}} \renewcommand{\thelemma}{\Alph{section}\arabic{lemma}} %
\renewcommand{\thetheorem}{\Alph{section}\arabic{theorem}} %
\renewcommand{\theproposition}{\Alph{section}\arabic{proposition}}
\renewcommand{\thepage}{S-\arabic{page}}%


\numberwithin{figure}{section} \numberwithin{table}{section}
\setcounter{page}{1}

\smallskip

Sections \ref{sec:notation}-\ref{sec:proofs_auxiliary} contain all proofs.
Section \ref{sec:appendix_numerical} contains some numerical implementation details.
Section \ref{sec:additional_figures} contains additional figures for the empirical application.

The proofs are organised as follows. 
Section \ref{sec:notation} introduces additional notation. Section \ref{sec:aux results} presents auxiliary theorems and lemmas 
used to prove Theorems \ref{thm:clt} and \ref{thm:avar} in the main paper.
Section \ref{sec:proof_thm_clt} proves Theorem \ref{thm:clt}. 
Section \ref{sec:proof_thm_avar} proves Theorem \ref{thm:avar}.
Section \ref{sec:proofs_auxiliary} collects the proofs of the auxiliary results of Section \ref{sec:aux results}.

\swpSmall

\section{Notation for Proofs}

\label{sec:notation} Our notation is similar to that of the proofs of \cite%
{jacodrosenbaum-sqrtn} whenever possible. Throughout, we denote by $K$ a
generic constant, which may change from line to line.
We let by convention $\sum_{i=a}^{a^{\prime }}=0$ when $a> a^{\prime }$. For
simplicity, we omit the subscript $r$ for results involving only one object
with this subscript.

\noindent By the usual localization argument, there exists a $\pi $%
-integrable function $J$ on $E$ and a constant such that the stochastic
processes in equations (\ref{eqn:LPP}) and (\ref{eqn:SVP}) satisfy
\begin{equation}
\Vert b\Vert ,\Vert \widetilde{b}\Vert ,\Vert c\Vert ,\Vert \widetilde{c}%
\Vert ,J\leq A,\Vert \delta (w,t,z)\Vert ^{r}\leq J(z).  \label{SA}
\end{equation}

\noindent We set%
\begin{equation*}
\mathcal{F}_{i}^{n}=\mathcal{F}_{i \Delta _{n}}\text{, }%
C_{i}^{n}=C_{i\Delta _{n}}\text{, }%
\overline{C}_{i}^{n}=\overline{C}_{i\Delta _{n}}
\text{, and }
\widehat{C}_{i}^{n}=\widehat{C}_{i\Delta _{n}}\text{.}
\end{equation*}

\noindent For any c\`{a}dl\`{a}g bounded process $Z$, we set
\begin{align*}
& \eta _{t,s}(Z)=\sqrt{\mathbb{E}\Big(\sup_{0<u\leq s}\Vert
Z_{t+u}-Z_{t}\Vert ^{2}|\mathcal{F}_{t}\Big)},~~\text{and} \\
& \eta _{i,j}^{n}(Z)=\sqrt{\mathbb{E}\Big(\sup_{0\leq u\leq j\Delta
_{n}}\Vert Z_{(i-1)\Delta _{n}+u}-Z_{(i-1)\Delta _{n}}\Vert ^{2}|\mathcal{F}%
_{(i-1)\Delta _{n}}\Big)}.
\end{align*}

\noindent For convenience, we decompose $Y_{t}$ as
\begin{equation*}
Y_{t}=Y_{0}+Y_{t}^{\prime }+\sum_{s\leq t}\Delta Y_{s}.
\end{equation*}%
where $Y_{t}^{\prime }=\int_{0}^{t}b_{s}^{^{\prime }}ds+\int_{0}^{t}\sigma
_{s}dW_{s}$ and $b_{t}^{\prime }=b_{t}-\int \delta (t,z)1_{\{\Vert \delta
(t,z)\Vert \leq 1\}}\pi (dz)$.\newline
Let $\widehat{C}_{i}^{\prime n}$ be the local estimator of the spot variance
of the unobservable process $Y^{\prime }$, i.e.,
\begin{equation}
\widehat{C}_{i}^{\prime n}=\frac{1}{k_{n}\Delta _{n}}\sum_{u=0}^{k_{n}-1}(%
\Delta _{i+u}^{n}Y^{\prime })(\Delta _{i+u}^{n}Y)^{\prime \top }=(\widehat{C}%
_{i}^{\prime n,gh})_{1\leq g,h\leq d}.  \label{eqn:CSP}
\end{equation}%
There is no price jump truncation applied in the definition of $\widehat{C}%
_{i}^{\prime n}$ since the process $Y^{\prime }$ is continuous. Hence, it is
more convenient to work with $\widehat{C}_{i}^{\prime n}$ rather than $%
\widehat{C}_{i}^{n}$ ($=\widehat{C}_{i\Delta _{n}}$, defined in equation (\ref{eqn:spotC_hat})).

\noindent We also define
\begin{equation}
\alpha _{i}^{n}
=(\Delta _{i}^{n}Y^{\prime })(\Delta _{i}^{n}Y^{\prime})^{\top }
-
C_{(i-1)\Delta _{n}}\Delta _{n},
\hspace{2mm}
\nu _{i}^{n}
=\widehat{C}_{i}^{^{\prime }n}-C_{(i-1)\Delta _n},
\hspace{2mm}
\text{and}\hspace{2mm}
\lambda_{i}^{n}
=\widehat{C}_{i+k_{n}}^{^{\prime }n}-\widehat{C}_{i}^{^{\prime }n},
\label{eq:not1}
\end{equation}%
which satisfy
\begin{equation}
\nu _{i}^{n}
=
\frac{1}{k_{n}\Delta _{n}}\sum_{j=0}^{k_{n}-1}(\alpha_{i+j}^{n}
+(C_{(i+j-1)\Delta _n}-C_{(i-1)\Delta _n})\Delta _{n})
\hspace{2mm}
\text{and}
\hspace{2mm}
\lambda _{i}^{n}
=
\nu _{i+k_{n}}-\nu _{i}^{n}
+C_{(i+k_n-1)\Delta _n}-C_{(i-1)\Delta _n}.
\label{eqn:BGD}
\end{equation}

\noindent The following multidimensional quantities will be used in the
sequel 

\begin{center}
\begin{tabular}{ll}
$\zeta (1)_{i}^{n}=\frac{1}{\Delta _{n}}\Delta _{i}^{n}Y^{\prime }(\Delta
_{i}^{n}Y^{\prime })^{\top }-C_{i-1}^{n},$ & $\zeta (2)_{i}^{n}=\Delta
_{i}^{n}c,$ \\
$\zeta ^{\prime }(u)_{i}^{n}=\mathbb{E}(\zeta (u)_{i}^{n}|\mathcal{F}%
_{i-1}^{n}),$ & $\zeta ^{\prime \prime }(u)_{i}^{n}=\zeta (u)_{i}^{n}-\zeta
^{\prime }(u)_{i}^{n},$ \\
$\zeta ^{r}(u)_{i}^{n}=\Big(\zeta ^{r}(u)_{i}^{n,gh}\Big)_{1\leq g,h\leq d}$
& $\text{with }r=^{\prime }\text{ or }^{\prime \prime }.$%
\end{tabular}
\end{center}

\noindent For $1\leq g,h\leq d$ and $u,v=1,2$, define%
\begin{equation*}
\rho _{gh}(u,v)_{i}^{n}=\sum_{m=1}^{2k_{n}-1}\lambda (u,v)_{m}^{n}\zeta
_{gh}(u)_{i-m}^{n}.
\end{equation*}

\noindent We also define, for $m\in \{0,\ldots ,2k_{n}-1\}$ and $j,l\in
\mathbb{Z}$,
\begin{equation*}
\varepsilon (1)_{m}^{n}=%
\begin{cases}
-1 & if\hspace{1mm}0\leq m<k_{n} \\
+1 & if\hspace{1mm}k_{n}\leq m<2k_{n},%
\end{cases}%
,\text{ }\varepsilon (2)_{m}^{n}=\sum_{q=m+1}^{2k_{n}-1}\varepsilon
(1)_{q}^{n}=(m+1)\wedge (2k_{n}-m-1),
\end{equation*}

\noindent For any $u,v,m,u^{\prime },v^{\prime }$, we set
\begin{equation*}
z_{u,v}^{n}=%
\begin{cases}
1/\Delta _{n} & \text{if\hspace{1mm}}u=v=1 \\
1 & \text{otherwise},%
\end{cases}%
\end{equation*}%
\begin{align*}
\lambda (u,v;m)_{j,l}^{n}& =\frac{3}{2k_{n}^{3}}\sum_{q=0\vee
(j-m)}^{(l-m-1)\vee (2k_{n}-m-1)}\varepsilon (u)_{q}^{n}\varepsilon
(u)_{q+m}^{n},\hspace{5mm}\lambda (u,v)_{m}^{n}=\lambda
(u,v;m)_{0,2k_{n}}^{n}, \\
M(u,v;u^{\prime },v^{\prime })_{n}& =z_{u,v}^{n}z_{u^{\prime },v^{\prime
}}^{n}\sum_{m=1}^{2k_{n}-1}\lambda (u,v)_{m}^{n}\lambda (u^{\prime
},v^{\prime })_{m}^{n}.
\end{align*}

\noindent We also need some notation for volatility jumps. Denote by $N_{s}$
the number of jumps in $C$ from time $0$ to $s$. Let %
\begin{eqnarray}
L\left( n\right) &=&\left\{ i=k_{n}+1,k_{n}+2,...:N_{\left( i+3\right)
k_{n}\Delta _{n}}-N_{\left( i-1\right) k_{n}\Delta _{n}}=0\right\} ,  \notag
\\
L\left( n,T\right) &=&\left\{ i=1,2,...,\left[ T/\Delta _{n}\right]
-3k_{n}+1\right\} \cap L\left( n\right) ,  \label{eqn:L_def} \\
L^{\prime }\left( n,T\right) &=&\left\{ i=1,2,...,\left[ T/\Delta _{n}\right]
:i-2k_{n}\in L\left( n,T\right) \right\} ,  \notag \\
\overline{L}\left( n,T\right) &=&\left\{ i=1,2,...,\left[ T/\Delta _{n}%
\right] -3k_{n}+1\right\} \backslash L\left( n\right) .  \notag
\end{eqnarray}

Additionally, set%
\begin{align}
\overline{A11}(H,gh,u;G,ab,v)_{T}^{n}& 
=
\frac{3}{2k_{n}^{3}}\sum_{i\in
L^{\prime }\left( n,T\right) }\Big(\sum_{j=0}^{2k_{n}-1}\varepsilon
(u)_{j}^{n}\varepsilon (v)_{j}^{n}\Big)(\partial _{gh}H\partial
_{ab}G)(C_{(i-2k_n-1)\Delta_n})\zeta (u)_{i}^{n,gh}\zeta (v)_{i}^{n,ab}  \notag \\
& 
=
\lambda (u,v)_{0}^{n}\sum_{i\in L^{\prime }\left( n,T\right) }(\partial
_{gh}H\partial _{ab}G)(C_{(i-2k_n-1)\Delta_n})\zeta (u)_{i}^{n,gh}\zeta
(v)_{i}^{n,ab},  \label{eqn:defA11}
\end{align}%
and%
\begin{align}
\overline{A12}(H,gh,u;G,ab,v)_{T}^{n}& =\frac{3}{2k_{n}^{3}}\sum_{i\in
L^{\prime }\left( n,T\right) }(\partial _{gh}H\partial
_{ab}G)(C_{(i-2k_n-1)\Delta_n})\sum_{m=1}^{(i-1)\wedge
(2k_{n}-1)}\sum_{j=0}^{(2k_{n}-m-1)}\varepsilon (u)_{j}^{n}\varepsilon
(v)_{j+m}^{n}  \notag \\
& \times \zeta _{gh}(u)_{i-m}^{n}\zeta _{ab}(v)_{i}^{n}.  \label{eqn:defA12}
\end{align}

Denote by $\vartheta _{i}^{AN}$ and $\vartheta _{i}^{LIN}$ the $i^{th}$
summand of $\widehat{\left[ H(C),G(C)\right] _{T}^{c}}^{AN}$ and $\widehat{%
\left[ H(C),G(C)\right] _{T}^{c}}^{LIN}$, without the volatility jump
truncation, so they satisfy%
\begin{eqnarray}
\widehat{\left[ H(C),G(C)\right] _{T}^{c}}^{AN}
&=&\sum_{i=k_{n}+1}^{[T/\Delta _{n}]-3k_{n}+1}\vartheta _{i}^{AN}1_{\left\{
A_{i}\cap A_{i+k_{n}}\right\} },\text{ and}  \label{eqn:v_AN_def} \\
\widehat{\left[ H(C),G(C)\right] _{T}^{c}}^{LIN}
&=&\sum_{i=k_{n}+1}^{[T/\Delta _{n}]-3k_{n}+1}\vartheta _{i}^{LIN}1_{\left\{
A_{i}\cap A_{i+k_{n}}\right\} }  \label{eqn:v_LIN_def}
\end{eqnarray}%
Let $\vartheta _{i}$ be either $\vartheta _{i}^{LIN}$ or $\vartheta
_{i}^{AN} $.

\section{Auxiliary Lemmas and Theorems}

\label{sec:aux results} This section presents useful auxiliary results,
which are used in the proofs of Theorems \ref{thm:clt} and \ref{thm:avar}.
The results of this section are proved in Section \ref{sec:proofs_auxiliary}
below.

First, we explain why we can assume, without loss of generality, that the
derivatives of functions $H_{r}$ and $G_{r}$ are bounded, for $r=1,\ldots
,\kappa $. Assumptions of Theorem \ref{thm:clt} imply Lemma 2 of \cite%
{litodorovtauchen17-adaptive}. Therefore, we can assume that the variables $%
\widehat{C}_{i\Delta _{n}}$ are bounded, uniformly over $i\in \left\{ 0,...,%
\left[ T/\Delta _{n}\right] -k_{n}+1\right\} $, with probability approaching
one. Using the spatial localization argument of \cite%
{litodorovtauchen-dependencies}, which in turn uses the spatial localization
argument of \cite{litodorovtauchen17-adaptive}, we can assume that $H_{r}$
and $G_{r}$ are compactly supported without loss of generality. Hence, the
derivatives of functions $H_{r}$ and $G_{r}$ are bounded, for $r=1,\ldots
,\kappa $.

We start with two auxiliary theorems for volatility jump truncation.

\begin{theorem}
\label{thm:vol_trunc_on_L} Under the assumptions of Theorem \ref{thm:clt},
we have
\begin{equation}
\sum_{i\in L\left( n,T\right) }\vartheta _{i}1_{\left\{ A_{i}\cap
A_{i+k_{n}}\right\} }-\sum_{i\in L\left( n,T\right) }\vartheta
_{i}=o_{p}\left( \Delta _{n}^{1/4}\right) .  \notag
\end{equation}
\end{theorem}

\begin{theorem}
\noindent \label{thm:vol_trunc_on_Lbar} Under the assumptions of Theorem \ref%
{thm:clt}, we have
\begin{equation}
\sum_{i\in \overline{L}\left( n,T\right) }\vartheta _{i}1_{\left\{ A_{i}\cap
A_{i+k_{n}}\right\} }=o_{p}\left( \Delta _{n}^{1/4}\right) .  \notag
\end{equation}
\end{theorem}

Theorems \ref{thm:vol_trunc_on_L} and \ref{thm:vol_trunc_on_Lbar} allow us
to focus on the simpler leading term $\sum_{i\in L\left( n,T\right)
}\vartheta _{i}$ instead of the original estimator(s) $\sum_{i=k_{n}+1}^{[T/%
\Delta _{n}]-3k_{n}+1}\vartheta _{i}1_{\left\{ A_{i}\cap A_{i+k_{n}}\right\}
}$ for the remaining proofs. Our next theorem shows negligibility of price
jump truncation.

\begin{theorem}
\label{thm:intermediate1} Let $\vartheta _{i}^{\prime LIN}$ and $\vartheta
_{i}^{\prime AN}$ be the modifications of $\vartheta _{i}^{LIN}$ and $%
\vartheta _{i}^{AN}$ obtained by replacing $\widehat{C}_{i}^{n}$ by $%
\widehat{C}_{i}^{^{\prime }n}$ in the definition of $\vartheta _{i}^{LIN}$
and $\vartheta _{i}^{AN}$ in equations (\ref{eqn:v_LIN_def}) and (\ref%
{eqn:v_AN_def}). Under the assumptions of Theorem \ref{thm:clt}, we have
\begin{align}
\Delta _{n}^{-1/4}\Big(\sum_{i\in L\left( n,T\right) }\vartheta
_{i}^{LIN}-\sum_{i\in L\left( n,T\right) }\vartheta _{i}^{\prime LIN}\Big)&
\overset{\mathbb{P}}{\longrightarrow }0  \notag \\
\text{and}\hspace{3mm}\Delta _{n}^{-1/4}\Big(\sum_{i\in L\left( n,T\right)
}\vartheta _{i}^{AN}-\sum_{i\in L\left( n,T\right) }\vartheta _{i}^{\prime
AN}\Big)& \overset{\mathbb{P}}{\longrightarrow }0.
\end{align}
\end{theorem}

\noindent Theorem \ref{thm:intermediate1} allows, in particular, to focus on
the derivation of the asymptotic distributions of $\sum_{i\in L\left(
n,T\right) }\vartheta _{i}^{\prime LIN}$ and $\sum_{i\in L\left( n,T\right)
}\vartheta _{i}^{\prime AN}$. The next theorem connects the \textit{LIN} and
\textit{AN} versions of these quantities. To state the theorem, define
\begin{align}
\vartheta _{i}^{\left( A\right) }& =\frac{3}{2k_{n}}\sum_{g,h,a,b=1}^{d}%
\Bigg(\Big(\partial _{gh}H\partial _{ab}G\big)
(C_{(i-1)\Delta_n})\Big[(\widehat{C}%
_{i+k_{n}}^{^{\prime }n,gh}-\widehat{C}_{i}^{^{\prime }n,gh})(\widehat{C}%
_{i+k_{n}}^{^{\prime }n,ab}-\widehat{C}_{i}^{^{\prime }n,ab})
\label{eqn:v_A_def} \\
& -\frac{2}{k_{n}}(\widehat{C}_{i}^{^{\prime }n,ga}\widehat{C}_{i}^{^{\prime
}n,hb}+\widehat{C}_{i}^{^{\prime }n,gb}\widehat{C}_{i}^{^{\prime }n,ha})\Big]%
\Bigg).  \notag
\end{align}%
where
superscript $\left( A\right) $ stands for \textquotedblleft approximated".
For simplicity, we do not index the above quantity by a prime although it
depends on $\widehat{C}_{i}^{^{\prime }n}$ instead of $\widehat{C}_{i}^{n}$.

\begin{theorem}
\label{thm:intermediate2} Under the assumptions of Theorem \ref{thm:clt}, we
have
\begin{align}
& \Delta _{n}^{-1/4}\Big(\sum_{i\in L\left( n,T\right) }\vartheta
_{i}^{\prime LIN}-\sum_{i\in L\left( n,T\right) }\vartheta _{i}^{\left(
A\right) }\Big)\overset{\mathbb{P}}{\longrightarrow }0\hspace{3mm}\text{and}
\notag \\
& \Delta _{n}^{-1/4}\Big(\sum_{i\in L\left( n,T\right) }\vartheta
_{i}^{\prime AN}-\sum_{i\in L\left( n,T\right) }\vartheta _{i}^{\left(
A\right) }\Big)\overset{\mathbb{P}}{\longrightarrow }0,
\end{align}%
where $\vartheta _{i}^{\left( A\right) }$ is defined in equation (\ref%
{eqn:v_A_def}).
\end{theorem}

\noindent Theorem \ref{thm:intermediate2} shows that the leading terms of
the the two estimators of $\widehat{[H(C),G(C)]}_{T}^{c}$, $\sum_{i\in
L\left( n,T\right) }\vartheta _{i}^{\prime LIN}$ and $\sum_{i\in L\left(
n,T\right) }\vartheta _{i}^{\prime AN}$ can be approximated by a certain
quantity with an error of approximation of order smaller than $\Delta
_{n}^{-1/4}$.

\noindent Now, we decompose the approximated estimator as follows%
\begin{equation}
\vartheta _{i}^{\left( A\right) }=\vartheta _{i}^{\left( A1\right)
}-\vartheta _{i}^{\left( A2\right) },  \label{eqn:v_A_decomposition}
\end{equation}%
with%
\begin{equation*}
\vartheta _{i}^{\left( A1\right) }=\frac{3}{2k_{n}}\sum_{g,h,a,b=1}^{d}\big(%
\partial _{gh}H\partial _{ab}G\big)(C_{i-1}^{n})(\widehat{C}%
_{i+k_{n}}^{^{\prime }n,gh}-\widehat{C}_{i}^{^{\prime }n,gh})(\widehat{C}%
_{i+k_{n}}^{^{\prime }n,ab}-\widehat{C}_{i}^{^{\prime }n,ab}),
\end{equation*}%
and
\begin{equation*}
\vartheta _{i}^{\left( A2\right) }=\frac{3}{k_{n}^{2}}\sum_{g,h,a,b=1}^{d}%
\big(\partial _{gh}H\partial _{ab}G\big)(C_{i-1}^{n})(\widehat{C}%
_{i}^{^{\prime }n,ga}\widehat{C}_{i}^{^{\prime }n,hb}+\widehat{C}%
_{i}^{^{\prime }n,gb}\widehat{C}_{i}^{^{\prime }n,ha}).
\end{equation*}%
The following theorem holds:

\begin{theorem}
\label{thm:intermediate3} Under the assumptions of Theorem \ref{thm:clt}, we
have
\begin{align*}
& \frac{1}{\Delta _{n}^{1/4}}\Bigg(\sum_{i\in L\left( n,T\right) }\vartheta
_{i}^{\left( A1\right) }-\sum_{g,h,a,b=1}^{d}\sum_{u,v=1}^{2}\overline{A11}%
(H,gh,u;G,ab,v)_{T}^{n}+\overline{A12}(H,gh,u;G,ab,v)_{T}^{n} \\
& \hspace{30mm}+\overline{A12}(G,ab,v;H,gh,u)_{T}^{n}\Bigg)\overset{\mathbb{P%
}}{\Longrightarrow }0.
\end{align*}
\end{theorem}

\begin{lemma}
\label{lemma:lem1} For any c\`{a}dl\`{a}g bounded process $Z$, for all $%
t,s>0 $, $j,k\geq 0$, set $\eta_{t,s}=\eta_{t,s}(Z)$. Then,
\begin{align*}
&\Delta_n \mathbb{E}\Bigg(\sum_{i=1}^{[t/\Delta_n]} \eta_{i,k_n}\Bigg)%
\longrightarrow 0,\hspace{5mm}\Delta_n \mathbb{E}\Bigg(\sum_{i=1}^{[t/%
\Delta_n]} \eta_{i,2k_n}\Bigg)\longrightarrow 0, \\
& \mathbb{E}\Bigg(\eta_{i+j,k}|\mathcal{F}_i^n\Bigg)\leq \eta_{i,j+k}\hspace{%
2mm}\text{and}\hspace{5mm}\Delta_n \mathbb{E}\Bigg(\sum_{i=1}^{[t/\Delta_n]}
\eta_{i,4k_n}\Bigg)\longrightarrow 0.
\end{align*}
\end{lemma}

\begin{lemma}
\noindent Let $Z$ be a continuous It\^o process with drift $b_t^Z$ and spot
variance process $C_t^{Z}$, and set $\eta_{t,s}=\eta_{t,s}(b^Z,c^Z)$. Then,
the following bounds hold: \label{lemma:ito}
\begin{align}
&\Big|\mathbb{E}(Z_t\Big|\mathcal{F}_0)-tb_0^Z\Big|\leq Kt\eta_{0,t}  \notag
\\
&\Big|\mathbb{E}(Z_t^{j}Z_t^{k}-tC_0^{Z,jk}\Big|\mathcal{F}_0)\Big|\leq
Kt^{3/2}(\sqrt{\Delta_n}+\eta_{0,t})  \notag \\
&\Big|\mathbb{E}\big((Z_t^{j}Z_t^{k}-tC_0^{Z,jk})(C_t^{Z,lm}-C_0^{Z,lm})\Big|%
\mathcal{F}_0\big)\Big|\leq Kt^2  \notag \\
&\Big|\mathbb{E}(Z_t^{j}Z_t^{k}Z_t^{l}Z_t^{m}\Big|\mathcal{F}%
_0)-%
\Delta_n^2(C_0^{Z,jk}C_0^{Z,lm}+C_0^{Z,jl}C_0^{Z,km}+C_0^{Z,jm}C_0^{Z,kl})%
\Big|\leq Kt^{5/2}  \notag \\
&\Big|\mathbb{E}(Z_t^{j}Z_t^{k}Z_t^{l}\Big|\mathcal{F}_0)\Big|\leq Kt^2
\notag \\
&\Big|\mathbb{E}(\prod_{l=1}^6 Z_t^{j_l}\Big|\mathcal{F}_0)-\frac{\Delta_n^3%
}{6}\sum_{l< l^{\prime }}\sum_{k< k^{\prime }}\sum_{m< m^{\prime
}}C_0^{Z,j_lj_{l^{\prime }}}C_0^{Z,j_kj_{k^{\prime }}}C_0^{Z,j_mj_{m^{\prime
}}}\Big|\leq Kt^{7/2}  \notag \\
&\mathbb{E}\Big(\sup_{w\in[0,s]}\Big\|Z_{t+w}-Z_t\Big\|^q \Big|\mathcal{F}_t%
\Big) \leq K_q s^{q/2},\hspace{1mm}\text{and}\hspace{1mm}\Big\|\mathbb{E}%
\Big(Z_{t+s}-Z_t\Big)\Big|\mathcal{F}_t\Big\|\leq Ks.  \label{eqn:SKCC} \\
\end{align}
\end{lemma}

\begin{lemma}
\label{lemma:key} Let $\zeta_i^n$ be a $r$-dimensional $\mathcal{F}_i^n $%
-measurable process satisfying $\|\mathbb{E}(\zeta_{i}^n|\mathcal{F}%
_{i-1}^n)\| \leq L^{\prime }$ and $\mathbb{E}\Big(\|\zeta_{i}^n\|^q \Big|%
\mathcal{F}_{i-1}^n\Big) \leq L_q $. Also, let $\varphi_i^n$ be a
real-valued $\mathcal{F}_{i}^n$-measurable process with $\mathbb{E}\Big(%
\|\varphi_{i+j-1}^n\|^q \Big|\mathcal{F}_{i-1}^n\Big) \leq L^q $ for $q \geq
2$ and $1\leq j \leq 2k_n-1$. Then,
\begin{align*}
\mathbb{E}\Bigg(\Bigg\|\sum_{j=1}^{2k_n-1}\varphi_{i+j-1}^n\zeta_{i+j}^n%
\Bigg\|^q \Bigg|\mathcal{F}_{i-1}^n\Bigg)\leq K_qL^q\Big(L_qk_n^{q/2}+L^{%
\prime q}k_n^q\Big).
\end{align*}
\end{lemma}

\begin{lemma}
Under the assumptions of Theorem \ref{thm:clt}, we have, for $i\in L\left(
n,T\right) $: \label{lem:avar}
\begin{align*}
& \Bigg|\mathbb{E}\Big(\left. \lambda _{i}^{n,jk}\lambda _{i}^{n,lm}\lambda
_{i+2k_{n}}^{n,gh}\lambda _{i+2k_{n}}^{n,ab}\right
\vert 
\mathcal{F}_{i-1}^{n}%
\Big)-\frac{4}{k_{n}^{2}}\Big(%
C_{i-1}^{n,ga}C_{i-1}^{n,hb}+C_{i-1}^{n,gb}C_{i-1}^{n,ha})(C_{i-1}^{n,jl}C_{i-1}^{n,km}+C_{i-1}^{n,jm}C_{i-1}^{n,kl}%
\Big) \\
& -\frac{4\Delta _{n}}{3}\Big(%
C_{i-1}^{n,jl}C_{i-1}^{n,km}+C_{i-1}^{n,jm}C_{i-1}^{n,kl}\Big)\overline{C}%
_{i-1}^{n,gh,ab}-\frac{4\Delta _{n}}{3}\Big(%
C_{i-1}^{n,ga}C_{i-1}^{n,hb}-C_{i-1}^{n,gb}C_{i-1}^{n,ha}\Big)\overline{C}%
_{i-1}^{n,jk,lm} \\
& -\frac{4(k_{n}\Delta _{n})^{2}}{9}\overline{C}_{i-1}^{n,gh,ab}\overline{C}%
_{i-1}^{n,jk,lm}\Bigg|\leq K\Delta _{n}(\Delta _{n}^{1/8}+\eta _{i,4k_{n}}^{n}%
\Big).
\end{align*}
\end{lemma}

\begin{lemma}
Under the assumptions of Theorem \ref{thm:clt}, we have, for $i\in L\left(
n,T\right) $: \label{lemma:approx1}
\begin{align}
& & \Big|\mathbb{E}\Big(\left. \nu _{i}^{n,jk}\nu _{i}^{n,lm}\nu
_{i}^{n,gh}\right\vert \mathcal{F}_{i-1}^{n}\Big)\Big|& \leq K\Delta _{n}^{3/4}%
\Big(\Delta _{n}^{1/4}+\eta _{i,k_{n}}^{n}\Big),  \label{eqn:R1} 
\\
& & \Big|\mathbb{E}\Big(\nu _{i}^{n,jk}\nu _{i}^{n,lm}\left. \Big(%
C_{i+k_{n}-1}^{n,gh}-C_{i-1}^{n,gh}\Big)\right\vert \mathcal{F}_{i-1}^{n}\Big)\Big|%
& \leq K\Delta _{n}^{3/4}\Big(\Delta _{n}^{1/4}+\eta _{i,k_{n}}^{n}\Big),
\label{eqn:R2} 
\\
& & \Big|\mathbb{E}\Big(\nu _{i}^{n,jk}\Big(C_{i+k_{n}-1}^{n,lm}-C_{i-1}^{n,lm}%
\Big)\left. \Big(C_{i+k_{n}-1}^{n,gh}-C_{i-1}^{n,gh}\Big)\right\vert \mathcal{F}%
_{i-1}^{n}\Big)\Big|& \leq K\Delta _{n}^{3/4}\Big(\Delta _{n}^{1/4}+\eta
_{i,k_{n}}^{n}\Big),  \label{eqn:R3} 
\\
& & \Big|\mathbb{E}\Big(\left. \nu _{i}^{n,jk}\lambda _{i}^{n,lm}\lambda
_{i}^{n,gh}\right\vert \mathcal{F}_{i-1}^{n}\Big)\Big|& \leq K\Delta _{n}^{3/4}%
\Big(\Delta _{n}^{1/4}+\eta _{i,2k_{n}}^{n}\Big),  \label{eqn:R4} 
\\
& & \Big|\mathbb{E}\Big(\left. \lambda _{i}^{n,jk}\lambda _{i}^{n,lm}\lambda
_{i}^{n,gh}\right\vert \mathcal{F}_{i-1}^{n}\Big)\Big|& \leq K\Delta _{n}^{3/4}%
\Big(\Delta _{n}^{1/4}+\eta _{i,2k_{n}}^{n}\Big).  \label{eqn:R5}
\end{align}
\end{lemma}

\begin{lemma}
Under the assumptions of Theorem \ref{thm:clt}, we have: \label%
{lemma:convresults}
\begin{align}
& \frac{1}{\Delta _{n}^{1/4}}\sum_{i\in L\left( n,T\right) }(\partial
_{gh}H\partial _{ab}G)(C_{(i-2k_{n}-1)\Delta_n})\rho _{gh}(u,v)_{i}^{n}\zeta
_{ab}^{^{\prime }}(v)_{i}^{n}\overset{\mathbb{P}}{\Longrightarrow }%
0,~~\forall ~~(u,v)  \label{eqn:convergence1} \\
& \frac{1}{\Delta _{n}^{1/4}}\Big(\overline{A11}(H,gh,u;G,ab,v)-%
\int_{0}^{T}(\partial _{gh}H\partial _{ab}G)(C_{t})\overline{C}_{t}^{gh,ab}dt%
\Big)\overset{\mathbb{P}}{\Longrightarrow }0\hspace{2mm}\text{when}%
~~(u,v)=(2,2)  \label{eqn:convergence2} \\
& \frac{1}{\Delta _{n}^{1/4}}\Big(\overline{A11}(H,gh,u;G,ab,v)-\frac{3}{%
\theta ^{2}}\int_{0}^{T}(\partial _{gh}H\partial
_{ab}G)(C_{t})(C_{t}^{ga}C_{t}^{hb}+C_{t}^{gb}C_{t}^{ha})dt\Big)\overset{%
\mathbb{P}}{\Longrightarrow }0\hspace{2mm}  \label{eqn:convergence3} \\
& \text{when}~~(u,v)=(1,1),  \notag \\
& \frac{1}{\Delta _{n}^{1/4}}\overline{A11}(H,gh,u;G,ab,v)\overset{\mathbb{P}%
}{\Longrightarrow }0\hspace{2mm}\text{when}~~(u,v)=(1,2),(2,1)
\label{eqn:convergence4}
\end{align}
\end{lemma}


\section{Proof of Theorem \protect\ref{thm:clt}\label{sec:proof_thm_clt}}

We now prove Theorem \ref{thm:clt}. By Theorem \ref{thm:intermediate3}, we
have
\begin{align*}
\frac{1}{\Delta _{n}^{1/4}}\Bigg(\sum_{i\in L\left( n,T\right) }\vartheta
_{i}^{\left( A1\right) }& -\sum_{g,h,a,b=1}^{d}\sum_{u,v=1}^{2}\overline{A11}%
(H,gh,u;G,ab,v)_{T}^{n}+\overline{A12}(H,gh,u;G,ab,v)_{T}^{n} \\
& +\overline{A12}(G,ab,v;H,gh,u)_{T}^{n}\Bigg)\overset{\mathbb{P}}{%
\Longrightarrow }0.
\end{align*}%
Recalling the definition of $\overline{A12}(H,gh,u;G,ab,v)_{T}^{n}$ from
equation (\ref{eqn:defA12}), Lemma \ref{lemma:convresults} implies that
\begin{align}
& \frac{1}{\Delta _{n}^{1/4}}\Bigg(\sum_{i\in L\left( n,T\right) }\vartheta
_{i}^{\left( A\right) }-[H(C),G(C)]_{T}-\frac{3}{2k_{n}^{3}}%
\sum_{g,h,a,b}^{d}\sum_{u,v=1}^{2}\sum_{i\in L^{\prime }\left( n,T\right) }
\label{eqn:important} \\
& \Big[(\partial _{gh}H\partial _{ab}G)(C_{(i-2k_{n}-1)\Delta_n})\rho
_{gh}(u,v)_{i}^{n}\zeta _{ab}^{^{\prime \prime }}(v)_{i}^{n}+(\partial
_{ab}H\partial _{gh}G)(C_{(i-2k_{n}-1)\Delta_n})\rho _{ab}(v,u)_{i}^{n}\zeta
_{gh}^{^{\prime \prime }}(v)_{i}^{n}\Big]\Bigg)\overset{\mathbb{P}}{%
\Longrightarrow }0.  \notag
\end{align}%
Next, define
\begin{align*}
& \xi (H,gh,u;G,ab,v)_{i}^{n}=\frac{1}{\Delta _{n}^{1/4}}(\partial
_{gh}H\partial _{ab}G)(C_{(i-2k_{n}-1)\Delta_n})\rho _{gh}(u,v)_{i}^{n}\zeta
_{ab}^{\prime \prime }(v)_{i}^{n}, \\
& Z(H,gh,u;G,ab,v)_{t}^{n}=\Delta _{n}^{1/4}\sum_{i=2k_{n}}^{[t/\Delta
_{n}]}\xi (H,gh,u;G,ab,v)_{i}^{n}.
\end{align*}%
Notice that (\ref{eqn:important}) implies
\begin{align}
& \frac{1}{\Delta _{n}^{1/4}}\Big(\sum_{i\in L\left( n,T\right) }\vartheta
_{i}^{\left( A\right) }-[H(C),G(C)]_{T}\Big)\overset{\mathcal{L}}{=}%
\sum_{g,h,a,b=1}^{d}\sum_{u,v=1}^{2}\frac{1}{\Delta _{n}^{1/4}}\Big(%
Z(H,gh,u;G,ab,v)_{T}^{n}  \notag \\
& +Z(H,ab,v;G,gh,u)_{T}^{n}\Big).  \label{eqn:mart}
\end{align}%
The term $\vartheta _{i}^{\left( A\right) }$ depends on functions $H$ and $G$%
, where we have so far suppressed the subscripts $r$, $r=1,..,\kappa $, in
the statement of Theorem \ref{thm:clt} for simplicity. Denote by $\vartheta
_{i,r}^{\left( A\right) }$ the term $\vartheta _{i}^{\left( A\right) }$ that
depends on functions $H_{r}$ and $G_{r}$. Observe that to derive the
asymptotic distribution of $\left( \sum_{i\in L\left( n,T\right) }\vartheta
_{i,1}^{\left( A\right) },...,\sum_{i\in L\left( n,T\right) }\vartheta
_{i,\kappa }^{\left( A\right) }\right) $ , it suffices to study the joint asymptotic
behavior of the family of processes $\frac{1}{\Delta _{n}^{1/4}}%
Z(H,gh,u;G,ab,v)_{T}^{n}$. Notice that $\xi (H,gh,u;G,ab,v)_{i}^{n}$ are
martingale increments relative to the discrete filtration $(\mathcal{F}%
_{i}^{n})$. Therefore, to obtain the joint asymptotic distribution of $\frac{%
1}{\Delta _{n}^{1/4}}Z(H,gh,u;G,ab,v)_{T}^{n}$, it is enough to prove the
following three properties:
\begin{align}
& A\Big((H,gh,u;G,ab,v),(H^{\prime },g^{\prime }h^{\prime },u^{\prime
};G^{\prime },a^{\prime }b^{\prime },v^{\prime })\Big)_{t}^{n}  \notag \\
& =\sum_{i\in L^{\prime }\left( n,T\right) }\mathbb{E}(\xi
(H,gh,u;G,ab,v)_{i}^{n}\xi (H^{\prime },g^{\prime }h^{\prime },u^{\prime
};G^{\prime },a^{\prime }b^{\prime },v^{\prime })_{i}^{n}|\mathcal{F}%
_{i-1}^{n}) \\
& \hspace{10mm}\overset{\mathbb{P}}{\Longrightarrow }A\Big(%
(H,gh,u;G,ab,v),(H^{\prime },g^{\prime }h^{\prime },u^{\prime };G^{\prime
},a^{\prime }b^{\prime },v^{\prime })\Big)_{t},  \label{eqn:cltP1} \\
& \sum_{i\in L^{\prime }\left( n,T\right) }\mathbb{E}\Big(\Big|\xi
(H,gh,u;G,ab,v\Big)_{i}^{n}\Big|^{4}\Big|\mathcal{F}_{i-1}^{n})\overset{%
\mathbb{P}}{\Longrightarrow }0,\text{ and}  \label{eqn:cltP2} \\
& B(N;H,gh,u;G,ab,v)_{t}^{n}:=\sum_{i\in L^{\prime }\left( n,T\right) }%
\mathbb{E}\Big(\xi (H,gh,u;G,ab,v)_{i}^{n}\Delta _{i}^{n}N|\mathcal{F}%
_{i-1}^{n}\Big)\overset{\mathbb{P}}{\Longrightarrow }0,  \label{eqn:cltP3}
\end{align}%
for all $t>0$, all $(H,gh,u;G,ab,v),(H^{\prime },g^{\prime }h^{\prime
},u^{\prime };G^{\prime },a^{\prime }b^{\prime },v^{\prime })$ and all
martingales $N$ which are either bounded and orthogonal to $W$, or equal to
one component $W^{j}$.\newline
Since the derivatives of $H_{r}$ and $G_{r}$ are bounded, equations (\ref%
{eqn:cltP2}) and (\ref{eqn:cltP3}) can be proved by an extension of (B.105)
and (B.106) in \cite{yacjacod14} to multivariate processes.\newline
Next, define
\begin{equation*}
V_{ab}^{a^{\prime }b^{\prime }}(v,v^{\prime })_{t}=%
\begin{cases}
(C_{t}^{aa^{\prime }}C_{t}^{bb^{\prime }}+C_{t}^{ab^{\prime
}}C_{t}^{ba^{\prime }}) & \text{if}\hspace{5mm}(v,v^{\prime })=(1,1) \\
\overline{C}_{t}^{ab,a^{\prime }b^{\prime }} & \text{if}\hspace{5mm}%
(v,v^{\prime })=(2,2) \\
0 & \text{otherwise}.%
\end{cases}%
\end{equation*}%
Using again the boundedness of the derivatives of $H_{r}$ and $G_{r}$, we
can show that
\begin{align*}
A\Big((H,gh,u;G,ab,v),& (H^{\prime },g^{\prime }h^{\prime },u^{\prime
};G^{\prime },a^{\prime }b^{\prime },v^{\prime })\Big)_{t}= \\
& M(u,v;u^{\prime },v^{\prime })\int_{0}^{t}(\partial _{gh}H\partial
_{ab}G\partial _{g^{\prime }h^{\prime }}H\partial _{a^{\prime }b^{\prime
}}G)(C_{s})V_{ab}^{a^{\prime }b^{\prime }}(v,v^{\prime })_{s}V%
_{gh}^{g^{\prime }h^{\prime }}(u,u^{\prime })_{s}ds,
\end{align*}%
with
\begin{equation*}
M(u,v;u^{\prime },v^{\prime })=%
\begin{cases}
3/\theta ^{3} & \text{if}\hspace{5mm}(u,v;u^{\prime },v^{\prime })=(1,1;1,1)
\\
3/4\theta & \text{if}\hspace{5mm}(u,v;u^{\prime },v^{\prime
})=(1,2;1,2),(2,1;2,1) \\
151\theta /280 & \text{if}\hspace{5mm}(u,v;u^{\prime },v^{\prime })=(2,2;2,2)
\\
0 & \text{otherwise}.%
\end{cases}%
\end{equation*}%
Therefore, we have \newline
$A\Big((H,gh,u;G,ab,v),(H^{\prime },g^{\prime }h^{\prime },u^{\prime
};G^{\prime },a^{\prime }b^{\prime },v^{\prime })\Big)_{T}$ =
\begin{equation*}
\begin{cases}
\frac{3}{\theta ^{3}}\int_{0}^{T}(\partial _{gh}H\partial _{ab}G\partial
_{g^{\prime }h^{\prime }}H^{\prime }\partial _{a^{\prime }b^{\prime
}}G^{\prime })(C_{t})(C_{t}^{gg^{\prime }}C_{t}^{hh^{\prime
}}+C_{t}^{gh^{\prime }}C_{t}^{hg^{\prime }})(C_{t}^{aa^{\prime
}}C_{t}^{bb^{\prime }}+C_{t}^{ab^{\prime }}C_{t}^{ba^{\prime }})dt, \\
\hfill \text{if}\hspace{2mm}(u,v;u^{\prime },v^{\prime })=(1,1;1,1) \\
\frac{3}{4\theta }\int_{0}^{T}(\partial _{gh}H\partial _{ab}G\partial
_{g^{\prime }h^{\prime }}H^{\prime }\partial _{a^{\prime }b^{\prime
}}G^{\prime })(C_{t})(C_{t}^{gg^{\prime }}C_{t}^{hh^{\prime
}}+C_{t}^{gh^{\prime }}C_{t}^{hg^{\prime }})\overline{C}_{t}^{ab,a^{\prime
}b^{\prime }}dt,\hfill \text{if}\hspace{2mm}(u,v;u^{\prime },v^{\prime
})=(1,2;1,2) \\
\frac{3}{4\theta }\int_{0}^{T}(\partial _{gh}H\partial _{ab}G\partial
_{g^{\prime }h^{\prime }}H^{\prime }\partial _{a^{\prime }b^{\prime
}}G^{\prime })(C_{t})(C_{t}^{aa^{\prime }}C_{t}^{bb^{\prime
}}+C_{t}^{ab^{\prime }}C_{s}^{ba^{\prime }})\overline{C}_{t}^{gh,g^{\prime
}h^{\prime }}dt,\hfill \text{if}\hspace{2mm}(u,v;u^{\prime },v^{\prime
})=(2,1;2,1) \\
\frac{151\theta }{280}\int_{0}^{T}(\partial _{gh}H\partial _{ab}G\partial
_{g^{\prime }h^{\prime }}H^{\prime }\partial _{a^{\prime }b^{\prime
}}G^{\prime })(C_{t})\overline{C}_{s}^{ab,a^{\prime }b^{\prime }}\overline{C}%
_{t}^{gh,g^{\prime }h^{\prime }}dt,\hfill \text{if}\hspace{2mm}%
(u,v;u^{\prime },v^{\prime })=(2,2;2,2) \\
0\hfill \text{otherwise}.%
\end{cases}%
\end{equation*}
Using equation (\ref{eqn:mart}), we deduce that the asymptotic covariance
between $\sum_{i\in L\left( n,T\right) }\vartheta _{i,r}^{\left( A\right) }$
and $\sum_{i\in L\left( n,T\right) }\vartheta _{i,s}^{\left( A\right) }$ %
is given by
\begin{eqnarray*}
&&\sum_{g,h,a,b=1}^{d}\sum_{g^{\prime },h^{\prime },a^{\prime },b^{\prime
}=1}^{d}\sum_{u,v,u^{\prime },v^{\prime }=1}^{2}\Bigg(A\Big(%
(H_{r},gh,u;G_{r},ab,v),(H_{s},g^{\prime }h^{\prime },u^{\prime
};G_{s},a^{\prime }b^{\prime },v^{\prime })\Big)_{T} \\
&&+A\Big((H_{r},gh,u;G_{r},ab,v),(H_{s},a^{\prime }b^{\prime },v^{\prime
};G_{s},g^{\prime }h^{\prime },u^{\prime })\Big)_{T} \\
&&+A\Big((H_{r},ab,v;G_{r},gh,u),(H_{s},g^{\prime }h^{\prime },u^{\prime
};G_{s},a^{\prime }b^{\prime },v^{\prime })\Big)_{T} \\
&&+A\Big((H_{r},ab,v;H_{r},gh,u),(H_{s},a^{\prime }b^{\prime },v^{\prime
};G_{s},g^{\prime }h^{\prime },u^{\prime })\Big)_{T}\Bigg).
\end{eqnarray*}%
The above expression can be rewritten as
\begin{eqnarray*}
&&\sum_{g,h,a,b=1}^{d}\sum_{j,k,l,m=1}^{d}\Bigg(\frac{6}{\theta ^{3}}%
\int_{0}^{T}\big(\partial _{gh}H_{r}\partial _{ab}G_{r}\partial
_{jk}H_{s}\partial _{lm}G_{s}(C_{t})\big)\Big[%
(C_{t}^{gj}C_{t}^{hk}+C_{t}^{gk}C_{t}^{hj})(C_{t}^{al}C_{t}^{bm}+C_{t}^{am}C_{t}^{bl})
\\
&&+(C_{t}^{aj}C_{t}^{bk}+C_{t}^{ak}C_{t}^{bj})(C_{t}^{gl}C_{t}^{hm}+C_{t}^{gm}C_{t}^{hl})%
\Big]dt \\
&&+\frac{151\theta }{140}\int_{0}^{t}\big(\partial _{gh}H_{r}\partial
_{ab}G_{r}\partial _{jk}H_{s}\partial _{lm}G_{s}(C_{t})\big)\Big[\overline{C}%
_{t}^{gh,jk}\overline{C}_{t}^{ab,lm}+\overline{C}_{t}^{ab,jk}\overline{C}%
_{t}^{gh,lm}\Big]dt \\
&&+\frac{3}{2\theta }\int_{0}^{t}\big(\partial _{gh}H_{r}\partial
_{ab}G_{r}\partial _{jk}H_{s}\partial _{lm}G_{s}(C_{t})\big)\Big[%
(C_{t}^{gj}C_{t}^{hk}+C_{t}^{gk}C_{t}^{hj})\overline{C}%
_{t}^{ab,lm}+(C_{t}^{al}C_{t}^{bm}+C_{t}^{am}C_{t}^{bl})\overline{C}%
_{t}^{gh,jk} \\
&&+(C_{t}^{gl}C_{s}^{hm}+C_{t}^{gm}C_{s}^{hl})\overline{C}%
_{t}^{ab,jk}+(C_{t}^{aj}C_{t}^{bk}+C_{t}^{ak}C_{t}^{bj})\overline{C}%
_{t}^{gh,lm}\Big]dt\Bigg),
\end{eqnarray*}%
which completes the proof.


\section{Proof of Theorem \protect\ref{thm:avar}}

\label{sec:proof_thm_avar}

Recall that $N_{s}$ is the number of jumps in $C$ from time $0$ to $s$. Let %
%
\begin{eqnarray*}
L^{\prime \prime }\left( n\right) &=&\left\{ i=k_{n}+1,k_{n}+2,...:N_{\left(
i+5\right) k_{n}\Delta _{n}}-N_{\left( i-1\right) k_{n}\Delta
_{n}}=0\right\} , \\
L^{\prime \prime }\left( n,T\right) &=&\left\{ i=1,2,...,\left[ T/\Delta _{n}%
\right] -5k_{n}+1\right\} \cap L^{\prime \prime }\left( n\right) , \\
\overline{L}^{\prime \prime }\left( n,T\right) &=&\left\{ i=1,2,...,\left[
T/\Delta _{n}\right] -5k_{n}+1\right\} \backslash L^{\prime \prime }\left(
n\right) .
\end{eqnarray*}

Denote by $\widehat{\omega }_{T}^{r,s,(1)}$, $\widehat{\omega }%
_{T}^{r,s,(2)} $, and $\widehat{\omega }_{T}^{r,s,(3)}$ the $i^{th}$ summand
of $\widehat{\Omega }_{T}^{r,s,(1)}$, $\widehat{\Omega }_{T}^{r,s,(2)}$, and
$\widehat{\Omega }_{T}^{r,s,(3)}$, without the volatility jump truncation,
so they satisfy%
\begin{equation*}
\widehat{\Omega }_{T}^{r,s,(m)}=\sum_{i=k_{n}+1}^{[T/\Delta _{n}]-5k_{n}+1}%
\widehat{\omega }_{T}^{r,s,(m)}1_{\left\{ A_{i}\cap A_{i+k_{n}}\cap
A_{i+2k_{n}}\cap A_{i+3k_{n}}\right\} }\text{ for }m=1,2\text{, and }3\text{.%
}
\end{equation*}

The same methods as in Theorems \ref{thm:vol_trunc_on_L} and \ref%
{thm:vol_trunc_on_Lbar} can be used to show%
\begin{eqnarray*}
\sum_{i\in L^{\prime \prime }\left( n,T\right) }\widehat{\omega }%
_{T}^{r,s,(m)}1_{\left\{ A_{i}\cap A_{i+k_{n}}\cap A_{i+2k_{n}}\cap
A_{i+3k_{n}}\right\} }-\sum_{i\in L^{\prime \prime }\left( n,T\right) }%
\widehat{\omega }_{T}^{r,s,(m)} &=&o_{p}\left( 1\right) \text{ and} \\
\sum_{i\in \overline{L}^{\prime \prime }\left( n,T\right) }\widehat{\omega }%
_{T}^{r,s,(m)}1_{\left\{ A_{i}\cap A_{i+k_{n}}\cap A_{i+2k_{n}}\cap
A_{i+3k_{n}}\right\} } &=&o_{p}\left( 1\right) .
\end{eqnarray*}%
We conclude that the probability limit of $\widehat{\Omega }_{T}^{r,s,(m)}$
is the same as $\sum_{i\in L^{\prime \prime }\left( n,T\right) }\widehat{%
\omega }_{T}^{r,s,(m)}$ for $m=1,2,3$.

Using boundedness of the derivatives of $H_{r},G_{r},H_{s}$ and $G_{s}$ and
Theorem 2.2 in \cite{jacodrosenbaum-sqrtn}, one can show that

\begin{equation*}
\frac{6}{\theta ^{3}}\sum_{i\in L^{\prime \prime }\left( n,T\right) }%
\widehat{\omega }_{T}^{r,s,(1)}\overset{\mathbb{P}}{\longrightarrow }\Sigma
_{T}^{r,s,(1)}.
\end{equation*}%
Next, by equation (3.27) in \cite{jacodrosenbaum-sqrtn}, we have
\begin{equation*}
\frac{3}{2\theta }\left( \sum_{i\in L^{\prime \prime }\left( n,T\right) }%
\widehat{\omega }_{T}^{r,s,(3)}-\frac{6}{\theta }\sum_{i\in L^{\prime \prime
}\left( n,T\right) }\widehat{\omega }_{T}^{r,s,(1)}\right) \overset{\mathbb{P%
}}{\longrightarrow }\Sigma _{T}^{r,s,(3)}.
\end{equation*}%
Finally, to show that
\begin{equation*}
\frac{151\theta }{140}\frac{9}{4\theta ^{2}}\left( \sum_{i\in L^{\prime
\prime }\left( n,T\right) }\widehat{\omega }_{T}^{r,s,(2)}+\frac{4}{\theta
^{2}}\sum_{i\in L^{\prime \prime }\left( n,T\right) }\widehat{\omega }%
_{T}^{r,s,(1)}-\frac{4}{3}\sum_{i\in L^{\prime \prime }\left( n,T\right) }%
\widehat{\omega }_{T}^{r,s,(3)}\right) \overset{\mathbb{P}}{\longrightarrow }%
\Sigma _{T}^{r,s,(2)},
\end{equation*}%
we first observe that the approximation error induced by replacing $\widehat{%
C}_{i}^{n}$ by $\widehat{C}_{i}^{^{\prime }n}$ in Theorem \ref{thm:avar} is
negligible.\newline
For $1\leq g,h,a,b,j,k,l,m\leq d$ and $1\leq r,s\leq d$, we define
\begin{align*}
\widehat{W}_{T}^{n}& =\sum_{i\in L^{\prime \prime }\left( n,T\right)
}(\partial _{gh}H_{r}\partial _{ab}G_{r}\partial _{gh}H_{s}\partial
_{lm}G_{s})(\widehat{C}_{i}^{n})\lambda _{i}^{n,gh}\lambda
_{i}^{n,jk}\lambda _{i+2k_{n}}^{n,ab}\lambda _{i+2k_{n}}^{n,lm}, \\
\widehat{w}(1)_{i}^{n}& =(\partial _{gh}H_{r}\partial _{ab}G_{r}\partial
_{jk}H_{s}\partial _{lm}G_{s})(C_{i-1}^{n})\mathbb{E}(\lambda
_{i}^{n,gh}\lambda _{i}^{n,jk}\lambda _{i+2k_{n}}^{n,ab}\lambda
_{i+2k_{n}}^{n,lm}|\mathcal{F}_{i}^{n}), \\
\widehat{w}(2)_{i}^{n}& =(\partial _{gh}H_{r}\partial _{ab}G_{r}\partial
_{jk}H_{s}\partial _{lm}G_{s})(C_{i-1}^{n})(\lambda _{i}^{n,gh}\lambda
_{i}^{n,jk}\lambda _{i+2k_{n}}^{n,ab}\lambda _{i+2k_{n}}^{n,lm}-\mathbb{E}%
(\lambda _{i}^{n,gh}\lambda _{i}^{n,jk}\lambda _{i+2k_{n}}^{n,ab}\lambda
_{i+2k_{n}}^{n,lm}|\mathcal{F}_{i}^{n})), \\
\widehat{w}(3)_{i}^{n}& =\Big((\partial _{gh}H_{r}\partial
_{ab}G_{r}\partial _{jk}H_{s}\partial _{lm}G_{s})(\widehat{C}%
_{i}^{n})-(\partial _{gh}H_{r}\partial _{ab}G_{r}\partial _{jk}H_{s}\partial
_{lm}G_{s})(C_{i-1}^{n})\Big)\lambda _{i}^{n,gh}\lambda _{i}^{n,jk}\lambda
_{i+2k_{n}}^{n,ab}\lambda _{i+2k_{n}}^{n,lm}, \\
\widehat{W}(u)_{t}^{n}& =\sum_{i\in L^{\prime \prime }\left( n,T\right) }%
\widehat{w}_{i}(u),\hspace{1mm}u=1,2,3.
\end{align*}%
Now, note that we also have $\widehat{W}_{t}^{n}=\widehat{W}(1)_{t}^{n}+%
\widehat{W}(2)_{t}^{n}+\widehat{W}(3)_{t}^{n}$. By Taylor expansion and
using repeatedly the boundedness of $C_{t}$, we obtain, for $i\in L^{\prime
\prime }\left( n,T\right) $
\begin{equation*}
|\widehat{w}(3)_{i}^{n}|\leq K\Vert \nu _{i}^{n}\Vert \Vert \lambda
_{i}^{n}\Vert ^{2}\Vert \lambda _{i+2k_{n}}^{n}\Vert ^{2},
\end{equation*}%
which implies $\mathbb{E}(|\widehat{w}(3)_{i}^{n}|)\leq K\Delta _{n}^{5/4}$
and hence $\widehat{W}(3)_{t}^{n}\overset{\mathbb{P}}{\longrightarrow }0$.
Using Cauchy-Schwartz inequality and the bound $\mathbb{E}(\Vert \lambda
_{i}^{n}\Vert ^{q}|\mathcal{F}_{i}^{n})\leq K\Delta _{n}^{q/4}$, we have $%
\mathbb{E}(|\widehat{w}(2)_{i}^{n}|^{2})\leq K\Delta _{n}^{2}$ for $i\in
L^{\prime \prime }\left( n,T\right) $. Observing furthermore that $\widehat{w%
}(2)_{i}^{n}$ is $\mathcal{F}_{i+4k_{n}}-$measurable, Lemma B.8 in \cite%
{yacjacod14} implies $\widehat{W}(2)_{t}^{n}\overset{\mathbb{P}}{%
\longrightarrow }0$. \newline
Next, define
\begin{align*}
& w_{i}^{n}=(\partial _{gh}H_{r}\partial _{ab}G_{r}\partial
_{jk}H_{s}\partial _{lm}G_{s})(C_{i-1}^{n})\Big[\frac{4}{k_{n}^{2}\Delta _{n}}%
(C_{i-1}^{n,ga}C_{i-1}^{n,hb}+C_{i-1}^{n,gb}C_{i-1}^{n,ha})(C_{i-1}^{n,jl}C_{i-1}^{n,km}+C_{i-1}^{n,jm}C_{i-1}^{n,kl})
\\
& \hspace{2mm}+\frac{4}{3}(C_{i-1}^{n,jl}C_{i-1}^{n,km}+C_{i-1}^{n,jm}C_{i-1}^{n,kl})%
\overline{C}_{i-1}^{n,gh,ab}+\frac{4}{3}%
(C_{i-1}^{n,ga}C_{i-1}^{n,hb}+C_{i-1}^{n,gb}C_{i-1}^{n,ha})\overline{C}_{i-1}^{n,jk,lm}
\\
& \hspace{2mm}+\frac{4(k_{n}^{2}\Delta _{n})}{9}\overline{C}_{i-1}^{n,gh,ab}%
\overline{C}_{i-1}^{n,jk,lm}\Big], \\
& W_{T}^{n}=\Delta _{n}\sum_{i\in L^{\prime \prime }\left( n,T\right)
}w_{i}^{n}.
\end{align*}%
Using the cadlag property of $c$ and $\overline{C}$, $k_{n}\sqrt{\Delta _{n}}%
\rightarrow \theta $, and the Riemann integral convergence, we conclude that
$W_{T}^{n}\overset{\mathbb{P}}{\longrightarrow }W_{T}$ where
\begin{align*}
& W_{T}=\int_{0}^{T}(\partial _{gh}H_{r}\partial _{ab}G_{r}\partial
_{jk}H_{s}\partial _{lm}G_{s})(C_{t})\Big[\frac{4}{\theta ^{2}}%
(C_{t}^{ga}C_{t}^{hb}+C_{t}^{gb}C_{t}^{ha})(C_{t}^{jl}C_{t}^{km}+C_{t}^{jm}C_{t}^{kl})
\\
& +\frac{4}{3}(C_{t}^{jl}C_{t}^{km}+C_{t}^{jm}C_{t}^{kl})\overline{C}%
_{t}^{gh,ab}+\frac{4}{3}(C_{t}^{ga}C_{i}^{hb}+C_{t}^{gb}C_{t}^{ha})\overline{%
C}_{t}^{jk,lm}+\frac{4\theta ^{2}}{9}\overline{C}_{t}^{gh,ab}\overline{C}%
_{t}^{jk,lm}\Big]dt.
\end{align*}%
In addition, by Lemma \ref{lem:avar}, it holds that
\begin{equation*}
\mathbb{E}(|\widehat{W}(1)_{T}^{n}-W_{T}^{n}|)\leq \Delta _{n}\mathbb{E}%
\Bigg(\sum_{i\in L^{\prime \prime }\left( n,T\right) }(\Delta
_{n}^{1/8}+\eta _{i,4k_{n}})\Bigg).
\end{equation*}%
Hence, by the third result of Lemma \ref{lemma:lem1} we have $\widehat{W}%
_{T}^{n}\overset{\mathbb{P}}{\longrightarrow }W_{t}$, from which it follows
that
\begin{align*}
& \frac{9}{4\theta ^{2}}\Big[\widehat{W}(1)_{T}^{n}+\frac{4}{k_{n}^{2}}%
\sum_{i\in L^{\prime \prime }\left( n,T\right) }(\partial _{gh}H_{r}\partial
_{ab}G_{r}\partial _{jk}H_{s}\partial _{lm}G_{s})(\widehat{C}%
_{i}^{n})[C_{i}^{n}(jk,lm)C_{i}^{n}(gh,ab)] \\
& -\frac{2}{k_{n}}\sum_{i\in L^{\prime \prime }\left( n,T\right) }(\partial
_{gh}H_{r}\partial _{ab}G_{r}\partial _{jk}H_{s}\partial _{lm}G_{s})(%
\widehat{C}_{i}^{n})C_{i}^{n}(gh,ab)\lambda _{i}^{n,jk}\lambda _{i}^{n,lm} \\
& -\frac{2}{k_{n}}\sum_{i\in L^{\prime \prime }\left( n,T\right) }(\partial
_{gh}H_{r}\partial _{ab}G_{r}\partial _{jk}H_{s}\partial _{lm}G_{s})(%
\widehat{C}_{i}^{n})C_{i}^{n}(jk,lm)\lambda _{i}^{n,gh}\lambda _{i}^{n,ab}%
\Big] \\
& \overset{\mathbb{P}}{\longrightarrow }\int_{0}^{T}(\partial
_{gh}H_{r}\partial _{ab}G_{r}\partial _{jk}H_{s}\partial _{lm}G_{s})(C_{t})%
\overline{C}_{t}^{gh,ab}\overline{C}_{t}^{jk,lm}dt.
\end{align*}%
The result follows from the above convergence, the already invoked symmetry
argument, and straightforward calculations.


\section{Proofs of Auxiliary Lemmas and Theorems}

\label{sec:proofs_auxiliary} This section is devoted to the proofs of the
auxiliary theorems and lemmas (listed in Section \ref{sec:aux results}) that
were used to prove Theorem \ref{thm:clt} and Theorem \ref{thm:avar}.

\subsection{Proof of Theorem \protect\ref{thm:vol_trunc_on_L}}

\bigskip 

The proof proceeds in three steps. In Step 1, we prove, for $i\in L\left(
n,T\right) $,%
\begin{equation}
P\left( \overline{A}_{i}\right) \leq Ka_{n}\Delta _{n}^{\left( 2-r\right)
\varpi -\varpi ^{\prime }},  \label{eqn:PA_on_L}
\end{equation}%
where $a_{n}$ is a sequence converging to zero, and $\overline{A}_{i}$ is
the complement of $A_{i}$. In Step 2, we prove, for $p\geq 1$ and $i\in
L\left( n,T\right) $,
\begin{equation}
E\left[ \left\vert \vartheta _{i}\right\vert ^{p}\right] \leq K\Delta
_{n}^{p}+Ka_{n}\Delta _{n}^{\left( 4p-r\right) \varpi +1-\frac{3}{2}p}\text{.%
}  \label{eqn:bound_on_vp_on_L}
\end{equation}%
Step 3 completes the proof of Theorem \ref{thm:vol_trunc_on_L}.

\textit{Step 1}. We now prove equation (\ref{eqn:PA_on_L}). Recall $\widehat{C}_{i}^{\prime n}$ notation in (\ref%
{eqn:CSP}). For $i\in L\left( n,T\right) $,%
%
\begin{eqnarray}
P\left( \overline{A}_{i}\right) &=&P\left( \left\Vert \widehat{C}%
_{i+k_{n}}^{n}-\widehat{C}_{i-k_{n}}^{n}\right\Vert \geq u_{n}^{\prime
}\right)  \notag \\
&\leq &P\left( \left\Vert \widehat{C}_{i+k_{n}}^{\prime n}-\widehat{C}%
_{i-k_{n}}^{\prime n}\right\Vert +\left\Vert \widehat{C}_{i+k_{n}}^{n}-%
\widehat{C}_{i+k_{n}}^{\prime n}\right\Vert +\left\Vert \widehat{C}%
_{i-k_{n}}^{n}-\widehat{C}_{i-k_{n}}^{\prime n}\right\Vert \geq
u_{n}^{\prime }\right)  \notag \\
&\leq &P\left( \left\Vert \widehat{C}_{i+k_{n}}^{\prime n}-\widehat{C}%
_{i-k_{n}}^{\prime n}\right\Vert \geq \frac{u_{n}^{\prime }}{2}\right)
+P\left( \left\Vert \widehat{C}_{i+k_{n}}^{n}-\widehat{C}_{i+k_{n}}^{\prime
n}\right\Vert +\left\Vert \widehat{C}_{i-k_{n}}^{n}-\widehat{C}%
_{i-k_{n}}^{\prime n}\right\Vert \geq \frac{u_{n}^{\prime }}{2}\right) .
\label{eqn:two_terms_PAbar_on_L}
\end{eqnarray}

Using standard results in the literature, we have for $q\geq 2$ and $i\in
L\left( n,T\right) $,%
\begin{equation}
E\left( \left\Vert \widehat{C}_{i+k_{n}}^{\prime n}-\widehat{C}%
_{i-k_{n}}^{\prime n}\right\Vert ^{q}\right) \leq K\Delta _{n}^{q/4},
\label{eqn:3-26_in_JR15}
\end{equation}%
see, for example, equation (3.26) in \cite{jacodrosenbaum-sqrtn}. Therefore,
the first term in (\ref{eqn:two_terms_PAbar_on_L}) satisfies, by Markov's inequality,
for $p\geq 2$,%
\begin{equation}
P\left( \left\Vert \widehat{C}_{i+k_{n}}^{\prime n}-\widehat{C}%
_{i-k_{n}}^{\prime n}\right\Vert \geq \frac{u_{n}^{\prime }}{2}\right) \leq
K\Delta _{n}^{p/4-\varpi ^{\prime }p}.  \label{eqn:PA1_on_L}
\end{equation}

By (4.8) in \cite{jacodrosenbaum13}, there exists a sequence of real numbers $%
a_{n} $ converging to zero such that
\begin{equation}
\mathbb{E}(\Vert \widehat{C}_{i}^{n}-\widehat{C}_{i}^{^{\prime }n}\Vert
^{q})\leq K_{q}a_{n}\Delta _{n}^{(2q-r)\varpi +1-q},~\text{for any}\hspace{%
2mm}q\geq 1,  \label{eqn:3-10_in_JR15_nojumps}
\end{equation}%
where for later use, we note that this result also holds in the presence of
volatility jumps. Therefore, the second term in (\ref%
{eqn:two_terms_PAbar_on_L}) satisfies, by Markov's inequality,%
\begin{eqnarray}
&&P\left( \left\Vert \widehat{C}_{i+k_{n}}^{n}-\widehat{C}_{i+k_{n}}^{\prime
n}\right\Vert +\left\Vert \widehat{C}_{i-k_{n}}^{n}-\widehat{C}%
_{i-k_{n}}^{\prime n}\right\Vert \geq \frac{u_{n}^{\prime }}{2}\right)
\notag \\
&\leq &\frac{1}{u_{n}^{\prime }/2}E\left( \left\Vert \widehat{C}%
_{i+k_{n}}^{n}-\widehat{C}_{i+k_{n}}^{\prime n}\right\Vert +\left\Vert
\widehat{C}_{i-k_{n}}^{n}-\widehat{C}_{i-k_{n}}^{\prime n}\right\Vert
\right) \leq Ka_{n}\Delta _{n}^{\left( 2-r\right) \varpi -\varpi ^{\prime }}.
\label{eqn:PA2_on_L}
\end{eqnarray}%
Since $\varpi ^{\prime }<\frac{1}{8}$ and by choosing sufficiently large
$p$ in (\ref{eqn:PA1_on_L}), equations (\ref{eqn:PA1_on_L}) and (\ref%
{eqn:PA2_on_L}) give (\ref{eqn:PA_on_L}).

\textit{Step 2}. We now prove equation (\ref{eqn:bound_on_vp_on_L}). First,
note that for $q\geq 1$, by (\ref{eqn:3-10_in_JR15_nojumps}),%
\begin{equation}
\mathrm{E}\left( \left\Vert \widehat{C}_{i\Delta _{n}}^{n}\right\Vert
^{q}\right) \leq K\mathrm{E}\left[ \left\vert \widehat{C}_{i\Delta _{n}}^{n}-%
\widehat{C}_{i\Delta _{n}}^{\prime n}\right\vert ^{q}\right] +K\mathrm{E}%
\left[ \left\vert \widehat{C}_{i\Delta _{n}}^{\prime n}\right\vert ^{q}%
\right] \leq Ka_{n}\Delta _{n}^{\left( 2q-r\right) \varpi +1-q}+K.
\label{eqn:Chat_bound}
\end{equation}
By Taylor expansion and $H$ and $G$ having bounded derivatives, for $%
i\in L\left( n,T\right) $ and $p\geq 1$,%
\begin{eqnarray}
&&E\left[ \left\vert \vartheta _{i}\right\vert ^{p}\right]  \notag \\
&\leq &K\frac{1}{%
k_{n}^{p}}\mathrm{E}\left( \left\Vert \widehat{C}_{\left( i+k_{n}\right)
\Delta _{n}}^{n}-\widehat{C}_{i\Delta _{n}}^{n}\right\Vert ^{2p}\right) +K%
\frac{1}{k_{n}^{2p}}\mathrm{E}\left( \left\Vert \widehat{C}_{i\Delta
_{n}}^{n}\right\Vert ^{2p}\right)  \notag \\
&\leq &K\Delta
_{n}^{p/2}\mathrm{E}\left( \left\Vert \widehat{C}_{\left( i+k_{n}\right)
\Delta _{n}}^{\prime n}-\widehat{C}_{i\Delta _{n}}^{\prime n}\right\Vert
^{2p}+\left\Vert \widehat{C}_{\left( i+k_{n}\right) \Delta _{n}}^{n}-%
\widehat{C}_{\left( i+k_{n}\right) \Delta _{n}}^{\prime n}\right\Vert
^{2p}+\left\Vert \widehat{C}_{i\Delta _{n}}^{n}-\widehat{C}_{i\Delta
_{n}}^{n\prime }\right\Vert ^{2p}\right) +K\Delta _{n}^{p}\mathrm{E}\left(
\left\Vert \widehat{C}_{i\Delta _{n}}^{n}\right\Vert ^{2p}\right)  \notag \\
&\leq &K\Delta _{n}^{p/2}\left( \Delta _{n}^{p/2}+a_{n}\Delta _{n}^{\left(
4p-r\right) \varpi +1-2p}\right) +K\Delta _{n}^{p}\left[ Ka_{n}\Delta
_{n}^{\left( 4p-r\right) \varpi +1-2p}+K\right]  \notag \\
&=&K\Delta _{n}^{p}+Ka_{n}\Delta _{n}^{\left( 4p-r\right) \varpi +1-\frac{3}{%
2}p},  \label{eqn:vip_bound_on_L}
\end{eqnarray}%
where the third inequality uses (\ref{eqn:Chat_bound}), (\ref%
{eqn:3-26_in_JR15}) and (\ref{eqn:3-10_in_JR15_nojumps}).%

\textit{Step 3}. We now complete the proof of Theorem \ref%
{thm:vol_trunc_on_L}. By the triangle
and Cauchy-Schwarz inequalities,%
\begin{eqnarray*}
&&E\left\vert \sum_{i\in L\left( n,T\right) }\vartheta _{i}1_{\left\{
A_{i}\cap A_{i+k_{n}}\right\} }-\sum_{i\in L\left( n,T\right) }\vartheta
_{i}\right\vert \\
&\leq &\sum_{i\in L\left( n,T\right) }E\left\vert \vartheta _{i}\left(
1_{\left\{ A_{i}\cap A_{i+k_{n}}\right\} }-1\right) \right\vert \\
&\leq &\sum_{i\in L\left( n,T\right) }\sqrt{E\left\vert \vartheta
_{i}^{2}\right\vert }\sqrt{P\left( \overline{A_{i}}\cup \overline{A_{i+k_{n}}%
}\right) } \\
&\leq &\sum_{i\in L\left( n,T\right) }\sqrt{E\left\vert \vartheta
_{i}^{2}\right\vert }\sqrt{P\left( \overline{A_{i}}\right) +P\left(
\overline{A_{i+k_{n}}}\right) } \\
&\leq &K\Delta _{n}^{-1}\left( \Delta _{n}^{\left( 8-r\right) \varpi
-2}\right) ^{1/2}\left( a_{n}\Delta _{n}^{\left( 2-r\right) \varpi -\varpi
^{\prime }}\right) ^{1/2}\\
&=&\Delta _{n}^{l\left( \varpi ,\varpi ^{\prime }\right) },
\end{eqnarray*}%
where 4th inequality follows by (\ref{eqn:bound_on_vp_on_L}) with $p=2$, and
(\ref{eqn:PA_on_L}). In the
above,%
\begin{equation*}
l\left( \varpi ,\varpi ^{\prime }\right) =-1+\frac{1}{2}\left[ \left(
8-r\right) \varpi -2\right] +\frac{1}{2}\left[ \left( 2-r\right) \varpi
-\varpi ^{\prime }\right] .
\end{equation*}%
A straightforward calculation shows that $\varpi >
\frac{2\varpi ^{\prime }+9}{4\left( 5-r\right) }$ implies $l\left( \varpi
,\varpi ^{\prime }\right) > \frac{1}{4}$, which completes the proof of
Theorem \ref{thm:vol_trunc_on_L}.

\bigskip

\subsection{Proof of Theorem \protect\ref{thm:vol_trunc_on_Lbar}}

\bigskip 

Without loss of generality, we can assume that there is at most one
volatility jump in $\left( \left( i-k_{n}\right) \Delta _{n},\left(
i+3k_{n}\right) \Delta _{n}\right] $ for any $i\in \overline{L}\left(
n,T\right) $. To study the behavior of $\vartheta _{i}1_{\left\{ A_{i}\cap
A_{i+k_{n}}\right\} }$ on $i\in \overline{L}\left( n,T\right) $, we will
distinguish between two cases, depending on whether or not there is a
volatility jump in $\left( i\Delta _{n},\left( i+2k_{n}\right) \Delta _{n}%
\right] $. So define $B_{i}$ as the event that there is a volatility jump in
$\left( i\Delta _{n},\left( i+2k_{n}\right) \Delta _{n}\right] $ (we omit
indexing $B_{i}$\ by $n$ for brevity). Denote by $\overline{B}_{i}$ the
complement of $B_{i}$. Intuitively, for $i\in \overline{L}\left( n,T\right) $%
, $\vartheta _{i}1_{\left\{ A_{i}\cap A_{i+k_{n}}\right\} }$ is small
because, on the one hand, $P\left( A_{i}\cap A_{i+k_{n}}\right) $ is small
on $B_{i}$, on the other hand, $\vartheta _{i}$ is small on $\overline{B}%
_{i} $.

We have%
\begin{eqnarray}
&&\mathrm{E}\left\vert \sum_{i\in \overline{L}\left( n,T\right) }\vartheta
_{i}1_{\left\{ A_{i}\cap A_{i+k_{n}}\right\} }\right\vert =\mathrm{E}%
\left\vert \sum_{i\in \overline{L}\left( n,T\right) :B_{i}}\vartheta
_{i}1_{\left\{ A_{i}\cap A_{i+k_{n}}\right\} }+\sum_{i\in \overline{L}\left(
n,T\right) :\overline{B}_{i}}\vartheta _{i}1_{\left\{ A_{i}\cap
A_{i+k_{n}}\right\} }\right\vert  \notag \\
&\leq &\sum_{i\in \overline{L}\left( n,T\right) :B_{i}}\mathrm{E}\left\vert
\vartheta _{i}1_{\left\{ A_{i}\cap A_{i+k_{n}}\right\} }\right\vert
+\sum_{i\in \overline{L}\left( n,T\right) :\overline{B}_{i}}\mathrm{E}%
\left\vert \vartheta _{i}1_{\left\{ A_{i}\cap A_{i+k_{n}}\right\}
}\right\vert ,  \label{eqn:thmB2_decomp}
\end{eqnarray}%
where \textquotedblleft $i\in \overline{L}\left( n,T\right) :B_{i}$%
\textquotedblright\ denotes those terms in $\overline{L}\left( n,T\right) $,
for which $B_{i}$ is true.

\bigskip First, we show that the second term in (\ref{eqn:thmB2_decomp}) is $%
o_{p}\left( \Delta _{n}^{1/4}\right) $. For $i\in \overline{L}\left(
n,T\right) $ such that $B_{i}$ if false, we can use the bound on $\mathrm{E}%
\left[ \left\vert \vartheta _{i}\right\vert ^{p}\right] $ in (\ref%
{eqn:vip_bound_on_L}) for $p\geq 1$. The second term in (%
\ref{eqn:thmB2_decomp}) satisfies%
\begin{eqnarray*}
&&\sum_{i\in \overline{L}\left( n,T\right) :\overline{B}_{i}}\mathrm{E}%
\left\vert \vartheta _{i}1_{\left\{ A_{i}\cap A_{i+k_{n}}\right\}
}\right\vert \\
&\leq &\sum_{i\in \overline{L}\left( n,T\right) :\overline{B}_{i}}\mathrm{E}%
\left\vert \vartheta _{i}\right\vert \\
&\leq &K\Delta _{n}^{-1/2}\left( \Delta _{n}+\Delta _{n}^{\left( 4-r\right)
\varpi -\frac{1}{2}}\right) \\
&=&K\Delta _{n}^{1/2}+K\Delta _{n}^{\left( 4-r\right) \varpi -1}.
\end{eqnarray*}%
Theorem 1 assumptions imply $\left( 4-r\right) \varpi -1> \frac{1}{4}$,
so the second term in (\ref{eqn:thmB2_decomp}) is $o_{p}\left( \Delta
_{n}^{1/4}\right) $.

\bigskip 

The rest of the proof is devoted to showing that the first term in (\ref%
{eqn:thmB2_decomp}) is $o_{p}\left( \Delta _{n}^{1/4}\right) $. This will
complete the proof of Theorem \ref{thm:vol_trunc_on_Lbar}.

The first term in (\ref{eqn:thmB2_decomp}) involves those $i\in \overline{L}%
\left( n,T\right) $, for which $B_{i}$ is true. We will show below that%
\begin{equation}
P\left( A_{i}\cap A_{i+k_{n}}\right) \leq K\Delta _{n}^{1/2}\text{ for }i\in
\overline{L}\left( n,T\right) \text{ such that }B_{i}\text{ holds.}
\label{eqn:PA_on_Lbar_and_B}
\end{equation}%
We use the following bound in the presence of the volatility jump,%
\begin{equation}
\mathrm{E}\left( \left\vert \vartheta _{i}\right\vert ^{p}\right) \leq K%
\frac{1}{k_{n}^{p}}\left[ \mathrm{E}\left[ \left\vert \widehat{C}%
_{i}^{n}\right\vert ^{p}\right] +\mathrm{E}\left[ \left\vert \widehat{C}%
_{i+k_{n}}^{n}\right\vert ^{p}\right] \right] \leq K\frac{1}{k_{n}^{p}}%
\left( a_{n}\Delta _{n}^{\left( 2p-r\right) \varpi +1-p}+K\right) ,
\label{eqn:vip_with_vol_jumps}
\end{equation}%
where the first inequality uses Taylor expansion and bounded derivatives of $%
H$ and $G$, and the last transition uses (\ref{eqn:Chat_bound}). 

The first term in (\ref{eqn:thmB2_decomp}) satisfies, for $p\geq 1$, by
Holder inequality, (\ref{eqn:PA_on_Lbar_and_B}) and (\ref%
{eqn:vip_with_vol_jumps}),
\begin{eqnarray*}
&&\sum_{i\in \overline{L}\left( n,T\right) :B_{i}}\mathrm{E}\left\vert
\vartheta _{i}1_{\left\{ A_{i}\cap A_{i+k_{n}}\right\} }\right\vert \\
&\leq &\sum_{i\in \overline{L}\left( n,T\right) :B_{i}}\left( \mathrm{E}%
\left[ \left\vert \vartheta _{i}\right\vert ^{p}\right] \right) ^{1/p}\left(
\mathrm{P}\left( A_{i}\cap A_{i+k_{n}}\right) \right) ^{\left( p-1\right) /p}
\\
&\leq &\sum_{i\in \overline{L}\left( n,T\right) :B_{i}}K\left[ \Delta
_{n}\times \left( \Delta _{n}^{\left( 2p-r\right) \varpi +1-p}+1\right) %
\right] ^{1/p}\left[ \Delta _{n}^{1/2}\right] ^{\left( p-1\right) /p} \\
&=&\sum_{i\in \overline{L}\left( n,T\right) :B_{i}}K\Delta _{n}^{l\left(
r,\varpi \right) }.
\end{eqnarray*}

Since the number of terms in $\overline{L}\left( n,T\right) $ is bounded by $%
Kk_{n}$ ($k_{n}$ arises due to overlapping blocks defining $\vartheta _{i}$%
), the first term in (\ref{eqn:thmB2_decomp}) is $o_{p}\left( \Delta
_{n}^{1/4}\right) $ if $l\left( r,\varpi \right) >\frac{3}{4}$. To study $%
l\left( r,\varpi \right) $, we distinguish two cases, depending on whether $%
\left( 2p-r\right) \varpi +1-p\geq 0$ holds.

Case 1. When $\left( 2p-r\right) \varpi +1-p\geq 0$, $l\left( r,\varpi
\right) =\frac{1}{2p}\left( p+1\right) $, so $l\left( r,\varpi \right) >%
\frac{3}{4}$ if $p<2.$

Case 2. When $\left( 2p-r\right) \varpi +1-p<0$, $l\left( r,\varpi
\right) =\frac{1}{p}\left( \left( 2p-r\right) \varpi +1-p\right) +\frac{p-1}{%
2p}$. We have $l\left( r,\varpi \right) >\frac{3}{4}$ if $\varpi >\frac{5p-6%
}{4\left( 2p-r\right) }$. This is satisfied if we choose, for example, $%
p=1.5 $.

The last step in the proof of Theorem \ref{thm:vol_trunc_on_Lbar} is to show
that (\ref{eqn:PA_on_Lbar_and_B}) is true. In order to do that, we first
prove that if there is a volatility jump on $\left( i\Delta _{n},\left(
i+k_{n}\right) \Delta _{n}\right] $, then%
%
\begin{equation}
P\left( \left\Vert \widehat{C}_{i+k_{n}}^{n}-\widehat{C}_{i-k_{n}}^{n}\right%
\Vert <u_{n}^{\prime }\right) =o_{p}\left( \Delta _{n}^{1/4}\right) .
\label{eqn:deltaChat_small_on_Lbar}
\end{equation}

Denote by $S$ the time of the volatility jump on $\left( i\Delta _{n},\left(
i+k_{n}\right) \Delta _{n}\right] $, so the jump is $\Delta C_{S}$. Denote $%
\xi _{n}\equiv \widehat{C}_{i+k_{n}}^{n}-\widehat{C}_{i-k_{n}}^{n}-\Delta
C_{S}$, so $\widehat{C}_{i+k_{n}}^{n}-\widehat{C}_{i-k_{n}}^{n}=\Delta
C_{S}+\xi _{n}$. We know $\xi _{n}=o_{p}\left( 1\right) $. 
We know that there exists $\epsilon $, independent of $i$ or $S$,
 such that $\left\Vert \Delta C\right\Vert > \epsilon $.%

We will first show that if there is a volatility jump on $\left( i\Delta
_{n},\left( i+k_{n}\right) \Delta _{n}\right] $, for $s\geq 0$, it follows
that %
\begin{equation}
P\left( \left\Vert \widehat{C}_{i+k_{n}}^{n}-\widehat{C}_{i-k_{n}}^{n}\right%
\Vert <u_{n}^{\prime }\right) \leq \frac{E\left( \left\Vert \widehat{C}%
_{i+k_{n}}^{n}-\widehat{C}_{i-k_{n}}^{n}-\Delta C_{S}\right\Vert ^{s}\right)
}{\left( \epsilon /2\right) ^{s}}.  \label{eqn:deltaChat_bound_by_moments}
\end{equation}%
To prove (\ref{eqn:deltaChat_bound_by_moments}), note that the reverse
triangle inequality gives $\left\Vert \widehat{C}_{i+k_{n}}^{n}-\widehat{C}%
_{i-k_{n}}^{n}\right\Vert =\left\Vert \Delta C+\xi _{n}\right\Vert \geq
\left\vert \left\Vert \Delta C\right\Vert -\left\Vert \xi _{n}\right\Vert
\right\vert $. Thus,%
\begin{eqnarray*}
&&P\left( \left\Vert \widehat{C}_{i+k_{n}}^{n}-\widehat{C}%
_{i-k_{n}}^{n}\right\Vert <u_{n}^{\prime }\right) \\
&\leq &P\left( \left\vert \left\Vert \Delta C\right\Vert -\left\Vert \xi
_{n}\right\Vert \right\vert <u_{n}^{\prime }\right) \\
&\leq &P\left( \left\Vert \xi _{n}\right\Vert >\frac{\epsilon }{2}\right) ,
\end{eqnarray*}%
where the second inequality follows by distinguishing two cases, depending
on whether $\left\Vert \Delta C\right\Vert \geq \left\Vert \xi
_{n}\right\Vert $. Case 1:\ if $\left\Vert \Delta C\right\Vert \geq
\left\Vert \xi _{n}\right\Vert $, $\left\{ \left\vert \left\Vert \Delta
C\right\Vert -\left\Vert \xi _{n}\right\Vert \right\vert <u_{n}^{\prime
}\right\} =\left\{ \left\Vert \Delta C\right\Vert -\left\Vert \xi
_{n}\right\Vert <u_{n}^{\prime }\right\} =\left\{ \left\Vert \Delta
C\right\Vert -u_{n}^{\prime }<\left\Vert \xi _{n}\right\Vert \right\} $, so
we deduce $\left\{ \epsilon -u_{n}^{\prime }<\left\Vert \xi _{n}\right\Vert
\right\} $. For $n\,\ $large enough, this implies $\left\{ \left\Vert \xi
_{n}\right\Vert >\frac{\epsilon }{2}\right\} $ since $u_{n}^{\prime
}\rightarrow 0$. Case 2:\ if $\left\Vert \Delta C\right\Vert <\left\Vert \xi
_{n}\right\Vert $, we have $P\left( \left\{ \left\vert \left\Vert \Delta
C\right\Vert -\left\Vert \xi _{n}\right\Vert \right\vert <u_{n}^{\prime
}\right\} \cap \left\{ \left\Vert \Delta C\right\Vert <\left\Vert \xi
_{n}\right\Vert \right\} \right) \leq P\left( \left\Vert \xi _{n}\right\Vert
>\left\Vert \Delta C\right\Vert \right) \leq P\left( \left\Vert \xi
_{n}\right\Vert >\epsilon \right) \leq P\left( \left\Vert \xi
_{n}\right\Vert >\frac{\epsilon }{2}\right) $. Finally, (\ref%
{eqn:deltaChat_bound_by_moments}) follows by Markov's inequality.

\bigskip By (\ref{eqn:deltaChat_bound_by_moments}), we obtain, for $s\geq 2$,%
\begin{eqnarray}
P\left( \left\Vert \widehat{C}_{i+k_{n}}^{n}-\widehat{C}_{i-k_{n}}^{n}\right%
\Vert <u_{n}^{\prime }\right) &\leq &\frac{E\left( \left\Vert \widehat{C}%
_{i+k_{n}}^{n}-\widehat{C}_{i-k_{n}}^{n}-\Delta C\right\Vert ^{s}\right) }{%
\left( \epsilon /2\right) ^{s}}  \notag \\
&\leq &KE\left( \left\Vert \widehat{C}_{i-k_{n}}^{n}-C_{S-}\right\Vert
^{s}\right) +KE\left( \left\Vert \widehat{C}_{i+k_{n}}^{n}-C_{S}\right\Vert
^{s}\right) .  \label{eqn:deltaChat_Lbar_decomp}
\end{eqnarray}%
The first term in (\ref{eqn:deltaChat_Lbar_decomp}) satisfies, for $s\geq 2,$
by (\ref{eqn:3-10_in_JR15_nojumps}) and (\ref{eq:estb})%
\begin{eqnarray*}
E\left( \left\Vert \widehat{C}_{i-k_{n}}^{n}-C_{S-}\right\Vert ^{s}\right)
&\leq &KE\left( \left\Vert \widehat{C}_{i-k_{n}}^{n}-\widehat{C}%
_{i-k_{n}}^{n\prime }\right\Vert ^{s}\right) +KE\left( \left\Vert \widehat{C}%
_{i-k_{n}}^{n\prime }-C_{S-}\right\Vert ^{s}\right) \\
&\leq &K_{q}a_{n}\Delta _{n}^{\left( 2s-r\right) \varpi +1-s}+K\Delta
_{n}^{s/4}.
\end{eqnarray*}%
The second term in (\ref{eqn:deltaChat_Lbar_decomp}) has the same bound by
the same arguments as the first term. Choosing $s=2$ in the above, and
taking into account that $\left( 2-r\right) \varpi \geq \frac{3}{4}$ and $%
\varpi \geq \frac{3}{8}$, we obtain (\ref{eqn:deltaChat_small_on_Lbar}).

\bigskip

Given (\ref{eqn:deltaChat_small_on_Lbar}), it is simple to obtain (\ref%
{eqn:PA_on_Lbar_and_B}) as follows. By (\ref{eqn:deltaChat_small_on_Lbar}),
if there is a jump on $\left( i\Delta _{n},\left( i+k_{n}\right) \Delta _{n}%
\right] $, we know $P\left( A_{i}\right) =o_{p}\left( \Delta
_{n}^{1/4}\right) $, thus $\left( A_{i}\cap A_{i+k_{n}}\right) \leq P\left(
A_{i}\right) =o_{p}\left( \Delta _{n}^{1/4}\right) $. Applying (\ref%
{eqn:deltaChat_small_on_Lbar}) with $i+k_{n}$ instead of $i$, if there is a
jump on $\left( \left( i+k_{n}\right) \Delta _{n},\left( i+2k_{n}\right)
\Delta _{n}\right] $, $P\left(
A_{i+k_{n}}\right) =o_{p}\left( \Delta _{n}^{1/4}\right) $. Thus, $P\left(
A_{i}\cap A_{i+k_{n}}\right) \leq P\left( A_{i+k_{n}}\right) =o_{p}\left(
\Delta _{n}^{1/4}\right) $. We conclude that if there is a jump on $\left(
i\Delta _{n},\left( i+2k_{n}\right) \Delta _{n}\right] $, i.e., event $B_{i}$
is true, then $\left( A_{i}\cap A_{i+k_{n}}\right) \leq P\left(
A_{i+k_{n}}\right) =o_{p}\left( \Delta _{n}^{1/4}\right) $. This concludes
the proof of (\ref{eqn:PA_on_Lbar_and_B}) and hence Theorem \ref%
{thm:vol_trunc_on_Lbar}.

\subsection{Proof of Theorem \protect\ref{thm:intermediate1}}

To show this result, let us define the functions
\begin{align*}
R(x,y)& =\sum_{g,h,a,b=1}^{d}\Big(\partial _{gh}H\partial _{ab}G\big)(x)\big(%
y^{gh}-x^{gh}\Big)\Big(y^{ab}-x^{ab}\Big) \\
S(x,y)& =\Big(H(y)-H(x)\Big)\Big(G(y)-G(x)\Big) \\
U(x)& =\sum_{g,h,a,b=1}^{d}\Big(\partial _{gh}H\partial _{ab}G\Big)(x)\Big(%
x^{ga}x^{hb}+x^{gb}x^{ha}\Big),
\end{align*}%
for any $\mathbb{R}^{d}\times \mathbb{R}^{d}$ matrices $x$ and $y$. The
following decompositions hold,%
\begin{align*}
& \sum_{i\in L\left( n,T\right) }\vartheta _{i}^{AN}-\sum_{i\in L\left(
n,T\right) }\vartheta _{i}^{\prime AN} \\
& \hspace{2mm}=\frac{3}{2k_{n}}\sum_{i\in L\left( n,T\right) }\Big[\big(S(%
\widehat{C}_{i}^{n},\widehat{C}_{i+k_{n}}^{n})-S(\widehat{C}_{i}^{^{\prime
}n},\widehat{C}_{i+k_{n}}^{^{\prime }n})\big)-\frac{2}{k_{n}}\big(U(\widehat{%
C}_{i}^{n})-U(\widehat{C}_{i}^{^{\prime }n})\big)\Big], \\
& \sum_{i\in L\left( n,T\right) }\vartheta _{i}^{LIN}-\sum_{i\in L\left(
n,T\right) }\vartheta _{i}^{\prime LIN} \\
& \hspace{2mm}=\frac{3}{2k_{n}}\sum_{i\in L\left( n,T\right) }\Big[\big(R(%
\widehat{C}_{i}^{n},\widehat{C}_{i+k_{n}}^{n})-R(\widehat{C}_{i}^{^{\prime
}n},\widehat{C}_{i+k_{n}}^{^{\prime }n})\big)-\frac{2}{k_{n}}\big(U(\widehat{%
C}_{i}^{n})-U(\widehat{C}_{i}^{^{\prime }n})\big)\Big].
\end{align*}%
Since $H$ and $G$ are three times continuously differentiable with bounded
derivatives, the functions $R$ and $S$ are continuously differentiable and
satisfy
\begin{eqnarray}
\Vert \partial J(x,y)\Vert &\leq &K\hspace{2mm}\text{for}\hspace{2mm}J\in
\{S,R\},  \label{eqn:taylor1} \\
\Vert \partial U(x)\Vert &\leq &K,  \label{eqn:taylor2}
\end{eqnarray}%
where $\partial J$ (respectively, $\partial U$) is a vector that collects
the first order partial derivatives of the function $J$ (respectively, $U$)
with respect to all the elements of $(x,y)$ (respectively, $x$). Using the
Taylor expansion, (\ref{eqn:taylor1}) and (\ref{eqn:taylor2}), it holds
that, for $J\in \{S,R\}$,
\begin{align*}
|J(\widehat{C}_{i}^{n},\widehat{C}_{i+k_{n}}^{n})-J(\widehat{C}%
_{i}^{^{\prime }n},\widehat{C}_{i+k_{n}}^{^{\prime }n})|& \leq K(\Vert
\widehat{C}_{i}^{n}-\widehat{C}_{i}^{^{\prime }n}\Vert +\Vert \widehat{C}%
_{i+k_{n}}^{n}-\widehat{C}_{i+k_{n}}^{^{\prime }n}\Vert )~\text{and} \\
|U(\widehat{C}_{i}^{n})-U(\widehat{C}_{i}^{^{\prime }n})|& \leq K(\Vert
\widehat{C}_{i}^{n}-\widehat{C}_{i}^{^{\prime }n}\Vert ).
\end{align*}%
By equation (\ref{eqn:3-10_in_JR15_nojumps}), the following condition is
sufficient for Theorem \ref{thm:intermediate1} to hold:
\begin{equation*}
(2-r)\varpi -\frac{3}{4}\geq 0.
\end{equation*}%
The above condition follows from our assumptions of Theorem \ref{thm:clt}.
Using the fact that $0<\varpi <\frac{1}{2}$, we can see that Theorem \ref%
{thm:intermediate1} holds when $3/4(2-r)\leq \varpi <\frac{1}{2}$, which
completes the proof.

\subsection{Proof of Theorem \protect\ref{thm:intermediate2}}

Note that we have
\begin{align*}
& \sum_{i\in L\left( n,T\right) }\vartheta _{i}^{\prime LIN}-\sum_{i\in
L\left( n,T\right) }\vartheta _{i}^{\left( A\right) }=\frac{3}{2k_{n}}%
\sum_{g,h,a,b=1}^{d}\sum_{i\in L\left( n,T\right) }\psi _{i}^{n}(g,h,a,b), \\
& \sum_{i\in L\left( n,T\right) }\vartheta _{i}^{\prime AN}-\sum_{i\in
L\left( n,T\right) }\vartheta _{i}^{\left( A\right) }=\frac{3}{2k_{n}}%
\sum_{i\in L\left( n,T\right) }\Big(\chi _{i}^{n}-\sum_{g,h,a,b=1}^{d}\big(%
\partial _{gh}H\partial _{ab}G\big)(C_{i}^{n})\lambda _{i}^{n,gh}\lambda
_{i}^{n,ab}\Big),
\end{align*}%
with
\begin{align*}
& \psi _{i}^{n}(g,h,a,b)=\Big(\big(\partial _{gh}H\partial _{ab}G\big)(%
\widehat{C}_{i}^{^{\prime }n})-\big(\partial _{gh}H\partial _{ab}G\big)%
(C_{i}^{n})\Big)\lambda _{i}^{n,gh}\lambda _{i}^{n,ab}, \\
& \chi _{i}^{n}=\Big(H(\widehat{C}_{i+k_{n}}^{^{\prime }n})-H(\widehat{C}%
_{i}^{^{\prime }n})\Big)\Big(G(\widehat{C}_{i+k_{n}}^{^{\prime }n})-G(%
\widehat{C}_{i}^{^{\prime }n})\Big).
\end{align*}%
By Taylor expansion, we have
\begin{align*}
& \big(\partial _{gh}S\partial _{ab}G\big)(\widehat{C}_{i}^{^{\prime }n})-%
\big(\partial _{gh}S\partial _{ab}G\big)(C_{i}^{n})=\sum_{x,y=1}^{d}\Big(%
\partial _{xy,gh}^{2}S\partial _{ab}G+\partial _{xy,ab}^{2}G\partial _{gh}S%
\Big)(C_{i}^{n})\nu _{i}^{n,xy} \\
& +\frac{1}{2}\sum_{j,k,x,y=1}^{d}\Big(\partial _{jk,xy,gh}^{3}S\partial
_{ab}G+\partial _{xy,gh}^{2}S\partial _{jk,ab}^{2}G+\partial
_{jk,xy,ab}^{3}G\partial _{gh}S+\partial _{xy,ab}^{2}G\partial _{jk,gh}^{2}S%
\Big)(\widetilde{c}_{i}^{n})\nu _{i}^{n,xy}\nu _{i}^{n,jk}
\end{align*}%
and
\begin{eqnarray*}
&&S(\widehat{C}_{i+k_{n}}^{^{\prime }n})-S(\widehat{C}_{i}^{^{\prime
}n})=\sum_{gh}\partial _{gh}S(C_{i}^{n})\lambda
_{i}^{n,gh}+\sum_{j,k,g,h}\partial _{jk,gh}^{2}S(C_{i}^{n})\lambda
_{i}^{n,gh}\nu _{i}^{n,jk} \\
&&+\frac{1}{2}\sum_{x,y,g,h}\partial _{xy,gh}^{2}S(C_{i}^{n})\lambda
_{i}^{n,gh}\lambda _{i}^{n,xy}+\frac{1}{2}\sum_{x,y,j,k,g,h}\partial
_{xy,jk,gh}^{3}S(CC_{i}^{n,S})\lambda _{i}^{n,gh}\nu _{i}^{n,xy}\nu
_{i}^{n,jk} \\
&&+\frac{1}{6}\sum_{j,k,x,y,g,h}\partial
_{jk,xy,gh}^{3}S(C_{i}^{n,S})\lambda _{i}^{n,jk}\lambda _{i}^{n,gh}\lambda
_{i}^{n,xy},
\end{eqnarray*}%
for $S\in \{H,G\}$, $\widetilde{c}_{i}^{n}=\pi C_{i}^{n}+(1-\pi )\widehat{C}%
_{i}^{^{\prime }n}$, $C_{i}^{n,S}=\pi _{S}\widehat{C}_{i}^{^{\prime
}n}+(1-\pi _{S})\widehat{C}_{i+k_{n}}^{^{\prime }n}$, $CC_{i}^{n,S}=\mu
_{S}C_{i}^{n}+(1-\mu _{S})\widehat{C}_{i}^{^{\prime }n}$ for $\pi ,\pi
_{H},\mu _{H},\pi _{G},\mu _{G}\in \lbrack 0,1]$. Although $\widetilde{c}%
_{i}^{n}$ and $\pi $ depend on $g,h,a,$ and $b$, we do not emphasize this in
our notation to simplify the exposition.\newline
By (4.10) in \cite%
{jacodrosenbaum13} we have
\begin{equation}
\mathbb{E}\Big(\Big\|\alpha _{i}^{n}\Big\|^{q}\Big|\mathcal{F}_{(i-1)\Delta_n}\Big)%
\leq K_{q}\Delta _{n}^{q}\hspace{1mm}\text{for all}\hspace{1mm}q\geq 0%
\hspace{1mm}\text{and}\hspace{1mm}\mathbb{E}\Big(\Big|\sum_{j=0}^{k_{n}-1}%
\alpha _{i+j}^{n}\Big|^{q}\big|\mathcal{F}_{(i-1)\Delta_n}\Big)\leq K_{q}\Delta
_{n}^{q}k_{n}^{q/2}\hspace{1mm}\text{for}\hspace{1mm}q\geq 2.
\label{eqn:alpha}
\end{equation}%
Combining (\ref{eqn:alpha}), (\ref{eqn:BGD}), (\ref{eqn:SKCC}) with $Z=C$
and the H\"{o}lder inequality yields for $q\geq 2$, for $i\in L\left(
n,T\right) $
\begin{equation}
\mathbb{E}\Big(\Big\|\nu _{i}^{n}\Big\|^{q}\Big|\mathcal{F}_{(i-1)\Delta_n}\Big)\leq
K_{q}\Delta ^{q/4},\hspace{2mm}\text{and}\hspace{2mm}\mathbb{E}\Big(\Big\|%
\lambda _{i}^{n}\Big\|^{q}\Big|\mathcal{F}_{(i-1)\Delta_n}\Big)\leq K_{q}\Delta
^{q/4}.  \label{eq:estb}
\end{equation}%
The bound in the first equation of (\ref{eq:estb}) is tighter than that in
(4.11) of \cite{jacodrosenbaum-sqrtn} due to the absence of volatility
jumps. This tighter bound will be useful later in deriving the asymptotic
distribution for the approximated estimator. By the boundedness of $C_{t}$
and the derivatives of $H$ and $G$,
\begin{equation}
\Big|\big(\partial _{jk,xy,ab}^{3}G\partial _{gh}H+\partial
_{xy,gh}^{2}H\partial _{jk,ab}^{2}G\big)(\widetilde{c}_{i}^{n})\nu
_{i}^{n,xy}\nu _{i}^{n,jk}\lambda _{i}^{n,gh}\lambda _{i}^{n,ab}\Big|\leq
K\Vert \nu _{i}^{n}\Vert ^{2}\Vert \lambda _{i}^{n}\Vert ^{2}.
\label{eqn:UPG}
\end{equation}%
Using the Taylor expansion, we have
\begin{align*}
& \chi _{i}^{n}-\sum_{g,h,a,b}(\partial _{gh}H\partial
_{ab}G)(C_{i}^{n})\lambda _{i}^{n,gh}\lambda _{i}^{n,ab}= \\
& \sum_{g,h,a,b,j,k}(\partial _{gh}H\partial _{jk,xy}^{2}G+\partial
_{gh}G\partial _{jk,xy}^{2}H)(C_{i}^{n})(\lambda _{i}^{n,gh}+\frac{1}{2}\nu
_{i}^{n,gh})\lambda _{i}^{n,ab}\lambda _{i}^{n,jk}+\varphi _{i}^{n},\text{ \
and} \\
& \sum_{g,h,a,b}\big(\partial _{gh}H\partial _{ab}G\big)(\widehat{C}%
_{i}^{^{\prime }n})-\big(\partial _{gh}H\partial _{ab}G\big)(C_{i}^{n})= \\
& \sum_{g,h,a,b,x,y}(\partial _{gh}H\partial _{ab,xy}^{2}G+\partial
_{ab}G\partial _{gh,xy}^{2}G)(C_{i}^{n})(\nu _{i}^{n,xy})\lambda
_{i}^{n,gh}\lambda _{i}^{n,ab}+\delta _{i}^{n}
\end{align*}%
with $\mathbb{E}(|\varphi _{i}^{n}|\big|\mathcal{F}_{i}^{n})\leq K\Delta
_{n} $ and $\mathbb{E}(|\delta _{i}^{n}|\big|\mathcal{F}_{i}^{n})\leq
K\Delta _{n} $ which follow by the Cauchy-Schwartz inequality together with
equation (\ref{eq:estb}). Given that $k_{n}=\theta (\Delta _{n})^{-1/2}$,
the previous inequalities imply
\begin{equation*}
\frac{3\Delta _{n}^{-1/4}}{2k_{n}}\sum_{i\in L\left( n,T\right) }\varphi
_{i}^{n}\overset{\mathbb{P}}{\Longrightarrow }0\hspace{3mm}\text{and}\hspace{%
3mm}\frac{3\Delta _{n}^{-1/4}}{2k_{n}}\sum_{i\in L\left( n,T\right) }\delta
_{i}^{n}\overset{\mathbb{P}}{\Longrightarrow }0.
\end{equation*}%
Therefore, it suffices to show that
\begin{equation}
\frac{3\Delta _{n}^{-1/4}}{2k_{n}}\sum_{i\in L\left( n,T\right)
}\sum_{g,h,a,b,j,k}(\partial _{gh}H\partial _{jk,ab}^{2}G+\partial
_{gh}H\partial _{jk,ab}^{2}G)(C_{i}^{n})\lambda _{i}^{n,gh}\lambda
_{i}^{n,ab}\lambda _{i}^{n,jk}\overset{\mathbb{P}}{\longrightarrow }0,
\label{eqn:res1}
\end{equation}%
\begin{equation}
\frac{3\Delta _{n}^{-1/4}}{2k_{n}}\sum_{i\in L\left( n,T\right)
}\sum_{g,h,a,b,j,k}(\partial _{gh}H\partial _{jk,ab}^{2}G+\partial
_{gh}H\partial _{jk,ab}^{2}G)(C_{i}^{n})\nu _{i}^{n,gh}\lambda
_{i}^{n,ab}\lambda _{i}^{n,jk}\overset{\mathbb{P}}{\longrightarrow }0.
\label{eqn:res2}
\end{equation}%
These results hold by the bounds in Lemma \ref{lemma:approx1}.

\subsection{Proof of Theorem \protect\ref{thm:intermediate3}\label%
{sec:Proof_of_A1}}

In Section \ref{sec:Proof_of_A1}, to simplify the notational burden, we
adopt the following strategy. Instead of studying $\sum_{i\in L\left(
n,T\right) }\vartheta _{i}^{\left( A\right) }$, we work with all indices $i$%
, i.e., $\sum_{i=k_{n}+1}^{[T/\Delta _{n}]-3k_{n}+1}\vartheta _{i}^{\left(
A\right) }$, together with the assumption that there are no volatility
jumps. The difference between the two quantities is $o_{p}\left( \Delta
_{n}^{1/4}\right) $ because in the absence of volatility jumps, $\vartheta
_{i}^{\left( A\right) }$ satisfies the bound in equation (\ref%
{eqn:vip_bound_on_L}).
Recall the decomposition from from \ref{eqn:v_A_decomposition}, %
\begin{equation}
\vartheta _{i}^{\left( A\right) }
=
\vartheta _{i}^{\left( A1\right)}
-
\vartheta _{i}^{\left( A2\right) }.
\end{equation}%

Given the boundedness of the derivatives of $H$ and $G$ and the fact that $%
k_{n}=\theta (\Delta _{n})^{-1/2}$, by Theorem 2.2 in \cite%
{jacodrosenbaum-sqrtn} we have
%
\begin{equation*}
\frac{1}{\sqrt{\Delta _{n}}}\Bigg(\sum_{i=k_{n}+1}^{[T/\Delta
_{n}]-3k_{n}+1}\vartheta _{i}^{\left( A2\right) }-\frac{3}{\theta ^{2}}%
\sum_{g,h,a,b=1}^{d}\int_{0}^{T}\big(\partial _{gh}H\partial _{ab}G\big)%
(C_{t})(C_{t}^{ga}C_{t}^{hb}+C_{t}^{gb}C_{t}^{ha})dt\Bigg)=O_{p}(1),
\end{equation*}%
which yields
\begin{equation*}
\frac{1}{\Delta _{n}^{1/4}}\Bigg(\sum_{i=k_{n}+1}^{[T/\Delta
_{n}]-3k_{n}+1}\vartheta _{i}^{\left( A2\right) }-\frac{3}{\theta ^{2}}%
\sum_{g,h,a,b=1}^{d}\int_{0}^{T}\big(\partial _{gh}H\partial _{ab}G\big)%
(C_{t})(C_{t}^{ga}C_{t}^{hb}+C_{t}^{gb}C_{t}^{ha})dt\Bigg)\overset{\mathbb{P}%
}{\longrightarrow }0.
\end{equation*}%
Using the multivariate quantities defined in Section \ref{sec:notation}, we
can show that the following decompositions hold:
\begin{align*}
\widehat{C}_{i}^{^{\prime }n}& =C_{i-1}^{n}+\frac{1}{k_{n}}%
\sum_{j=0}^{k_{n}-1}\sum_{u=1}^{2}\overline{\varepsilon }(u)_{j}^{n}\zeta
(u)_{i+j}^{n},\hspace{5mm}\widehat{C}_{i+k_{n}}^{^{\prime }n}-\widehat{C}%
_{i}^{^{\prime }n}=\frac{1}{k_{n}}\sum_{j=0}^{2k_{n}-1}\sum_{u=1}^{2}%
\varepsilon (u)_{j}^{n}\zeta (u)_{i+j}^{n}, \\
\lambda _{i}^{n,gh}\lambda _{i}^{n,ab}& =\frac{1}{k_{n}^{2}}%
\sum_{u=1}^{2}\sum_{v=1}^{2}\Bigg(\sum_{j=0}^{2k_{n}-1}\varepsilon
(u)_{j}^{n}\varepsilon (v)_{j}^{n}\zeta (u)_{i+j}^{n,gh}\zeta
(v)_{i+j}^{n,ab} \\
& +\sum_{j=0}^{2k_{n}-2}\sum_{q=j+1}^{2k_{n}-1}\varepsilon
(u)_{j}^{n}\varepsilon (v)_{q}^{n}\zeta (u)_{i+j}^{n,gh}\zeta
(v)_{i+q}^{n,ab}+\sum_{j=1}^{2k_{n}-1}\sum_{q=0}^{j-1}\varepsilon
(u)_{j}^{n}\varepsilon (v)_{q}^{n}\zeta (u)_{i+j}^{n,gh}\zeta
(v)_{i+q}^{n,ab}\Bigg).
\end{align*}%
Changing the order of the summation in the last term yields
\begin{align*}
\lambda _{i}^{n,gh}\lambda _{i}^{n,ab}& =\frac{1}{k_{n}^{2}}%
\sum_{u=1}^{2}\sum_{v=1}^{2}\Bigg(\sum_{j=0}^{2k_{n}-1}\varepsilon
(u)_{j}^{n}\varepsilon (v)_{j}^{n}\zeta (u)_{i+j}^{n,gh}\zeta
(v)_{i+j}^{n,ab} \\
& +\sum_{j=0}^{2k_{n}-2}\sum_{q=j+1}^{2k_{n}-1}\varepsilon
(u)_{j}^{n}\varepsilon (v)_{q}^{n}\zeta (u)_{i+j}^{n,gh}\zeta
(v)_{i+q}^{n,ab}+\sum_{j=0}^{2k_{n}-2}\sum_{q=j+1}^{2k_{n}-1}\varepsilon
(v)_{j}^{n}\varepsilon (u)_{q}^{n}\zeta (v)_{i+j}^{n,ab}\zeta
(u)_{i+q}^{n,gh}\Bigg).
\end{align*}%
Therefore, we can further rewrite $\sum_{i=k_{n}+1}^{[T/\Delta
_{n}]-3k_{n}+1}\vartheta _{i}^{\left( A1\right) }$ as %
\begin{align*}
& \sum_{i=k_{n}+1}^{[T/\Delta _{n}]-3k_{n}+1}\vartheta _{i}^{\left(
A1\right) }=\sum_{i\in L\left( n,T\right) }\vartheta _{i}^{\left( A11\right)
}+\sum_{i\in L\left( n,T\right) }\vartheta _{i}^{\left( A12\right)
}+\sum_{i\in L\left( n,T\right) }\vartheta _{i}^{\left( A13\right) },\text{%
with} \\
& \sum_{i=k_{n}+1}^{[T/\Delta _{n}]-3k_{n}+1}\vartheta _{i}^{\left(
A1w\right) }=\sum_{g,h,a,b=1}^{d}\sum_{u,v=1}^{2}\widehat{A1w}%
(H,gh,u;G,ab,v)_{T}^{n},\hspace{3mm}w=1,2,3,
\end{align*}%
where

\begin{align*}
& \widehat{A11}(H,gh,u;G,ab,v)_{T}^{n}=\frac{3}{2k_{n}^{3}}%
\sum_{i=k_{n}+1}^{[T/\Delta _{n}]-3k_{n}+1}\sum_{j=0}^{2k_{n}-1}(\partial
_{gh}H\partial _{ab}G)(C_{i-1}^{n})\varepsilon (u)_{j}^{n}\varepsilon
(v)_{j}^{n}\zeta (u)_{i+j}^{n,gh}\zeta (v)_{i+j}^{n,ab}, \\
& \widehat{A12}(H,gh,u;G,ab,v)_{T}^{n}=\frac{3}{2k_{n}^{3}}%
\sum_{i=k_{n}+1}^{[T/\Delta
_{n}]-3k_{n}+1}\sum_{j=0}^{2k_{n}-2}\sum_{q=j+1}^{2k_{n}-1}(\partial
_{gh}H\partial _{ab}G)(C_{i-1}^{n})\varepsilon (u)_{j}^{n}\varepsilon
(v)_{q}^{n}\zeta (u)_{i+j}^{n,gh}\zeta (v)_{i+q}^{n,ab}, \\
& \widehat{A13}(H,gh,u;G,ab,v)_{T}^{n}=\frac{3}{2k_{n}^{3}}%
\sum_{i=k_{n}+1}^{[T/\Delta
_{n}]-3k_{n}+1}\sum_{j=0}^{2k_{n}-2}\sum_{q=j+1}^{2k_{n}-1}(\partial
_{gh}H\partial _{ab}G)(C_{i-1}^{n})\varepsilon (v)_{j}^{n}\varepsilon
(u)_{q}^{n}\zeta (v)_{i+j}^{n,ab}\zeta (u)_{i+q}^{n,gh},
\end{align*}%
where we clearly have $\widehat{A13}(H,gh,u;G,ab,v)_{T}^{n}=\widehat{A12}%
(G,ab,v;H,gh,u)_{T}^{n}.$ By a change of the order of the summation,%
\begin{align*}
\widehat{A11}(H,gh,u;G,ab,v)_{T}^{n}& =\frac{3}{2k_{n}^{3}}%
\sum_{i=1}^{[T/\Delta _{n}]}\sum_{j=0\vee (i+2k_{n}-1-[T/\Delta
_{n}])}^{(2k_{n}-1)\wedge (i-1)}(\partial _{gh}H\partial _{ab}G) \\
& \times (C_{i-j-1}^{n})\varepsilon (u)_{j}^{n}\varepsilon (v)_{j}^{n}\zeta
(u)_{i}^{n,gh}\zeta (v)_{i}^{n,ab}, \\
\widehat{A12}(H,gh,u;G,ab,v)_{T}^{n}& =\frac{3}{2k_{n}^{3}}%
\sum_{i=2}^{[T/\Delta _{n}]}\sum_{m=1}^{(i-1)\wedge (2k_{n}-1)}\sum_{j=0\vee
(i+2k_{n}-1-m-[T/\Delta _{n}])}^{(2k_{n}-m-1)\wedge (i-m-1)}(\partial
_{gh}H\partial _{ab}G)(C_{i-1-j-m}^{n}) \\
& \times \varepsilon (u)_{j}^{n}\varepsilon (v)_{j+m}^{n}\zeta
_{gh}(u)_{i-m}^{n}\zeta _{ab}(v)_{i}^{n}.
\end{align*}%
Now, set%
\begin{align*}
\widetilde{A11}(H,gh,u;G,ab,v)_{T}^{n}& =\frac{3}{2k_{n}^{3}}%
\sum_{i=3k_{n}}^{[T/\Delta _{n}]-k_{n}}\sum_{j=0}^{2k_{n}-1}(\partial
_{gh}H\partial _{ab}G)(C_{i-j-1}^{n})\varepsilon (u)_{j}^{n}\varepsilon
(v)_{j}^{n}\zeta (u)_{i}^{n,gh}\zeta (v)_{i}^{n,ab}, \\
\widetilde{A12}(H,gh,u;G,ab,v)_{T}^{n}& =\frac{3}{2k_{n}^{3}}%
\sum_{i=3k_{n}}^{[T/\Delta _{n}]-k_{n}}\sum_{m=1}^{(i-1)\wedge
(2k_{n}-1)}\sum_{j=0}^{(2k_{n}-m-1)}(\partial _{gh}H\partial
_{ab}G)(C_{i-j-1-m}^{n})\varepsilon (u)_{j}^{n}\varepsilon (v)_{j+m}^{n} \\
& \times \zeta _{gh}(u)_{i-m}^{n}\zeta _{ab}(v)_{i}^{n}.
\end{align*}%
We show below that the following results hold:
\begin{equation}
\frac{1}{\Delta _{n}^{1/4}}\Big(\widehat{A1w}(H,gh,u;G,ab,v)_{T}^{n}-%
\widetilde{A1w}(H,gh,u;G,ab,v)_{T}^{n}\Big)\overset{\mathbb{P}}{%
\longrightarrow }0  \label{eqn:approx1}
\end{equation}%
\begin{equation}
\frac{1}{\Delta _{n}^{1/4}}\Big(\widetilde{A1w}(H,gh,u;G,ab,v)_{T}^{n}-%
\overline{A1w}(H,gh,u;G,ab,v)_{T}^{n}\Big)\overset{\mathbb{P}}{%
\longrightarrow }0  \label{eqn:approx2}
\end{equation}%
\text{for all} \hspace{1mm}$(H,gh,u,G,ab,v)~\text{and}~w=1,2.$

\subsubsection{Proof of Equation (\protect\ref{eqn:approx1}) for $w=1$}

To prove this result, first, notice that the $\zeta (u)_{i}^{n,gh}\zeta
(v)_{i}^{n,ab}$ are scaled by random variables rather that constant real
numbers. Next, observe that we can write%
\begin{align*}
& \widehat{A11}-\widetilde{A11}=\widetilde{\widehat{A11}}(1)+\widetilde{%
\widehat{A11}}(2)+\widetilde{\widehat{A11}}(3)\hspace{5mm}\text{with} \\
& \widetilde{\widehat{A11}}(1)=\sum_{i=1}^{(2k_{n}-1)\wedge \lbrack T/\Delta
_{n}]}\Bigg(\frac{3}{2k_{n}^{3}}\sum_{j=0\vee (i+2k_{n}-1-[T/\Delta
_{n}])}^{(2k_{n}-1)\wedge (i-1)}(\partial _{gh}H\partial
_{ab}G)(C_{i-j-1}^{n})\varepsilon (u)_{j}^{n}\varepsilon (v)_{j}^{n}\Bigg)%
\zeta (u)_{i}^{n,gh}\zeta (v)_{i}^{n,ab}, \\
& \widetilde{\widehat{A11}}(2)=\sum_{i=[T/\Delta _{n}]-2k_{n}+2}^{[T/\Delta
_{n}]}\frac{3}{2k_{n}^{3}}\Bigg(\sum_{j=0\vee (i+2k_{n}-1-[T/\Delta
_{n}])}^{(2k_{n}-1)\wedge (i-1)}(\partial _{gh}H\partial
_{ab}G)(C_{i-j-1}^{n})\varepsilon (u)_{j}^{n}\varepsilon (v)_{j}^{n} \\
& \hspace{40mm}-\sum_{j=0}^{(2k_{n}-1)}(\partial _{gh}H\partial
_{ab}G)(C_{i-j-1}^{n})\varepsilon (u)_{j}^{n}\varepsilon (v)_{j}^{n}\Bigg)%
\zeta (u)_{i}^{n,gh}\zeta (v)_{i}^{n,ab}, \\
& \widetilde{\widehat{A11}}(3)=\sum_{i=2k_{n}}^{[T/\Delta _{n}]-2k_{n}+1}%
\frac{3}{2k_{n}^{3}}\Bigg(\sum_{j=0\vee (i+2k_{n}-1-[T/\Delta
_{n}])}^{(2k_{n}-1)\wedge (i-1)}(\partial _{gh}H\partial
_{ab}G)(C_{i-j-1}^{n})\varepsilon (u)_{j}^{n}\varepsilon (v)_{j}^{n} \\
& \hspace{40mm}-\sum_{j=0}^{(2k_{n}-1)}(\partial _{gh}H\partial
_{ab}G)(C_{i-j-1}^{n})\varepsilon (u)_{j}^{n}\varepsilon (v)_{j}^{n}\Bigg)%
\zeta (u)_{i}^{n,gh}\zeta (v)_{i}^{n,ab}.
\end{align*}%
It is easy to see that $\widetilde{\widehat{A12}}(3)=0$. Using equation (\ref%
{eqn:SKCC}) with $Z=c$ and equation (\ref{eqn:alpha}), we obtain
\begin{equation}
\mathbb{E}(\Vert \zeta (1)_{i}^{n}\Vert ^{q}|\mathcal{F}_{i-1}^{n})\leq
K_{q},\hspace{2mm}\mathbb{E}(\Vert \zeta (2)_{i}^{n}\Vert ^{q}|\mathcal{F}%
_{i-1}^{n})\leq K_{q}\Delta _{n}^{q/2}.  \label{eqn:powers}
\end{equation}

\noindent By the boundedness of the derivatives of $H$ and $G$, the random
quantities $\Big(\frac{3}{2k_{n}^{3}}\sum_{j=0\vee (i+2k_{n}-1-[T/\Delta
_{n}])}^{(2k_{n}-1)\wedge (i-1)}(\partial _{gh}H\partial
_{ab}G)(C_{i-j-1}^{n})\varepsilon (u)_{j}^{n}\varepsilon (v)_{j}^{n}\Big)$
and\newline
$\frac{3}{2k_{n}^{3}}\sum_{j=0}^{(2k_{n}-1)}(\partial _{gh}H\partial
_{ab}G)(C_{i-j-1}^{n})\varepsilon (u)_{j}^{n}\varepsilon (v)_{j}^{n}$ are $%
\mathcal{F}_{i-1}^{n}-$ measurable and are bounded by $\widetilde{\lambda }%
_{u,v}^{n}$ defined as
\begin{equation*}
\widetilde{\lambda }_{u,v}^{n}=%
\begin{cases}
K & \text{if}\hspace{1mm}(u,v)=(2,2) \\
K/k_{n} & \text{if}\hspace{1mm}(u,v)=(1,2),(2,1) \\
K/k_{n}^{2} & \text{if}\hspace{1mm}(u,v)=(1,1).%
\end{cases}%
\end{equation*}%
Similarly, the quantity
\begin{equation*}
\frac{3}{2k_{n}^{3}}\Bigg(\sum_{j=0\vee (i+2k_{n}-1-[T/\Delta
_{n}])}^{(2k_{n}-1)\wedge (i-1)}\hspace{-1mm}(\partial _{gh}H\partial
_{ab}G)(C_{i-j-1}^{n})\varepsilon (u)_{j}^{n}\varepsilon (v)_{j}^{n}-\hspace{%
-2mm}\sum_{j=0}^{(2k_{n}-1)}(\partial _{gh}H\partial
_{ab}G)(C_{i-j-1}^{n})\varepsilon (u)_{j}^{n}\varepsilon (v)_{j}^{n}\Bigg),
\end{equation*}%
is $\mathcal{F}_{i-1}^{n}-$ measurable and bounded by $2\widetilde{\lambda }%
_{u,v}^{n}$. Note also that, by equation (\ref{eqn:powers}) and the Cauchy
Schwartz inequality, we have
\begin{align*}
\mathbb{E}(|\zeta (u)_{i}^{n,gh}\zeta (v)_{i}^{n,ab}|\big|\mathcal{F}%
_{i-1}^{n})& \leq \mathbb{E}(\Vert \zeta (u)_{i}^{n}\Vert ^{2}|\mathcal{F}%
_{i-1}^{n})^{1/2}\mathbb{E}(\Vert \zeta (v)_{i}^{n}\Vert ^{2}|\mathcal{F}%
_{i-1}^{n})^{1/2} \\
& \leq
\begin{cases}
K\Delta _{n} & \text{if}\hspace{1mm}(u,v)=(2,2) \\
K\Delta _{n}^{1/2} & \text{if}\hspace{1mm}(u,v)=(1,2),(2,1) \\
K & \text{if}\hspace{1mm}(u,v)=(1,1).%
\end{cases}%
\end{align*}%
The above bounds, together with the fact that $k_{n}=\theta \Delta
_{n}^{-1/2}$, imply $\mathbb{E}(|\widetilde{\widehat{A11}}(1)|)\leq K\Delta
_{n}^{1/2}$ and $\mathbb{E}(|\widetilde{\widehat{A11}}(2)|)\leq K\Delta
_{n}^{1/2}$ for all $(u,v)$. These two results together imply $\widetilde{%
\widehat{A11}}(1)=o(\Delta _{n}^{-1/4})$ and $\widetilde{\widehat{A11}}%
(2)=o(\Delta _{n}^{-1/4})$, which yields the result.

\subsubsection{Proof of Equation (\protect\ref{eqn:approx1}) for $w=2$}

First, observe that $\widehat{A12}-\widetilde{A12}=\widetilde{\widehat{A12}}%
(1)+\widetilde{\widehat{A12}}(2)$, with%
\begin{align*}
& \widetilde{\widehat{A12}}(1)=\sum_{i=2}^{(2k_{n}-1)\wedge \lbrack T/\Delta
_{n}]}\Bigg(\sum_{m=1}^{(i-1)}\frac{3}{2k_{n}^{3}}\Big(\sum_{j=0\vee
(i+2k_{n}-1-m-[T/\Delta _{n}])}^{(2k_{n}-m-1)\wedge (i-m-1)}(\partial
_{gh}H\partial _{ab}G)(C_{i-1-j-m}^{n})\varepsilon (u)_{j}^{n}\varepsilon
(v)_{j+m}^{n}\Big) \\
& \hspace{20mm}\times \zeta _{gh}(u)_{i-m}^{n}\Bigg)\zeta _{ab}(v)_{i}^{n},
\\
& \widetilde{\widehat{A12}}(2)=\sum_{i=[T/\Delta _{n}]-2k_{n}+2}^{[T/\Delta
_{n}]}\Bigg(\sum_{m=1}^{(i-1)\wedge (2k_{n}-1)}\Big(\frac{3}{2k_{n}^{3}}%
\sum_{j=0\vee (i+2k_{n}-1-m-[T/\Delta _{n}])}^{(2k_{n}-m-1)\wedge
(i-m-1)}(\partial _{gh}H\partial _{ab}G)(C_{i-1-j-m}^{n})\varepsilon
(u)_{j}^{n} \\
& \hspace{20mm}\times \varepsilon (v)_{j+m}^{n}\Big)%
-\sum_{j=0}^{(2k_{n}-m-1)}(\partial _{gh}H\partial
_{ab}G)(C_{i-1-j-m}^{n})\varepsilon (u)_{j}^{n}\varepsilon (v)_{j+m}^{n}\Big)%
\zeta _{gh}(u)_{i-m}^{n}\Bigg)\zeta _{ab}(v)_{i}^{n}.
\end{align*}%
Notice that the quantity
\begin{equation*}
\kappa _{i}^{m,n}=\frac{3}{2k_{n}^{3}}\Big(\sum_{j=0\vee
(i+2k_{n}-1-m-[T/\Delta _{n}])}^{(2k_{n}-m-1)\wedge (i-m-1)}(\partial
_{gh}H\partial _{ab}G)(C_{i-1-j-m}^{n})\varepsilon (u)_{j}^{n}\varepsilon
(v)_{j+m}^{n}\Big)
\end{equation*}%
is $\mathcal{F}_{i-m-1}^{n}$ measurable and bounded by $\widetilde{\lambda }%
_{u,v}^{n}$. Let
\begin{equation*}
\kappa _{i}^{n}=\sum_{m=1}^{(i-1)}\frac{3}{2k_{n}^{3}}\Big(\sum_{j=0\vee
(i+2k_{n}-1-m-[T/\Delta _{n}])}^{(2k_{n}-m-1)\wedge (i-m-1)}(\partial
_{gh}H\partial _{ab}G)(C_{i-1-j-m}^{n})\varepsilon (u)_{j}^{n}\varepsilon
(v)_{j+m}^{n}\Big)\zeta _{gh}(u)_{i-m}^{n}.
\end{equation*}%
It follows that $\kappa _{i}^{n}$ is $\mathcal{F}_{i-1}^{n}$-measurable and
we have
\begin{eqnarray*}
\mathbb{E}(|\kappa _{i}^{m,n}|^{z}\big|\mathcal{F}_{0}) &\leq &(\widetilde{%
\lambda }_{u,v}^{n})^{z}, \\
|\mathbb{E}(\zeta (u)_{i-m}^{n}|\mathcal{F}_{i-m-1})| &\leq &%
\begin{cases}
K\sqrt{\Delta _{n}} & \text{if}\hspace{1mm}u=1 \\
K\Delta _{n} & \text{if}\hspace{1mm}u=2%
\end{cases}%
, \\
\mathbb{E}(\Vert \zeta (u)_{i-m}^{n}\Vert ^{z}|\mathcal{F}_{i-m-1}) &\leq &%
\begin{cases}
K_{z} & \text{if}\hspace{1mm}u=1 \\
K_{z}\Delta _{n}^{z/2} & \text{if}\hspace{1mm}u=2%
\end{cases}%
.
\end{eqnarray*}%
Using Lemma \ref{lemma:key}, we deduce that for $z\geq 2$,
\begin{equation*}
\mathbb{E}(|\kappa _{i}^{n}|^{z})\leq
\begin{cases}
K_{z}(\widetilde{\lambda }_{u,v}^{n})^{z}k_{n}^{z/2} & \text{if}\hspace{1mm}%
u=1 \\
K_{z}(\widetilde{\lambda }_{u,v}^{n})^{z}/k_{n}^{z/2} & \text{if}\hspace{1mm}%
u=2%
\end{cases}%
\leq
\begin{cases}
K_{z}/k_{n}^{-3z/2} & \text{if}\hspace{1mm}v=1 \\
K_{z}k_{n}^{-z/2} & \text{if}\hspace{1mm}v=2%
\end{cases}%
.
\end{equation*}%
Using the above result, we obtain $\frac{1}{\Delta _{n}^{1/4}}\widetilde{%
\widehat{A12}}(1)\overset{\mathbb{P}}{\Rightarrow }0$. A similar argument
yields $\frac{1}{\Delta _{n}^{1/4}}\widetilde{\widehat{A12}}(2)\overset{%
\mathbb{P}}{\Rightarrow }0$, which completes the proof of the equation (\ref%
{eqn:approx1}) for $w=2$.

\subsubsection{Proof of Equation (\protect\ref{eqn:approx2}) for $w=1$}

Define
\begin{equation*}
\hspace{3mm}\Theta (u,v)_{0}^{(C),i,n}=\frac{3}{2k_{n}^{3}}%
\sum_{j=0}^{2k_{n}-1}\Big((\partial _{gh}H\partial
_{ab}G)(C_{i-j-1}^{n})-(\partial _{gh}H\partial _{ab}G)(C_{i-2k_{n}}^{n})%
\Big)\varepsilon (u)_{j}^{n}\varepsilon (v)_{j}^{n}.
\end{equation*}%
By Taylor expansion, boundedness of the derivatives of $H$ and $G$,
and using (\ref{eqn:SKCC}) with $Z=c$, we have
\begin{align*}
& \Big|\mathbb{E}\Big((\partial _{gh}H\partial
_{ab}G)(C_{i-j-1}^{n})-(\partial _{gh}H\partial _{ab}G)(C_{i-2k_{n}}^{n})%
\big|\mathcal{F}_{i-2k_{n}}^{n}\Big)\Big|\leq K(k_{n}\Delta _{n})\leq K\sqrt{%
\Delta _{n}} \\
& \mathbb{E}(|(\partial _{gh}H\partial _{ab}G)(C_{i-j-1}^{n})-(\partial
_{gh}H\partial _{ab}G)(C_{i-2k_{n}}^{n})|^{q}|\mathcal{F}_{i-2k_{n}}^{n})|%
\leq K(k_{n}\Delta _{n})^{q/2}\leq K\Delta _{n}^{q/4},
\end{align*}%
for $q\geq 2$ and for $j=0,\ldots ,2k_{n}-1$. Next, observe that $\Theta
(u,v)_{0}^{(C),i,n}$ is $\mathcal{F}_{i-1}^{n}$ -measurable and satisfies $%
|\Theta (u,v)_{0}^{(C),i,n}|\leq \widetilde{\lambda }_{u,v}^{n}$, $|\mathbb{E%
}\Big(\Theta (u,v)_{0}^{(C),i,n}|\mathcal{F}_{i-2k_{n}}^{n}\Big)|\leq
K\Delta _{n}^{1/2}\widetilde{\lambda }_{u,v}^{n}$ and $\mathbb{E}\Big(%
|\Theta (u,v)_{0}^{(C),i,n}|^{q}\big|\mathcal{F}_{i-2k_{n}}^{n}\Big)\leq
K_{q}\Delta _{n}^{q/4}(\widetilde{\lambda }_{u,v}^{n})^{q}$ where the latter
follows from the H\"{o}lder inequality. We aim to prove that
\begin{equation*}
\widehat{E}=\frac{1}{\Delta _{n}^{1/4}}\Bigg[\sum_{i=2k_{n}}^{[T/\Delta
_{n}]}\Theta (u,v)_{0}^{(C),i,n}\zeta (u)_{i}^{n,gh}\zeta (v)_{i}^{n,ab}%
\Bigg]
\end{equation*}%
converges to zero in probability for any $H$, $G$, $g$, $h$, $a$, and $b$
with $u,v=1,2$.
\newline
To show this result, we first introduce the following quantities:%
\begin{align*}
& \widehat{E}(1)=\frac{1}{\Delta _{n}^{1/4}}\Bigg[\sum_{i=3k_{n}}^{[T/\Delta
_{n}]-k_{n}}\Theta (u,v)_{0}^{(C),i,n}\mathbb{E}(\zeta (u)_{i}^{n,gh}\zeta
(v)_{i}^{n,ab}|\mathcal{F}_{i-1}^{n})\Bigg] \\
& \widehat{E}(2)=\frac{1}{\Delta _{n}^{1/4}}\Bigg[\sum_{i=3k_{n}}^{[T/\Delta
_{n}]-k_{n}}\Theta (u,v)_{0}^{(C),i,n}\big(\zeta (u)_{i}^{n,gh}\zeta
(v)_{i}^{n,ab}-\mathbb{E}(\zeta (u)_{i}^{n,gh}\zeta (v)_{i}^{n,ab}|\mathcal{F%
}_{i-1}^{n})\big)\Bigg],
\end{align*}%
with $\widehat{E}=\widehat{E}(1)+\widehat{E}(2)$. By Cauchy-Schwartz
inequality, we have
\begin{equation*}
\mathbb{E}(|\zeta (u)_{i}^{n,gh}\zeta (v)_{i}^{n,ab}|^{q})\leq (\widehat{%
\lambda }_{u,v}^{n})^{q/2},\text{where}\hspace{2mm}\widehat{\lambda }%
_{u,v}^{n}=%
\begin{cases}
K & \text{if}\hspace{1mm}(u,v)=(1,1) \\
K\Delta _{n} & \text{if}\hspace{1mm}(u,v)=(1,2),(2,1) \\
K\Delta _{n}^{2} & \text{if}\hspace{1mm}(u,v)=(2,2)%
\end{cases}%
\end{equation*}%
Since $\zeta (u)_{i}^{n,gh}\zeta (v)_{i}^{n,ab}$ is $\mathcal{F}_{i}^{n}$%
-measurable,\newline
the martingale property of $\zeta (u)_{i}^{n,gh}\zeta (v)_{i}^{n,ab}-\mathbb{%
E}(\zeta (u)_{i}^{n,gh}\zeta (v)_{i}^{n,ab}|\mathcal{F}_{i-1}^{n})$ implies,
for all $(u,v)$,
\begin{equation*}
\mathbb{E}(|\widehat{E}(2)|^{2})\leq K\Delta _{n}^{-3/2}(\Delta _{n}^{1/4}%
\widetilde{\lambda }_{u,v}^{n})^{2}\widehat{\lambda }_{u,v}^{n}\leq K\Delta
_{n}.
\end{equation*}%
The latter inequality implies $\widehat{E}(2)\overset{\mathbb{P}}{%
\Rightarrow }0$ for all $(u,v)$. It remains to show that $\widehat{E}(1)%
\overset{\mathbb{P}}{\Rightarrow }0$.\newline
Here, we recall some bounds under Assumption \ref{ass:volvol},
\begin{align}
& |\mathbb{E}(\zeta (1)_{i}^{n,gh}\zeta (2)_{i}^{n,ab}|\mathcal{F}%
_{i-1}^{n})|\leq K\Delta _{n},  \label{eqn:skcc0} \\
& |\mathbb{E}(\zeta (1)_{i}^{n,gh}\zeta (1)_{i}^{n,ab}|\mathcal{F}%
_{i-1}^{n})-\big(C_{i-1}^{n,ga}C_{i-1}^{n,hb}+C_{i-1}^{n,gb}C_{i-1}^{n,ha}%
\big)|\leq K\Delta _{n}^{1/2},  \label{eqn:skcc1} \\
& |\mathbb{E}(\zeta (2)_{i}^{n,gh}\zeta (2)_{i}^{n,ab}|\mathcal{F}_{i-1}^{n}-%
\overline{C}_{i-1}^{n,gh,ab}\Delta _{n})|\leq K\Delta _{n}^{3/2}(\sqrt{%
\Delta _{n}}+\eta _{i}^{n}).  \label{eqn:skcc2}
\end{align}%
\noindent \textbf{Case} $(u,v)\in \{(1,2),(2,1)\}$. By equation (\ref%
{eqn:skcc0}) we have
\begin{equation*}
\mathbb{E}(|\widehat{E}(1)|)\leq K\frac{T}{\Delta _{n}}\frac{1}{\Delta
_{n}^{1/4}}(\Delta _{n}^{1/4}\widetilde{\lambda }_{u,v}^{n}\Delta _{n})\leq
K\Delta _{n}^{1/2}\hspace{3mm}\text{so}\hspace{3mm}\widehat{E}(1)\overset{%
\mathbb{P}}{\Rightarrow }0.
\end{equation*}%
\noindent \textbf{Case} $(u,v)\in \{(1,1),(2,2)\}$. Set%
\begin{align*}
\widehat{E}^{\prime }(1)& =\frac{1}{\Delta _{n}^{1/4}}\Bigg[%
\sum_{i=3k_{n}}^{[T/\Delta _{n}]-k_{n}}\Theta
(u,v)_{0}^{(C),i,n}V_{i-2k_{n}}^{n}\Bigg] \\
\widehat{E}^{\prime \prime }(1)& =\frac{1}{\Delta _{n}^{1/4}}\Bigg[%
\sum_{i=3k_{n}}^{[T/\Delta _{n}]-k_{n}}\Theta (u,v)_{0}^{(C),i,n}\big(%
V_{i-1}^{n}-V_{i-2k_{n}}^{n}\big)\Bigg] \\
\widehat{E}^{\prime \prime \prime }(1)& =\frac{1}{\Delta _{n}^{1/4}}\Bigg[%
\sum_{i=3k_{n}}^{[T/\Delta _{n}]-k_{n}}\Theta (u,v)_{0}^{(C),i,n}\Big(%
\mathbb{E}(\zeta (u)_{i}^{n,gh}\zeta (v)_{i}^{n,ab}|\mathcal{F}%
_{i-1}^{n})-V_{i-1}^{n}\Big)\Bigg]
\end{align*}%
where
\begin{equation*}
V_{i-1}^{n}=%
\begin{cases}
C_{i-1}^{n,ga}C_{i-1}^{n,hb}+C_{i-1}^{n,gb}C_{i-1}^{n,ha} & \text{if}\hspace{%
1mm}(u,v)=(2,2) \\
\overline{C}_{i-1}^{n,gh,ab}\Delta _{n} & \text{if}\hspace{1mm}(u,v)=(1,1)
\\
0 & \text{otherwise}%
\end{cases}%
\end{equation*}%
Note that we have $\widehat{E}(1)=\widehat{E}^{\prime }(1)+\widehat{E}%
^{\prime \prime }(1)+\widehat{E}^{\prime \prime \prime }(1)$. Using
equations (\ref{eqn:skcc1}) and (\ref{eqn:skcc2}), it can be shown that
\begin{equation*}
\mathbb{E}(|\widehat{E}^{\prime \prime \prime }(1)|)\leq
\begin{cases}
K\frac{1}{\Delta _{n}^{5/4}}(\Delta _{n}^{1/4}\widetilde{\lambda }%
_{u,v}^{n})\Delta _{n}^{1/2} & \text{if}\hspace{1mm}(u,v)=(1,1) \\
K\frac{1}{\Delta _{n}^{5/4}}(\Delta _{n}^{1/4}\widetilde{\lambda }%
_{u,v}^{n})\Delta _{n}^{3/2} & \text{if}\hspace{1mm}(u,v)=(2,2)%
\end{cases}%
\leq K\Delta _{n}^{1/2}\hspace{3mm}\text{in all cases.}
\end{equation*}%
Next, we prove 
$\widehat{E}^{\prime }(1)\overset{\mathbb{P}}{\Rightarrow }0$. To this end,
write
\begin{equation*}
\widehat{E}^{\prime }(1)=\frac{1}{\Delta _{n}^{1/4}}\Bigg[%
\sum_{i=1}^{[T/\Delta _{n}]-2k_{n}+1}\Theta
(u,v)_{0}^{(C),i-1+2k_{n},n}V_{i-1}^{n}\Bigg].
\end{equation*}%
Using the $\mathcal{F}_{i+2k_{n}-2}^{n}$-measurability of the last sum, we
are able to show
%
\begin{align*}
\frac{1}{\Delta _{n}^{1/4}}\Bigg[\sum_{i=k_{n}+1}^{[T/\Delta _{n}]-3k_{n}+1}|%
\mathbb{E}(\Theta (u,v)_{0}^{(C),i-1+2k_{n},n}V_{i-1}^{n}|\mathcal{F}%
_{i-1}^{n})|\Bigg]& \overset{\mathbb{P}}{\Rightarrow }0\hspace{3mm}\text{and}
\\
\frac{2k_{n}-2}{\Delta _{n}^{1/2}}\Bigg[\sum_{i=k_{n}+1}^{[T/\Delta
_{n}]-3k_{n}+1}\mathbb{E}\Big(|\Theta
(u,v)_{0}^{(C),i-1+2k_{n},n}V_{i-1}^{n})|^{2}\Big)\Bigg]& \Rightarrow 0.
\end{align*}%
The first result readily follows from the inequality
\begin{equation*}
|\mathbb{E}(\Theta (u,v)_{0}^{(C),i-1+2k_{n},n}V_{i-1}^{n}|\mathcal{F}%
_{i-1}^{n})|\leq
\begin{cases}
K\Delta _{n}^{1/2}\widetilde{\lambda }_{u,v}^{n} & \text{if}\hspace{1mm}%
(u,v)=(1,1) \\
K\Delta _{n}^{1/2}\widetilde{\lambda }_{u,v}^{n}\Delta _{n} & \text{if}%
\hspace{1mm}(u,v)=(2,2)%
\end{cases}%
\leq K\Delta _{n}^{3/2}\hspace{3mm}\text{in all cases,}
\end{equation*}%
while the second is a direct consequence of
\begin{equation*}
\mathbb{E}(|\Theta (u,v)_{0}^{(C),i-1+2k_{n},n}V_{i-1}^{n}|^{2})\leq
\begin{cases}
K\Delta _{n}^{1/2}(\widetilde{\lambda }_{u,v}^{n})^{2} & \text{if}\hspace{1mm%
}(u,v)=(1,1) \\
K\Delta _{n}^{1/2}(\widetilde{\lambda }_{u,v}^{n})^{2}\Delta _{n}^{2} &
\text{if}\hspace{1mm}(u,v)=(2,2)%
\end{cases}%
\leq K\Delta _{n}^{5/2}\hspace{3mm}\text{in all cases.}
\end{equation*}%
Finally, to prove that $\widehat{E}^{\prime \prime }(1)\overset{\mathbb{P}}{%
\Longrightarrow }0$, we use the fact that
\begin{align*}
\mathbb{E}(|\Theta (u,v)_{0}^{(C),i,n}\big(V_{i-1}^{n}-V_{i-2k_{n}}^{n}\big)%
|)& \leq \mathbb{E}(|\Theta (u,v)_{0}^{(C),i,n}|^{2})^{1/2}\mathbb{E}%
(|V_{i-1}^{n}-V_{i-2k_{n}}^{n}|^{2})^{1/2} \\
& \leq
\begin{cases}
K\Delta _{n}^{1/2}\widetilde{\lambda }_{u,v}^{n} & \text{if}\hspace{1mm}%
(u,v)=(1,1) \\
K\Delta _{n}^{1/4}\widetilde{\lambda }_{u,v}^{n}\Delta _{n}\Delta _{n}^{1/4}
& \text{if}\hspace{1mm}(u,v)=(2,2)%
\end{cases}%
,
\end{align*}%
which follows from the Cauchy-Schwartz inequality and earlier bounds. In
particular, successive conditioning together with Assumption \ref{ass:volvol}
imply that for $(u,v)=(1,1)$ and $(2,2)$,\newline
$\mathbb{E}(|V_{i-1}^{n}-V_{i-2k_{n}}^{n}|^{2})\leq \Delta _{n}^{1/2}$.

\subsubsection{Proof of Equation (\protect\ref{eqn:approx2}) for $w=2$}

Our aim here is to show that
\begin{align*}
& \widehat{E}(2)=\frac{1}{\Delta _{n}^{1/4}}\sum_{i=3k_{n}}^{[T/\Delta
_{n}]-k_{n}}\Bigg(\sum_{m=1}^{2k_{n}-1}\Big(\frac{3}{2k_{n}^{3}}%
\sum_{j=0}^{2k_{n}-m-1}\big[(\partial _{gh}H\partial
_{ab}G)(C_{i-j-m-1}^{n})-(\partial _{gh}H\partial _{ab}G)(C_{i-2k_{n}}^{n})%
\big]\varepsilon (u)_{j}^{n}\varepsilon (v)_{j+m}^{n}\Big)\times \\
& \zeta (u)_{i-m}^{n,gh}\Bigg)\zeta (v)_{i}^{n,ab}\overset{\mathbb{P}}{%
\Longrightarrow }0.
\end{align*}%
For this purpose, we introduce some new notation. For any $0\leq m\leq
2k_{n}-1$, set
\begin{align*}
& \Theta (u,v)_{m}^{(C),i,n}=\frac{3}{2k_{n}^{3}}\sum_{j=0}^{2k_{n}-m-1}\big[%
(\partial _{gh}H\partial _{ab}G)(C_{i-j-m-1}^{n})-(\partial _{gh}H\partial
_{ab}G)(C_{i-2k_{n}}^{n})\big]\varepsilon (u)_{j}^{n}\varepsilon
(v)_{j+m}^{n} \\
& \rho (u,v)^{(C),i,n,gh}=\sum_{m=1}^{2k_{n}-1}\Theta
(u,v)_{m}^{(C),i,n}\zeta (u)_{i-m}^{n,gh}.
\end{align*}%
It is easy to see that $\Theta (u,v)_{m}^{(C),i,n}$ is $\mathcal{F}%
_{i-m-1}^{n}$ measurable and satisfies, by H\"{o}lder inequality,
\begin{equation*}
|\Theta (u,v)_{m}^{(C),i,n}|\leq \widetilde{\lambda }_{u,v}^{n}\hspace{2mm}%
\text{and}\hspace{2mm}\mathbb{E}\Big(|\Theta (u,v)_{m}^{(C),i,n}|^{q}\big|%
\mathcal{F}_{i-2k_{n}}^{n}\Big)\leq K_{q}\Delta _{n}^{q/4}(\widetilde{%
\lambda }_{u,v}^{n})^{q}.
\end{equation*}%
Lemma \ref{lemma:key} implies that for $q\geq 2$,
\begin{equation}
\mathbb{E}(|\rho (u,v)^{(C),i,n,gh}|^{q})\leq
\begin{cases}
K_{q}(\Delta _{n}^{1/4}\widetilde{\lambda }_{u,v}^{n})^{q}k_{n}^{q/2} &
\text{if}\hspace{1mm}u=1 \\
K_{q}(\Delta _{n}^{1/4}\widetilde{\lambda }_{u,v}^{n})^{q}/k_{n}^{q/2} &
\text{if}\hspace{1mm}u=2%
\end{cases}%
\leq
\begin{cases}
K_{q}/k_{n}^{2q} & \text{if}\hspace{1mm}v=1 \\
K_{q}k_{n}^{q} & \text{if}\hspace{1mm}v=2%
\end{cases}%
.  \label{eqn:rho_bound} 
\end{equation}%
Set
\begin{align*}
& \widehat{E}^{\prime }(2)=\frac{1}{\Delta _{n}^{1/4}}\sum_{i=3k_{n}}^{[T/%
\Delta _{n}]-k_{n}}\rho (u,v)^{(C),i,n,gh}\mathbb{E}(\zeta (v)_{i}^{n,ab}|%
\mathcal{F}_{i-1}^{n}), \\
& \widehat{E}^{\prime \prime }(2)=\frac{1}{\Delta _{n}^{1/4}}%
\sum_{i=3k_{n}}^{[T/\Delta _{n}]-k_{n}}\rho (u,v)^{(C),i,n,gh}(\zeta
(v)_{i}^{n,ab}-\mathbb{E}(\zeta (v)_{i}^{n,ab}|\mathcal{F}_{i-1}^{n})).
\end{align*}%
The martingale increments property implies $\mathbb{E}(|\widehat{E}^{\prime
\prime }(2)|^{2})\leq K\Delta _{n}^{1/2}$ in all the cases, which in turn
implies $\widehat{E}^{\prime \prime }(2)\overset{\mathbb{P}}{\Longrightarrow
}0$. Next, using the bounds on $\rho (u,v)^{(C),i,n,gh}$, we obtain that $%
\widehat{E}^{\prime }(2)\overset{\mathbb{P}}{\Longrightarrow }0$.

We refer to \cite{jacodrosenbaum-sqrtn} for the proofs of Lemma \ref%
{lemma:lem1} and Lemma \ref{lemma:ito}.

\subsection{Proof of Lemma \protect\ref{lemma:key}}

\noindent Set
\begin{align*}
& \hspace{5mm} \xi_i^n=\varphi^n_{i-1}\zeta_i^n, \hspace{5mm}
\xi_i^{^{\prime }n}=\mathbb{E}(\xi_i|\mathcal{F}_{i-1}^n)=\mathbb{E}%
(\varphi^n_{i-1}\zeta_i^n|\mathcal{F}_{i-1}^n)=\varphi^n_{i-1}\mathbb{E}%
(\zeta_i^n|\mathcal{F}_{i-1}^n),\hspace{2mm}\text{and}\hspace{2mm}%
\xi_i^{^{\prime \prime }n}=\xi_i^{n}-\xi_i^{^{\prime }n}.
\end{align*}
Given that $\|\mathbb{E}(\zeta_{i}^n|\mathcal{F}_{i-1}^n)\| \leq L^{\prime}$%
, we have $\|\xi_i^{^{\prime }n}\|\leq L^{\prime} |\varphi^n_{i-1}|$. By the
convexity of the function $x^q$, which holds for $q\geq 2$, we have
\begin{align*}
\|\sum_{j=1}^{2k_n-1}\xi_{i+j}^{n}\|^q \leq K\Big(\|\sum_{j=1}^{2k_n-1}%
\xi_{i+j}^{^{\prime }n}\|^q+\|\sum_{j=1}^{2k_n-1}\xi_{i+j}^{^{\prime \prime
}n}\|^q\Big).
\end{align*}
Therefore, on the one hand we have
\begin{align*}
&\|\sum_{j=1}^{2k_n-1}\xi_{i+j}^{^{\prime }n}\|^q \leq Kk_n^{q-1}
\sum_{j=1}^{2k_n-1}\|\xi_{i+j}^{^{\prime }n}\|^q \leq Kk_n^{q-1}L^{\prime q
}\sum_{j=1}^{2k_n-1} |\varphi_{i+j-1}^n|^q,
\end{align*}
which by $\mathbb{E}\Big(\|\varphi_{i+j-1}^n\|^q \Big|\mathcal{F}_{i-1}^n%
\Big) \leq L^q$, satisfies
\begin{align*}
&\mathbb{E}(\|\sum_{j=1}^{2k_n-1}\xi_{i+j}^{^{\prime }n}\|^q | \mathcal{F}%
_{i-1}^n)\leq KL^{\prime q}k_n^{q-1} \sum_{j=1}^{2k_n-1} \mathbb{E}%
(|\varphi_{i+j-1}^n|^q|\mathcal{F}_{i-1}^n)\leq KL^{\prime q}k_n^qL^q.
\end{align*}
On the other hand, we have $\mathbb{E}(\|\xi_{i+j}^{^{\prime \prime }n}\|^q
| \mathcal{F}_{i-1}^n)\leq \mathbb{E}(\|\xi_{i+j}^{n}\|^q | \mathcal{F}%
_{i-1}^n) \leq L_qL^q$ and $\mathbb{E}(\xi_{i+j}^{^{\prime \prime }n} |
\mathcal{F}_{i-1}^n)=0$, where the first inequality is a consequence of $%
\mathbb{E}(\|\xi_{i+j}^{^{\prime }n}\|^q | \mathcal{F}_{i-1}^n)\leq \mathbb{E%
}(\|\xi_{i+j}^{n}\|^q | \mathcal{F}_{i-1}^n) \leq L_qL^q$, which follows
from the Jensen's inequality and the law of iterated expectations. Hence, by
Lemma B.2 of \cite{yacjacod14} we have
\begin{align*}
\mathbb{E}(\|\sum_{j=1}^{2k_n-1}\xi_{i+j}^{^{\prime \prime }n}\|^q |
\mathcal{F}_{i-1}^n) \leq K_qL^qL_qk_n^{q/2}.
\end{align*}
To see the latter, we first prove that the required condition $\mathbb{E}%
(\|\xi_{i}^{n}\|^q | \mathcal{F}_{i-1}^n) \leq L_qL^q$) in the Lemma B.2 of
\cite{yacjacod14} can be replaced by $\mathbb{E}(\|\xi_{i+j}^{n}\|^q |
\mathcal{F}_{i-1}^n) \leq L_qL^q$) for $1\leq j \leq 2k_n-1$ without
altering the result.\newline


\subsection{Proof of Lemma \protect\ref{lem:avar}}

\noindent We use $i\in L\left( n,T\right) $ throughout the proof of Lemma %
\ref{lem:avar}. We use the terminology \textquotedblleft successive
conditioning" to refer to either of the following two equalities,
\begin{eqnarray*}
x_{1}y_{1}-x_{0}y_{0}
&=&x_{0}(y_{1}-y_{0})+y_{0}(x_{1}-x_{0})+(x_{1}-x_{0})(y_{1}-y_{0}), \\
x_{1}y_{1}z_{1}-x_{0}y_{0}z_{0}
&=&x_{0}y_{0}(z_{1}-z_{0})+x_{0}z_{0}(y_{1}-y_{0})+y_{0}z_{0}(x_{1}-x_{0})+x_{0}(y_{0}-y_{1})(z_{0}-z_{1})
\\
&&+y_{0}(x_{0}-x_{1})(z_{0}-z_{1})+z_{0}(x_{0}-x_{1})(y_{0}-y_{1})+(x_{1}-x_{0})(y_{1}-y_{0})(z_{1}-z_{0}),
\end{eqnarray*}%
which hold for any real numbers $x_{0},y_{0},z_{0},x_{1},y_{1},$ and $z_{1}$%
. \newline
\noindent To prove Lemma \ref{lem:avar}, we first note that $\lambda
_{i}^{n,jk}\lambda _{i}^{n,lm}$ is $\mathcal{F}_{i+2k_{n}}^{n}$-measurable.
Therefore, by the law of iterated expectations, we have
\begin{equation*}
\mathbb{E}\Big(\lambda _{i}^{n,jk}\lambda _{i}^{n,lm}\lambda
_{i+2k_{n}}^{n,gh}\lambda _{i+2k_{n}}^{n,ab}|\mathcal{F}_{i}^{n}\Big)=%
\mathbb{E}\Big(\lambda _{i}^{n,jk}\lambda _{i}^{n,lm}\mathbb{E}\big(\lambda
_{i+2k_{n}}^{n,gh}\lambda _{i+2k_{n}}^{n,ab}|\mathcal{F}_{i+2k_{n}}^{n}\big)|%
\mathcal{F}_{i}^{n}\Big).
\end{equation*}%
By equation (3.27) in \cite{jacodrosenbaum-sqrtn}, we have
\begin{align*}
& |\mathbb{E}(\lambda _{i+2k_{n}}^{n,gh}\lambda _{i+2k_{n}}^{n,ab}|\mathcal{F%
}_{i+2k_{n}}^{n})-\frac{2}{k_{n}}%
(C_{i+2k_{n}}^{n,ga}C_{i+2k_{n}}^{n,hb}+C_{i+2k_{n}}^{n,gb}C_{i+2k_{n}}^{n,ha})-%
\frac{2k_{n}\Delta _{n}}{3}\overline{C}_{i+2k_{n}}^{n,gh,ab}| \\
& \hspace{2mm}\leq K\sqrt{\Delta _{n}}(\Delta _{n}^{1/8}+\eta
_{i+2k_{n},2k_{n}}^{n}),~~\text{and} \\
& |\mathbb{E}(\lambda _{i}^{n,jk}\lambda _{i}^{n,lm}|\mathcal{F}_{i}^{n})-%
\frac{2}{k_{n}}(C_{i}^{n,jl}C_{i}^{n,km}+C_{i}^{n,jm}C_{i}^{n,kl})-\frac{%
2k_{n}\Delta _{n}}{3}\overline{C}_{i}^{n,jk,lm}|\leq K\sqrt{\Delta _{n}}%
(\Delta _{n}^{1/8}+\eta _{i,2k_{n}}^{n}).
\end{align*}%
From the above, it follows that
\begin{align*}
& |\mathbb{E}\Big(\lambda _{i}^{n,jk}\lambda _{i}^{n,lm}\Big[\mathbb{E}%
(\lambda _{i+2k_{n}}^{n,gh}\lambda _{i+2k_{n}}^{n,ab}\Big|\mathcal{F}%
_{i+2k_{n}}^{n})-\frac{2}{k_{n}}%
(C_{i+2k_{n}}^{n,ga}C_{i+2k_{n}}^{n,hb}+C_{i+2k_{n}}^{n,gb}C_{i+2k_{n}}^{n,ha})-%
\frac{2k_{n}\Delta _{n}}{3}\overline{C}_{i+2k_{n}}^{n,gh,ab}\Big]\Bigg|%
\mathcal{F}_{i}^{n}\Big)| \\
& \hspace{2mm}\leq \sqrt{\Delta _{n}}\mathbb{E}(|\lambda
_{i}^{n,jk}||\lambda _{i}^{n,lm}|(\Delta _{n}^{1/8}+\eta
_{i+2k_{n},2k_{n}}^{n})|\Big|\mathcal{F}_{i}^{n})\leq K\sqrt{\Delta _{n}}%
\Delta _{n}^{1/8}\mathbb{E}(|\lambda _{i}^{n,jk}||\lambda _{i}^{n,lm}|\Big|%
\mathcal{F}_{i}^{n}) \\
& \hspace{2mm}+K\sqrt{\Delta _{n}}\mathbb{E}(|\lambda _{i}^{n,jk}||\lambda
_{i}^{n,lm}|\eta _{i+2k_{n},2k_{n}}^{n}|\Big|\mathcal{F}_{i}^{n})\leq
K\Delta _{n}(\Delta _{n}^{1/8}+\eta _{i,4k_{n}}^{n}),
\end{align*}%
where the last inequality follows from Lemma \ref{lemma:lem1}. \newline
Now, using equation (\ref{eqn:SKCC}) successively with $Z=C$ and $Z=%
\overline{C}$ (recall that the latter holds under Assumption \ref{ass:volvol}%
), together with the successive conditioning, we also have
\begin{align*}
& |\mathbb{E}\Big(\lambda _{i}^{n,jk}\lambda _{i}^{n,lm}\Big[\frac{2}{k_{n}}%
(C_{i+2k_{n}}^{n,ga}C_{i+2k_{n}}^{n,hb}+C_{i+2k_{n}}^{n,gb}C_{i+2k_{n}}^{n,ha})+%
\frac{2k_{n}\Delta _{n}}{3}\overline{C}_{i+2k_{n}}^{n,gh,ab}-\frac{2}{k_{n}}%
(C_{i}^{n,ga}C_{i}^{n,hb}+C_{i}^{n,gb}C_{i}^{n,ha}) \\
& \hspace{2mm}-\frac{2k_{n}\Delta _{n}}{3}\overline{C}_{i}^{n,gh,ab}\Big]%
\Big|\mathcal{F}_{i}^{n}\Big)|\leq K\Delta _{n}\Delta _{n}^{1/4}, \\
& |\mathbb{E}\Big(\lambda _{i}^{n,jk}\lambda _{i}^{n,lm}\Big[\frac{2}{k_{n}}%
(C_{i}^{n,ga}C_{i}^{n,hb}+C_{i}^{n,gb}C_{i}^{n,ha}) \\
& \hspace{2mm}+\frac{2k_{n}\Delta _{n}}{3}\overline{C}_{i}^{n,gh,ab}\Big]-%
\Big[\frac{2}{k_{n}}(C_{i}^{n,jl}C_{i}^{n,km}+C_{i}^{n,jm}C_{i}^{n,kl})+%
\frac{2k_{n}\Delta _{n}}{3}\overline{C}_{i}^{n,jk,lm}\Big] \\
& \hspace{2mm}\times \Big[\frac{2}{k_{n}}%
(C_{i}^{n,ga}C_{i}^{n,hb}+C_{i}^{n,gb}C_{i}^{n,ha})+\frac{2k_{n}\Delta _{n}}{%
3}\overline{C}_{i}^{n,gh,ab}\Big]\Big|\mathcal{F}_{i}^{n}\Big)|\leq K\Delta
_{n}(\Delta _{n}^{1/8}+\eta _{i,2k_{n}}^{n}).
\end{align*}%
The result derives from the last inequality.\newline


\subsection{Proof of Lemma \ref{lemma:approx1}}

\subsubsection{Proof of Equation (\ref{eqn:R1}) in Lemma \ref{lemma:approx1}}

We start by obtaining some useful bounds for some important quantities.
First, using the second statement in Lemma \ref{lemma:ito} applied to $%
Z=Y^{\prime }$, we have
\begin{equation}  \label{eqn:alpha1}
|\mathbb{E}(\alpha_i^{n,jk}|\mathcal{F}_{i-1}^n)|\leq K\Delta_n^{3/2}(\sqrt{%
\Delta_n}+\eta^n_{i,1}).
\end{equation}
Second, by repeated application of the Cauchy-Schwartz inequality and making
use of the third and last statements in Lemma \ref{lemma:ito} as well as
equation (\ref{eqn:SKCC}) with $Z=C$, it can be shown that
\begin{align}
\Big|\mathbb{E}(\alpha_i^{n,jk}\alpha_i^{n,lm}&|\mathcal{F}_{i-1}^n)-\Delta_n^2%
\Big(C_i^{n,jl}C_i^{n,km}+C_i^{n,jm}C_i^{n,kl}\Big)\Big|\leq K\Delta_n^{5/2}.
\label{alpha2}
\end{align}
Next, by successive conditioning and using the bound in equation (\ref%
{eqn:SKCC}) for $Z=C$ as well as equations (\ref{eqn:alpha1}) and (\ref%
{alpha2}), we have for $0\leq u \leq k_n-1$,
\begin{align}
\Big|\mathbb{E}(\alpha_{i+u}^{n,jk}\big|\mathcal{F}_{i-1}^n)\Big|\leq
K\Delta_n^{3/2}(\sqrt{\Delta_n}+\eta^n_{i,u}),  \label{eqn:alpha1u}
\end{align}
\begin{align}
\Big|\mathbb{E}(\alpha_{i+u}^{n,jk}\alpha_{i+u}^{n,lm}&|\mathcal{F}%
_{i-1}^n)-\Delta_n^2\Big(C_i^{n,jl}C_i^{n,km}+C_i^{n,jm}C_i^{n,kl}\Big)\Big|\leq
K\Delta_n^{5/2}.  \label{eqn:alpha2u}
\end{align}
\noindent To prove equation (\ref{eqn:R1}), we first observe that $%
\nu_i^{n,jk}\nu_i^{n,lm}\nu_i^{n,gh}$ can be decomposed as
\begin{align*}
&\nu_i^{n,jk}\nu_i^{n,lm}\nu_i^{n,gh}=\frac{1}{k_n^3\Delta_n^3}%
\sum_{u=0}^{k_n-1}\zeta_{i,u}^{n,jk}\zeta_{i,u}^{n,lm}\zeta_{i,u}^{n,gh}+%
\frac{1}{k_n^3\Delta_n^3}\sum_{u=0}^{k_n-2}\sum_{v=u+1}^{k_n-1}\Big[%
\zeta_{i,u}^{n,jk}\zeta_{i,v}^{n,lm}\zeta_{i,v}^{n,gh}+
\zeta_{i,u}^{n,gh}\zeta_{i,v}^{n,jk}\zeta_{i,v}^{n,lm} \\
& \hspace{2mm} +\zeta_{i,u}^{n,lm}\zeta_{i,v}^{n,gh}\zeta_{i,v}^{n,jk}\Big]+%
\frac{1}{k_n^3\Delta_n^3}\sum_{u=0}^{k_n-2}\sum_{v=u+1}^{k_n-1}[%
\zeta_{i,u}^{n,jk}\zeta_{i,u}^{n,lm}\zeta_{i,v}^{n,gh}+
\zeta_{i,u}^{n,gh}\zeta_{i,u}^{n,jk}\zeta_{i,v}^{n,lm}+\zeta_{i,u}^{n,lm}%
\zeta_{i,u}^{n,gh}\zeta_{i,v}^{n,jk}\Big] \\
&\hspace{2mm} +\frac{1}{k_n^3\Delta_n^3}\sum_{u=0}^{k_n-3}%
\sum_{v=u+1}^{k_n-2}\sum_{w=v+1}^{k_n-1}\Big[\zeta_{i,u}^{n,jk}%
\zeta_{i,v}^{n,lm}\zeta_{i,w}^{n,gh}+\zeta_{i,u}^{n,jk}\zeta_{i,v}^{n,gh}%
\zeta_{i,w}^{n,lm}+\zeta_{i,u}^{n,lm}\zeta_{i,v}^{n,jk}\zeta_{i,w}^{n,gh}+%
\zeta_{i,u}^{n,lm}\zeta_{i,v}^{n,gh}\zeta_{i,w}^{n,jk} \\
&\hspace{2mm} +\zeta_{i,u}^{n,gh}\zeta_{i,v}^{n,lm}\zeta_{i,w}^{n,jk}+%
\zeta_{i,u}^{n,gh}\zeta_{i,v}^{n,jk}\zeta_{i,w}^{n,lm}\Big],
\end{align*}
with $\zeta_{i,u}^{n}=\alpha_{i+u}^{n}+(C_{i+u}^{n}-C_i^{n})\Delta_n$, which
satisfies $\mathbb{E}(\|\zeta_{i,u}^n\|^q|\mathcal{F}_{i-1}^n)\leq K\Delta_n^q$
for $q\geq 2$.\newline
Set
\begin{align*}
&\xi_i^n(1)=\frac{1}{k_n^3\Delta_n^3}\sum_{u=0}^{k_n-1}\zeta_{i,u}^{n,jk}%
\zeta_{i,u}^{n,lm}\zeta_{i,u}^{n,gh},\hspace{3mm}\xi_i^n(2)=\frac{1}{%
k_n^3\Delta_n^3}\sum_{u=0}^{k_n-2}\sum_{v=u+1}^{k_n-1}\zeta_{i,u}^{n,jk}%
\zeta_{i,v}^{n,lm}\zeta_{i,v}^{n,gh} \\
&\xi_i^n(3)=\frac{1}{k_n^3\Delta_n^3}\sum_{u=0}^{k_n-2}\sum_{v=u+1}^{k_n-1}%
\zeta_{i,u}^{n,jk}\zeta_{i,u}^{n,lm}\zeta_{i,v}^{n,gh}~~\text{and}%
~~\xi_i^n(4)=\frac{1}{k_n^3\Delta_n^3}\sum_{u=0}^{k_n-3}\sum_{v=u+1}^{k_n-2}%
\sum_{w=v+1}^{k_n-1}\zeta_{i,u}^{n,jk}\zeta_{i,v}^{n,lm}\zeta_{i,w}^{n,gh}.
\end{align*}
The following bounds complete the proof of equation (\ref{eqn:R1}),
\begin{align}
& |\mathbb{E}(\xi_i^n(1)|\mathcal{F}_{i-1}^n) |\leq K\Delta_n  \label{B1} \\
& |\mathbb{E}(\xi_i^n(2)|\mathcal{F}_{i-1}^n) |\leq K\Delta_n  \label{B2} \\
&|\mathbb{E}(\xi_i^n(3)|\mathcal{F}_{i-1}^n) |\leq K\Delta_n  \label{B3} \\
& |\mathbb{E}(\xi_i^n(4)|\mathcal{F}_{i-1}^n) |\leq
K\Delta_n^{3/4}(\Delta_n^{1/4}+\eta_{i,k_n}) .  \label{B4}
\end{align}
These bounds are proved below.

\bigskip
\noindent \large{\textbf{Proof of Equation (\ref{B1})}}
\medskip

The result readily follows from an application of the Cauchy Schwartz
inequality coupled with the bound $\mathbb{E}(\|\zeta^n_{i+u}\|^q|\mathcal{F}%
_{i-1}^n)\leq K_q\Delta_n^q$ for $q\geq 2$.

\bigskip
\noindent \large{\textbf{Proof of Equation (\protect\ref{B2})}}
\medskip

Using the law of iterated expectation, we have, for $u<v$,
\begin{equation}
\mathbb{E}(\zeta_{i+u}^{n,jk}\zeta_{i+v}^{n,lm}\zeta_{i+v}^{n,gh}|\mathcal{F}%
_{i-1}^n)=\mathbb{E}(\zeta_{i+u}^{n,jk}\mathbb{E}(\zeta_{i+v}^{n,lm}%
\zeta_{i+v}^{n,gh}|\mathcal{F}_{i+u+1}^n)\big|\mathcal{F}_{i-1}^n).
\label{eqn:scond2}
\end{equation}
By successive conditioning, equation (\ref{alpha2}), and the Cauchy-Schwartz
inequality, we also have
\begin{equation*}
|\mathbb{E}(\zeta_{i,v}^{n,lm}\zeta_{i,v}^{n,gh}|\mathcal{F}%
_{i+u}^n)-%
\Delta_n^2(C_{i+u+1}^{n,lg}C_{i+u+1}^{n,mh}+C_{i+u+1}^{n,lh}C_{i+u+1}^{n,mg})
\\
-\Delta_n^2(C_{i+u+1}^{n,gh}-C_{i}^{n,gh})(C_{i+u+1}^{n,lm}-C_{i}^{n,lm})|
\leq K\Delta_n^{5/2}.
\end{equation*}
Given that $\mathbb{E}(|\zeta_{i+u}^{n,jk}|^q\big|\mathcal{F}_{i-1}^n)\leq
\Delta^q_n$, the approximation error involved in replacing $\mathbb{E}%
(\zeta_{i+v}^{n,lm}\zeta_{i+v}^{n,gh}|\mathcal{F}_{i+u+1}^n)$ by\newline
$%
\Delta_n^2(C_{i+u+1}^{n,lg}C_{i+u+1}^{n,mh}+C_{i+u+1}^{n,lh}C_{i+u+1}^{n,mg})+\Delta_n^2(C_{i+u+1}^{n,gh}-C_{i}^{n,gh})(C_{i+u+1}^{n,lm}-C_{i}^{n,lm})
$ in equation (\ref{eqn:scond2}) is smaller than $\Delta_n^{7/2}$. \newline
We can also easily show that
\begin{equation}
|\mathbb{E}(\alpha_{i+u}^{n,jk}(C_{i+u+1}^{n,lm}-C_{i+u}^{n,lm})|\mathcal{F}%
_{i-1}^n)|\leq K\Delta_n^{3/2}(\sqrt{\Delta_n}+\eta_{i,k_n}^n).
\label{eqn:alphanew}
\end{equation}
Since $(C_{i+u}^n-C_i^n)$ is $\mathcal{F}_{i+u}^n$-measurable, we use the
successive conditioning, the Cauchy-Schwartz inequality, equation (\ref%
{eqn:alpha1}), equation (\ref{alpha2}), and the fifth statement in Lemma \ref%
{lemma:ito} applied to $Z=c$ to obtain
\begin{eqnarray}
|\mathbb{E}%
(\alpha_{i+u}^{n,gh}(C_{i+u}^{n,lm}-C_i^{n,lm})(C_{i+u}^{n,jk}-C_i^{n,jk})|%
\mathcal{F}_{i-1}^n)| & \leq & K\Delta_n^{5/2}  \notag \\
|\mathbb{E}(\alpha_{i+u}^{n,jk}%
\alpha_{i+u}^{n,lm}(C_{i+u}^{n,gh}-C_i^{n,gh})|\mathcal{F}_{i-1}^n)| & \leq &
K\Delta_n^{5/2}  \label{eqn:eps1} \\
|\mathbb{E}\big(%
(C_{i+u}^{n,lm}-C_i^{n,lm})(C_{i+u}^{n,jk}-C_i^{n,jk})(C_{i+u}^{n,gh}-C_i^{n,gh})%
\big)|\mathcal{F}_{i-1}^n)| & \leq & K\Delta_n.  \notag
\end{eqnarray}
The following inequalities can be established 
using equation (\ref{eqn:alpha1}), the successive conditioning together with
equation (\ref{eqn:SKCC}) for $Z=C$,
\begin{eqnarray*}
\Big|\mathbb{E}%
(\alpha_{i+u}^{n,jk}(C_{i+u+1}^{n,lg}C_{i+u+1}^{n,mh}+C_{i+u+1}^{n,lh}C_{i+u+1}^{n,mg})|%
\mathcal{F}_{i-1}^n)\Big| &\leq & K\Delta_n^{3/2} \\
\Big|\mathbb{E}\Big((C_{i+u}^{n,jk}-C_{i}^{n,jk})\big(%
C_{i+u+1}^{n,lg}C_{i+u+1}^{n,mh}+C_{i+u+1}^{n,lh}C_{i+u+1}^{n,mg}\big)|%
\mathcal{F}_{i-1}^n\Big)\Big| & \leq & K\Delta_n^{1/2} \\
\Big|\mathbb{E}%
(\alpha_{i+u}^{n,jk}(C_{i+u+1}^{n,gh}-C_{i}^{n,gh})(C_{i+u+1}^{n,lm}-C_{i}^{n,lm})|%
\mathcal{F}_{i-1}^n)\Big| & \leq & K\Delta_n^{3/2}(\sqrt{\Delta_n}%
+\eta_{i,k_n}^n).
\end{eqnarray*}
The last three inequalities together yield $|\mathbb{E}(\xi_i^n(2)|\mathcal{F%
}_{i-1}^n) |\leq K\Delta_n$.

\bigskip
\noindent \large{\textbf{Proof of Equation (\ref{B3})}}
\medskip

First, note that, for $u<v$, we have
\begin{equation}
\mathbb{E}(\zeta _{i+u}^{n,jk}\zeta _{i+u}^{n,lm}\zeta _{i+v}^{n,gh}|%
\mathcal{F}_{i-1}^{n})=\mathbb{E}(\zeta _{i+u}^{n,jk}\zeta _{i+u}^{n,lm}%
\mathbb{E}(\zeta _{i+v}^{n,gh}|\mathcal{F}_{i+u}^{n})\big|\mathcal{F}%
_{i-1}^{n}).  \label{eqn:scond3}
\end{equation}%
By successive conditioning and equation (\ref{eqn:alpha1}), we have
\begin{equation}
|\mathbb{E}(\alpha _{i+w}^{n,gh}|\mathcal{F}_{i+v}^{n})|\leq K\Delta
_{n}^{3/2}(\sqrt{\Delta _{n}}+\eta _{i+v+1,w-v}).  \label{eqn:newapprox}
\end{equation}%
Using the first statement of Lemma applied to $Z=C$, it can be shown that
\begin{eqnarray*}
&&|\mathbb{E}\big((C_{i+w}^{n,gh}-C_{i+v+1}^{n,gh}))|\mathcal{F}_{i-1}^{n}\big)%
-\Delta _{n}(w-v-1)\widetilde{b}_{i+v+1}^{n,gh}| \\
&\leq &K(w-v-1)\Delta _{n}\eta _{i+v+1,w-v}\leq K\Delta _{n}^{1/2}\eta
_{i+v+1,w-v}.
\end{eqnarray*}%
The last two inequalities together imply
\begin{equation}
\Big|\mathbb{E}\Big(\zeta _{i+w}^{n,gh}|\mathcal{F}_{i+v}^{n}\Big)%
-(C_{i+v+1}^{n,gh}-C_{i}^{n,gh})\Delta _{n}-\Delta _{n}^{2}(w-v-1)\widetilde{%
b}_{i+v+1}^{n,gh}\Big|\leq K\Delta _{n}^{3/2}(\sqrt{\Delta _{n}}+\eta
_{i+v+1,w-v}).  \label{eqn:scond}
\end{equation}%
Since $\mathbb{E}(|\zeta _{i,u}^{n,jk}|^{q}|\mathcal{F}_{i-1}^{n})\leq \Delta
_{n}^{q}$, the error induced by replacing $\mathbb{E}(\zeta _{i+v}^{n,gh}|%
\mathcal{F}_{i+u}^{n})$ by $(C_{i+v+1}^{n,gh}-C_{i}^{n,gh})\Delta
_{n}+\Delta _{n}^{2}(w-v-1)\widetilde{b}_{i+v+1}^{n,gh}$ in equation (\ref%
{eqn:scond3}) is smaller that $\Delta _{n}^{7/2}$.\newline
Using Cauchy Schwartz inequality, successive conditioning, equation (\ref%
{eqn:eps1}), equation (\ref{eqn:SKCC}) for $Z=C$ and the boundedness of $%
\widetilde{b}_{t}$ and $C_{t}$ we obtain
\begin{eqnarray*}
\Big|\mathbb{E}\Big(\alpha _{i+u}^{n,jk}\alpha
_{i+u}^{n,lm}(C_{i+u+1}^{n,jk}-C_{i}^{n,gh})|\mathcal{F}_{i+u-1}^{n}\Big)\Big| %
&\leq &K\Delta _{n}^{5/2} \\
\Big|\mathbb{E}\Big(\alpha _{i+u}^{n,jk}\alpha _{i+u}^{n,lm}\widetilde{b}%
_{i+u+1}^{n,gh}|\mathcal{F}_{i+u-1}^{n}\Big)\Big| &\leq &K\Delta _{n}^{2} \\
\Big|\mathbb{E}\Big(\alpha
_{i+u}^{n,jk}(C_{i+u}^{n,lm}-C_{i}^{n,lm})(C_{i+u+1}^{n,gh}-C_{i}^{n,gh})|%
\mathcal{F}_{i-1}^{n}\Big)\Big| &\leq &K\Delta _{n}^{1/4}\Delta _{n}^{3/2}(%
\sqrt{\Delta _{n}}+\eta _{i,k_{n}}^{n}) \\
\Big|\mathbb{E}\Big(\alpha _{i+u}^{n,jk}(C_{i+u}^{n,lm}-C_{i}^{n,lm})%
\widetilde{b}_{i+u+1}^{n,gh}|\mathcal{F}_{i-1}^{n}\Big)\Big| &\leq &\Delta
_{n}^{5/4} \\
\Big|\mathbb{E}\Big(%
(C_{i+u}^{n,jk}-C_{i}^{n,gh})(C_{i+u}^{n,lm}-C_{i}^{n,lm})\widetilde{b}%
_{i+u+1}^{n,gh}|\mathcal{F}_{i-1}^{n}\Big)\Big| &\leq &K\Delta _{n}^{1/2} \\
\Big|\mathbb{E}\Big(%
(C_{i+u}^{n,jk}-C_{i}^{n,jk})(C_{i+u}^{n,lm}-C_{i}^{n,lm})(C_{i+u+1}^{n,gh}-C_{i}^{n,gh})|%
\mathcal{F}_{i-1}^{n}\Big)\Big| &\leq &K\Delta _{n}.
\end{eqnarray*}%
The above inequalities together yield $|\mathbb{E}(\xi _{i}^{n}(3)|\mathcal{F%
}_{i-1}^{n})|\leq K\Delta _{n}$.

\bigskip
\noindent \large{\textbf{Proof of Equation (\ref{B4})}}
\medskip

We first observe that $\xi_i^n(4)$ can be rewritten as
\begin{align*}
\xi_i^n(4)=\frac{1}{(k_n\Delta_n)^3}\sum_{w=2}^{k_n-1}\sum_{v=0}^{w-1}%
\sum_{u=0}^{v-1}\zeta_{i+u}^{n,jk}\zeta_{i+v}^{n,lm}\zeta_{i+w}^{n,gh},
\end{align*}
where
\begin{align*}
&\zeta_{i+u}^{n,jk}\zeta_{i+v}^{n,lm}\zeta_{i+w}^{n,gh}=\Bigg[%
\alpha_{i+u}^{n,jk}\alpha_{i+v}^{n,lm}\alpha_{i+w}^{n,gh}+%
\alpha_{i+u}^{n,jk}\Delta_n\alpha_{i+v}^{n,lm}(C_{i+w}^{n,gh}-C_i^{n,gh})+%
\alpha_{i+u}^{n,jk}\Delta_n(C_{i+v}^{n,lm}-C_i^{n,lm})\alpha_{i+w}^{n,gh} \\
&\hspace{5mm}+\Delta_n^2%
\alpha_{i+u}^{n,jk}(C_{i+v}^{n,lm}-C_i^{n,lm})(C_{i+w}^{n,gh}-C_i^{n,gh})+%
\Delta_n(C_{i+u}^{n,jk}-C_i^{n,jk})\alpha_{i+v}^{n,lm}\alpha_{i+w}^{n,gh} \\
&\hspace{5mm}+\Delta_n^2(C_{i+u}^{n,jk}-C_i^{n,jk})%
\alpha_{i+v}^{n,lm}(C_{i+w}^{n,gh}-C_i^{n,gh})
+\Delta_n^2(C_{i+u}^{n,jk}-C_i^{n,jk})(C_{i+v}^{n,lm}-C_i^{n,lm})%
\alpha_{i+w}^{n,gh} \\
&\hspace{5mm}%
+\Delta_n^3(C_{i+u}^{n,jk}-C_i^{n,jk})(C_{i+v}^{n,lm}-C_i^{n,lm})(C_{i+w}^{n,gh}-C_i^{n,gh})%
\Bigg].
\end{align*}
Based on the above decomposition, we set
\begin{align*}
\xi_i^n(4)=\sum_{j=1}^8 \chi(j),
\end{align*}
with $\chi(j)$ defined below. We aim to show that $|\mathbb{E}(\chi(j)\big|%
\mathcal{F}_{i-1}^n)|\leq K\Delta_n^{3/4}(\Delta_n^{1/4}+\eta_{i,k_n}^n)$, $%
j=1,\ldots,8$.\newline
First, set
\begin{align*}
\chi(1)=\frac{1}{(k_n\Delta_n)^3}\sum_{w=2}^{k_n-1}\sum_{v=0}^{w-1}%
\sum_{u=0}^{v-1} \alpha_{i+u}^{n,jk}\alpha_{i+v}^{n,lm}\alpha_{i+w}^{n,gh}.
\end{align*}
Upon changing the order of the summation, we have
\begin{align*}
&\chi(1)=\frac{1}{(k_n\Delta_n)^3}\sum_{w=2}^{k_n-1}\sum_{v=0}^{w-1}\Big(%
\sum_{u=0}^{v-1} \alpha_{i+u}^{n,jk}\Big)\alpha_{i+v}^{n,lm}%
\alpha_{i+w}^{n,gh}.
\end{align*}
Define also
\begin{align*}
\chi^{\prime}(1)=\frac{1}{(k_n\Delta_n)^3}\sum_{w=2}^{k_n-1}\sum_{v=0}^{w-1}%
\Big(\sum_{u=0}^{v-1} \alpha_{i+u}^{n,jk}\Big)\alpha_{i+v}^{n,lm}\mathbb{E}%
(\alpha_{i+w}^{n,gh}|\mathcal{F}_{i+v}^n).
\end{align*}
Note that $\mathbb{E}(\chi(1)|\mathcal{F}_{i-1}^n)=\mathbb{E}(\chi^{\prime}(1)|%
\mathcal{F}_{i-1}^n)$.\newline
By Lemma \ref{lemma:key}, we have for $q\geq 2$,
\begin{align*}
&\mathbb{E}\Big(\Big\|\sum_{u=0}^{v-1}\alpha_{i+u}^{n,jk}\Big\|^q \Big|%
\mathcal{F}^n_{i-1}\Big)\leq K_q\Delta_n^{3q/4}.
\end{align*}
The Cauchy-Schwartz inequality yields
\begin{align*}
&\mathbb{E}\Bigg(\Big|\sum_{w=2}^{k_n-1}\sum_{v=0}^{w-1}\Big(%
\sum_{u=0}^{v-1} \alpha_{i+u}^{n,jk}\Big)\alpha_{i+v}^{n,lm}\mathbb{E}%
(\alpha_{i+w}^{n,gh}|\mathcal{F}_{i+v}^n)\Big|\Bigg|\mathcal{F}_{i-1}^n\Bigg)%
\leq Kk_n^2 \Big[\mathbb{E}\Big(\Big|\sum_{u=0}^{v-1} \alpha_{i+u}^{n,jk}%
\Big|^{4}\Big|\mathcal{F}_{i-1}^n\Big)\Big]^{1/4} \\
&\times\Big[\mathbb{E}\Big(\Big|\alpha_{i+v}^{n,lm}\Big|^{4}\Big|\mathcal{F}%
_{i-1}^n\Big)\Big]^{1/4}\times\Big[\mathbb{E}\Big(\Big|\mathbb{E}%
(\alpha_{i+w}^{n,gh}|\mathcal{F}_{i+v}^n)\Big|^{2}\Big|\mathcal{F}_{i-1}^n\Big)%
\Big]^{1/2}\leq K\Delta_n k_n^2 \Delta_n^{3/4} \Delta_n^{3/2}(\sqrt{\Delta_n}%
+\eta_{i,k_n}^{n}),
\end{align*}
where the last iteration is obtained using equation (\ref{eqn:newapprox}) as
well as the inequality $(a+b)^{1/2}\leq a^{1/2}+b^{1/2}$, which holds for
positive real numbers $a$ and $b$, and the third statement in Lemma \ref%
{lemma:lem1}. It follows that
\begin{align*}
|\mathbb{E}\Big(\chi(1)\big|\mathcal{F}_{i-1}^n\Big)|\leq K\Delta_n^{3/4} (\sqrt{%
\Delta_n}+\eta_{i,k_n}^n).
\end{align*}
Next, we introduce
\begin{align*}
&\chi(2)=\frac{1}{(k_n\Delta_n)^3}\sum_{w=2}^{k_n-1}\sum_{v=0}^{w-1}\Big(%
\sum_{u=0}^{v-1}\Delta_n(C_{i+u}^{n,jk}-C_{i}^{n,jk})\Big)%
\alpha_{i+v}^{n,lm}\alpha_{i+w}^{n,gh}, \\
&\chi(3)=\frac{1}{(k_n\Delta_n)^3}\sum_{w=2}^{k_n-1}\sum_{v=0}^{w-1}\Big(%
\sum_{u=0}^{v-1} \alpha_{i+v}^{n,jk}\Big)%
\Delta_n(C_{i+u}^{n,lm}-C_{i}^{n,lm})\alpha_{i+w}^{n,gh}, \\
&\chi(4)=\frac{1}{(k_n\Delta_n)^3}\sum_{w=2}^{k_n-1}\sum_{v=0}^{w-1}\Big(%
\sum_{u=0}^{v-1} \Delta_n(C_{i+u}^{n,jk}-C_{i}^{n,jk})\Big)%
\Delta_n(C_{i+u}^{n,lm}-C_{i}^{n,lm})\alpha_{i+w}^{n,gh}.
\end{align*}
Given that for $q \geq 2$, we have
\begin{align*}
&\mathbb{E}\Big(\Big\|\sum_{u=0}^{v-1}\Delta_n(C_{i+u}^{n,jk}-C_{i}^{n,jk})%
\Big\|^q \Big|\mathcal{F}^n_{i-1}\Big)\leq K_q\Delta_n^{3q/4}~~\text{and}~~
\mathbb{E}(\|C_{i+u}^{n,jk}-C_{i}^{n,jk}\|^q\big|\mathcal{F}_{i-1}^n)\leq K_q
\Delta_n^{q/4}.
\end{align*}
Similar steps to $\chi(1)$ lead to
\begin{align*}
|\mathbb{E}(\chi(2)\big|\mathcal{F}_{i-1}^n)|\leq K\Delta_n^{3/4} (\sqrt{\Delta_n%
}+\eta_{i,k_n}^n)~~\text{and}~~|\mathbb{E}(\chi(j)\big|\mathcal{F}_{i-1}^n)|\leq
K\Delta_n(\sqrt{\Delta_n}+\eta_{i,k_n}^n)~~\text{for}~~j=3,4.
\end{align*}
Define
\begin{align*}
&\chi(5)=\frac{1}{(k_n\Delta_n)^3}\sum_{w=2}^{k_n-1}\sum_{v=0}^{w-1}\Big(%
\sum_{u=0}^{v-1} \alpha_{i+u}^{n,jk}\Big)\alpha_{i+v}^{n,lm}%
\Delta_n(C_{i+w}^{n,gh}-C_i^{n,gh}) \\
&\chi^{\prime}(5)=\frac{1}{(k_n\Delta_n)^3}\sum_{w=2}^{k_n-1}\sum_{v=0}^{w-1}%
\Big(\sum_{u=0}^{v-1} \alpha_{i+u}^{n,jk}\Big)\alpha_{i+v}^{n,lm}\Delta_n%
\mathbb{E}\big((C_{i+w}^{n,gh}-C_i^{n,gh})\big|\mathcal{F}_{i+v}^n) \\
&\chi(6)=\frac{1}{(k_n\Delta_n)^3}\sum_{w=2}^{k_n-1}\sum_{v=0}^{w-1}\Big(%
\sum_{u=0}^{v-1} \Delta_n(C_{i+u}^{n,jk}-C_i^{n,jk})\Big)\alpha_{i+v}^{n,lm}%
\Delta_n(C_{i+w}^{n,gh}-C_i^{n,gh}) \\
&\chi(7)=\frac{1}{(k_n\Delta_n)^3}\sum_{w=2}^{k_n-1}\sum_{v=0}^{w-1}\Big(%
\sum_{u=0}^{v-1} \alpha_{i+u}^{n,jk}\Big)\Delta_n(C_{i+v}^{n,lm}-C_i^{n,lm})%
\Delta_n(C_{i+w}^{n,gh}-C_{i}^{n,gh}),
\end{align*}
where we have $\mathbb{E}(\chi(5)|\mathcal{F}_{i-1}^n)=\mathbb{E}%
(\chi^{\prime}(5)|\mathcal{F}_{i-1}^n)$. Recalling equation (\ref{eqn:scond}),
we further decompose $\chi^{\prime}(5)$ as,
\begin{align*}
\chi^{\prime}(5)=\sum_{j=1}^{5}\chi(5)[j],
\end{align*}
with 
\begin{align*}
\chi{\prime}(5)[1]=&\frac{1}{(k_n\Delta_n)^3}\sum_{w=2}^{k_n-1}%
\sum_{v=0}^{w-1}\Big(\sum_{u=0}^{v-1}\alpha_{i+u}^{n,jk}\Big)%
\alpha_{i+v}^{n,lm}\Big(\mathbb{E}\Big(C_{i+w}^{n,gh}-C_{i}^{n,gh}|\mathcal{F%
}_{i+v}^n\Big) \\
&-(C_{i+v+1}^{n,gh}-C_{i}^{n,gh})\Delta_n -\widetilde{b}_{i+v+1}^{n,gh}
\Delta_n^2(w-v-1)\Big) \\
\chi{\prime}(5)[2]=&\frac{1}{(k_n\Delta_n)^3}\sum_{w=2}^{k_n-1}%
\sum_{v=0}^{w-1}\Delta_n(C_{i+v}^{n,gh}-C_{i}^{n,gh})\Big(%
\sum_{u=0}^{v-1}\alpha_{i+u}^{n,jk}\Big)\alpha_{i+v}^{n,lm} \\
\chi{\prime}(5)[3]=&\frac{1}{(k_n\Delta_n)^3}\sum_{w=2}^{k_n-1}%
\sum_{v=0}^{w-1}\Big(\sum_{u=0}^{v-1} \alpha_{i+u}^{n,jk}\Big)%
\Delta_n(C_{i+v+1}^{n,gh}-C_{i+v}^{n,gh})\alpha_{i+v}^{n,lm} \\
\chi{\prime}(5)[4]=&\frac{1}{(k_n\Delta_n)^3}\sum_{w=2}^{k_n-1}%
\sum_{v=0}^{w-1}\Big(\sum_{u=0}^{v-1} \alpha_{i+u}^{n,jk}\Big)%
\Delta_n^2(w-v-1)(\widetilde{b}_{i+v+1}^{n,gh}-\widetilde{b}%
_{i+v}^{n,gh})\alpha_{i+v}^{n,lm} \\
\chi{\prime}(5)[5]=&\frac{1}{(k_n\Delta_n)^3}\sum_{w=2}^{k_n-1}%
\sum_{v=0}^{w-1}\Delta_n^2(w-v-1)\widetilde{b}_{i+v}^{n,gh}\Big(%
\sum_{u=0}^{v-1} \alpha_{i+u}^{n,jk}\Big)\alpha_{i+v}^{n,lm}.
\end{align*}
Using equations (\ref{eqn:scond}), (\ref{eqn:newapprox}), and (\ref%
{eqn:alphanew}) and following the same strategy proof as for $\chi(1)$, it
can be shown that
\begin{align*}
|\mathbb{E}\Big(\chi{\prime}(5)[j]\big|\mathcal{F}_{i-1}^n\Big)|\leq
K\Delta_n^{3/4} (\sqrt{\Delta_n}+\eta_{i,k_n}^n),~~\text{for}~~j=1,\ldots,5,
\end{align*}
which in turn implies
\begin{align*}
|\mathbb{E}\Big(\chi(5)\big|\mathcal{F}_{i-1}^n\Big)|\leq K\Delta_n^{3/4} (\sqrt{%
\Delta_n}+\eta_{i,k_n}^n),~~\text{for}~~j=1,\ldots,5.
\end{align*}
The term $\chi(6)$ can be handled similarly to $\chi(5)$, hence we conclude
that
\begin{align*}
|\mathbb{E}\Big(\chi(6)\big|\mathcal{F}_{i-1}^n\Big)|\leq K\Delta_n^{3/4} (\sqrt{%
\Delta_n}+\eta_{i,k_n}^n).
\end{align*}

Next, we set
\begin{align*}
\chi(7)=\frac{1}{(k_n\Delta_n)^3}\sum_{w=2}^{k_n-1}\Bigg(\sum_{v=0}^{w-1}%
\Big(\sum_{u=0}^{v-1} \alpha_{i+u}^{n,jk}\Big)%
\Delta_n(C_{i+v}^{n,lm}-C_i^{n,lm})\Delta_n(C_{i+w}^{n,gh}-C_{i}^{n,gh})%
\Bigg).
\end{align*}
Define
\begin{align*}
&\chi(7)[1]=\frac{1}{(k_n\Delta_n)^3}\sum_{w=2}^{k_n-1}\Bigg(\sum_{v=0}^{w-1}%
\Big(\sum_{u=0}^{v-1} \alpha_{i+u}^{n,jk}\Big)%
\Delta_n(C_{i+v}^{n,lm}-C_i^{n,lm})\Delta_n(C_{i+v+1}^{n,gh}-C_{i+v}^{n,gh})%
\Bigg) \\
&\chi(7)[2]=\frac{1}{(k_n\Delta_n)^3}\sum_{w=2}^{k_n-1}\Bigg(\sum_{v=0}^{w-1}%
\Big(\sum_{u=0}^{v-1} \alpha_{i+u}^{n,jk}\Big)%
\Delta_n(C_{i+v}^{n,lm}-C_i^{n,lm})\Delta_n(C_{i+v}^{n,gh}-C_i^{n,gh})\Bigg)
\\
&\chi(7)[3]=\frac{1}{(k_n\Delta_n)^3}\sum_{w=2}^{k_n-1}\Bigg(\sum_{v=0}^{w-1}%
\Big(\sum_{u=0}^{v-1} \alpha_{i+u}^{n,jk}\Big)%
\Delta_n(C_{i+v}^{n,lm}-C_i^{n,lm})\Delta_n^2(w-v-1)(\widetilde{b}%
_{i+v+1}^{n,gh}-\widetilde{b}_{i+v}^{n,gh})\Bigg) \\
&\chi(7)[4]=\frac{1}{(k_n\Delta_n)^3}\sum_{w=2}^{k_n-1}\Bigg(%
\sum_{v=0}^{w-1}\Delta_n^2(w-v-1)\widetilde{b}_{i+v}^{n,gh}\Big(%
\sum_{u=0}^{v-1} \alpha_{i+u}^{n,jk}\Big)\Delta_n(C_{i+v}^{n,lm}-C_i^{n,lm})%
\Bigg).
\end{align*}
It is easy to see that
\begin{align*}
\chi(7)=\sum_{j=1}^4 \chi(7)[j].
\end{align*}
Similarly to calculations used for $\chi(1)$, it can be shown that
\begin{align*}
|\mathbb{E}(\chi(7)[j]\big|\mathcal{F}_{i-1}^n)|\leq
K\Delta_n^{1/4}(\Delta_n^{1/4}+\eta_{i,k_n}),~~\text{for}~~ j=1,\ldots,3.
\end{align*}
To handle the remaining term $\chi(7)[4]$, we decompose it $%
\chi(7)[4]=\sum_{j=1}^9 \chi(7)[4][j]$, where
\begin{align*}
&\chi(7)[4][1]=\frac{\Delta_n^2}{(k_n\Delta_n)^3}\sum_{w=2}^{k_n-1}%
\sum_{v=0}^{w-1}\sum_{u=0}^{v-1}
\alpha_{i+u}^{n,jk}(C_{i+u+1}^{n,lm}-C_{i+u}^{n,lm})(C_{i+u+1}^{n,gh}-C_{i+u}^{n,gh})
\\
&\chi(7)[4][2]=\frac{\Delta_n^2}{(k_n\Delta_n)^3}\sum_{w=2}^{k_n-1}%
\sum_{v=0}^{w-1}\sum_{u=0}^{v-1}
(C_{i+u}^{n,gh}-C_{i}^{n,gh})%
\alpha_{i+u}^{n,jk}(C_{i+u+1}^{n,lm}-C_{i+u}^{n,lm}) \\
&\chi^{\prime}(7)[4][2]=\frac{\Delta_n^2}{(k_n\Delta_n)^3}%
\sum_{w=2}^{k_n-1}\sum_{v=0}^{w-1}\sum_{u=0}^{v-1}
(C_{i+u}^{n,gh}-C_{i}^{n,gh})\mathbb{E}%
(\alpha_{i+u}^{n,jk}(C_{i+u+1}^{n,lm}-C_{i+u}^{n,lm})|\mathcal{F}_{i+u-1}^n) \\
&\chi(7)[4][3]=\frac{\Delta_n^2}{(k_n\Delta_n)^3}\sum_{w=2}^{k_n-1}%
\sum_{v=0}^{w-1}\sum_{u=0}^{v-1}
(C_{i+u}^{n,lm}-C_{i}^{n,lm})%
\alpha_{i+u}^{n,jk}(C_{i+u+1}^{n,gh}-C_{i+u}^{n,gh}) \\
&\chi(7)[4][4]=\frac{\Delta_n^2}{(k_n\Delta_n)^3}\sum_{w=2}^{k_n-1}%
\sum_{v=0}^{w-1}\sum_{u=0}^{v-1}
(C_{i+u}^{n,lm}-C_{i}^{n,lm})(C_{i+u}^{n,gh}-C_{i}^{n,gh})\alpha_{i+u}^{n,jk}
\\
&\chi(7)[4][5]=\frac{\Delta_n^2}{(k_n\Delta_n)^3}\sum_{w=2}^{k_n-1}%
\sum_{v=0}^{w-1}\sum_{u=0}^{v-1}
(C_{i+u}^{n,lm}-C_{i}^{n,lm})%
\alpha_{i+u}^{n,jk}(C_{i+v}^{n,gh}-C_{i+u+1}^{n,gh}) \\
&\chi^{\prime}(7)[2][5]=\frac{\Delta_n^2}{(k_n\Delta_n)^3}%
\sum_{w=2}^{k_n-1}\sum_{v=0}^{w-1}\sum_{u=0}^{v-1}
(C_{i+u}^{n,lm}-C_{i}^{n,lm})\alpha_{i+u}^{n,jk}\mathbb{E}%
((C_{i+v}^{n,gh}-C_{i+u+1}^{n,gh}|\mathcal{F}_{i+u-1}^n) \\
&\chi(7)[4][6]=\frac{\Delta_n^2}{(k_n\Delta_n)^3}\sum_{w=2}^{k_n-1}%
\sum_{v=0}^{w-1}\sum_{u=0}^{v-1}
\alpha_{i+u}^{n,jk}(C_{i+u+1}^{n,lm}-C_{i+u}^{n,lm})(C_{i+v}^{n,gh}-C_{i+u+1}^{n,gh})
\\
&\chi(7)[4][7]=\frac{\Delta_n^2}{(k_n\Delta_n)^3}\sum_{w=2}^{k_n-1}%
\sum_{v=0}^{w-1}\sum_{u=0}^{v-1}
(C_{i+u}^{n,gh}-C_{i}^{n,gh})%
\alpha_{i+u}^{n,jk}(C_{i+v}^{n,lm}-C_{i+u+1}^{n,lm}) \\
&\chi(7)[4][8]=\frac{\Delta_n^2}{(k_n\Delta_n)^3}\sum_{w=2}^{k_n-1}%
\sum_{v=0}^{w-1}\sum_{u=0}^{v-1}
\alpha_{i+u}^{n,jk}(C_{i+u+1}^{n,gh}-C_{i+u}^{n,gh})(C_{i+v}^{n,lm}-C_{i+u+1}^{n,lm})
\\
&\chi(7)[4][9]=\frac{\Delta_n^2}{(k_n\Delta_n)^3}\sum_{w=2}^{k_n-1}%
\sum_{v=0}^{w-1}\sum_{u=0}^{v-1}
\alpha_{i+u}^{n,jk}(C_{i+v}^{n,lm}-C_{i+u+1}^{n,lm})(C_{i+v}^{n,gh}-C_{i+u+1}^{n,gh}).
\\
\end{align*}
Using arguments similar to those involved for the treatment of $\chi(1)$, it
can be shown that
\begin{align*}
|\mathbb{E}(\chi(7)[4][j]\big|\mathcal{F}_{i-1}^n)|\leq
K\Delta_n^{1/4}(\Delta_n^{1/4}+\eta_{i,k_n}),~~\text{for}~~ j=1,\ldots,8,
\end{align*}
which yields
\begin{align*}
|\mathbb{E}(\chi(7)\big|\mathcal{F}_{i-1}^n)|\leq
K\Delta_n^{1/4}(\Delta_n^{1/4}+\eta_{i,k_n}).
\end{align*}

Next, define
\begin{align*}
\chi(8)=\frac{1}{k_n^3}\sum_{w=2}^{k_n-1}\sum_{v=0}^{w-1}%
\sum_{u=0}^{v-1}(C_{i+u}^{n,jk}-C_{i}^{n,jk})(C_{i+v}^{n,lm}-C_{i}^{n,lm})(C_{i+w}^{n,gh}-C_{i}^{n,gh}).
\end{align*}
This term can be further decomposed into six 
components. Successive conditioning and existing bounds give
\begin{align*}
&|\mathbb{E}\Big(%
(C_{i+u}^{n,jk}-C_i^{n,jk})(C_{i+v}^{n,lm}-C_{i+u}^{n,lm})(C_{i+w}^{n,gh}-C_{i+v}^{n,gh})%
\big|\mathcal{F}_{i-1}^n\Big)|\leq K\Delta_n \\
&|\mathbb{E}\Big(%
(C_{i+u}^{n,jk}-C_i^{n,jk})(C_{i+v}^{n,lm}-C_{i+u}^{n,lm})(C_{i+v}^{n,gh}-C_{i+u}^{n,gh})%
\big|\mathcal{F}_{i-1}^n\Big)|\leq K\Delta_n^{3/4}(\Delta_n^{1/4}+\eta_{i,k_n})
\\
&|\mathbb{E}\Big(%
(C_{i+u}^{n,jk}-C_i^{n,jk})(C_{i+v}^{n,lm}-C_{i+u}^{n,lm})(C_{i+u}^{n,gh}-C_{i}^{n,gh})%
\big|\mathcal{F}_{i-1}^n\Big)|\leq K\Delta_n \\
&|\mathbb{E}\Big(%
(C_{i+u}^{n,jk}-C_i^{n,jk})(C_{i+u}^{n,lm}-C_{i}^{n,lm})(C_{i+w}^{n,gh}-C_{i+v}^{n,gh})%
\big|\mathcal{F}_{i-1}^n\Big)|\leq K\Delta_n \\
&|\mathbb{E}\Big(%
(C_{i+u}^{n,jk}-C_i^{n,jk})(C_{i+u}^{n,lm}-C_{i}^{n,lm})(C_{i+v}^{n,gh}-C_{i+u}^{n,gh})%
\big|\mathcal{F}_{i-1}^n\Big)|\leq K\Delta_n \\
&|\mathbb{E}\Big(%
(C_{i+u}^{n,jk}-C_i^{n,jk})(C_{i+u}^{n,lm}-C_{i}^{n,lm})(C_{i+u}^{n,gh}-C_{i}^{n,gh})%
\big|\mathcal{F}_{i-1}^n\Big)|\leq K\Delta_n
\end{align*}
These bounds can be used to deduce
\begin{align*}
|\mathbb{E}(\chi(8)\big|\mathcal{F}_{i-1}^n)|\leq K\Delta_n.
\end{align*}
This completes the proof. 

\subsubsection{Proof of Equations (\protect\ref{eqn:R2}) and (\protect\ref%
{eqn:R3}) in Lemma \protect\ref{lemma:approx1}}

Observe that
\begin{align*}
&\nu_i^{n,jk}(C_{i+k_n}^{n,lm}-C_{i}^{n,lm})(C_{i+k_n}^{n,gh}-C_{i}^{n,gh})=%
\frac{1}{k_n\Delta_n}\sum_{u=0}^{k_n-1}%
\zeta_{i,u}^{n,jk}(C_{i+k_n}^{n,lm}-C_{i}^{n,lm})(C_{i+k_n}^{n,gh}-C_{i}^{n,gh}),
\\
&\nu_i^{n,jk}\nu_i^{n,lm}(C_{i+k_n}^{n,gh}-C_{i}^{n,gh})=\frac{1}{%
k_n^2\Delta_n^2} \sum_{u=0}^{k_n-1}\zeta_{i,u}^{n,jk}%
\zeta_{i,u}^{n,lm}(C_{i+k_n}^{n,gh}-C_{i}^{n,gh}) \\
&+\frac{1}{k_n^2\Delta_n^2} \sum_{u=0}^{k_n-2}\sum_{v=0}^{k_n-1}%
\zeta_{i,u}^{n,jk}\zeta_{i,v}^{n,lm}(C_{i+k_n}^{n,gh}-C_{i}^{n,gh}) +\frac{1%
}{k_n^2\Delta_n^2}\sum_{u=0}^{k_n-2}\sum_{v=0}^{k_n-1}\zeta_{i,u}^{n,lm}%
\zeta_{i,v}^{n,jk}(C_{i+k_n}^{n,gh}-C_{i}^{n,gh}).
\end{align*}
Hence, equations (\ref{eqn:R2}) and (\ref{eqn:R3}) can be proved using the
same strategy as for (\ref{eqn:R1}).

\subsubsection{Proof of Equations (\protect\ref{eqn:R4}) and (\protect\ref%
{eqn:R5}) in Lemma \protect\ref{lemma:approx1}}

Note that we have
\begin{align*}
&\lambda_i^{n,jk}\lambda_i^{n,lm}\nu_i^{n,gh}=\nu_i^{n,gh}\nu_{i+k_n}^{n,jk}%
\nu_{i+k_n}^{n,lm}+\nu_i^{n,gh}\nu_{i}^{n,jk}\nu_{i}^{n,lm}-\nu_i^{n,gh}%
\nu_{i}^{n,lm}\nu_{i+k_n}^{n,jk}-\nu_i^{n,gh}\nu_{i}^{n,lm}\nu_{i+k_n}^{n,jk}
\\
&+\nu_i^{n,gh}\nu_{i+k_n}^{n,jk}(C_{i+k_n}^{n,lm}-C_{i}^{n,lm})-\nu_i^{n,gh}%
\nu_{i}^{n,jk}(C_{i+k_n}^{n,lm}-C_{i}^{n,lm})+\nu_i^{n,gh}%
\nu_{i+k_n}^{n,lm}(C_{i+k_n}^{n,jk}-C_{i}^{n,jk}) \\
&-\nu_i^{n,gh}\nu_{i}^{n,lm}(C_{i+k_n}^{n,jk}-C_{i}^{n,jk})
+\nu_i^{n,gh}(C_{i+k_n}^{n,jk}-C_{i}^{n,jk})(C_{i+k_n}^{n,lm}-C_{i}^{n,lm}),
\end{align*}
and
\begin{eqnarray*}
&&\lambda_i^{n,gh}\lambda_i^{n,jk}\lambda_i^{n,lm}=\nu_{i+k_n}^{n,gh}%
\nu_{i+k_n}^{n,jk}\nu_{i+k_n}^{n,lm}+\nu_{i+k_n}^{n,gh}\nu_{i}^{n,jk}%
\nu_{i}^{n,lm}-\nu_{i+k_n}^{n,gh}\nu_{i}^{n,lm}\nu_{i+k_n}^{n,jk}-%
\nu_{i+k_n}^{n,gh}\nu_{i}^{n,lm}\nu_{i+k_n}^{n,jk} \\
&&+\nu_{i+k_n}^{n,gh}\nu_{i+k_n}^{n,jk}(C_{i+k_n}^{n,lm}-C_{i}^{n,lm})-%
\nu_{i+k_n}^{n,gh}\nu_{i}^{n,jk}(C_{i+k_n}^{n,lm}-C_{i}^{n,lm})+%
\nu_{i+k_n}^{n,gh}\nu_{i+k_n}^{n,lm}(C_{i+k_n}^{n,jk}-C_{i}^{n,jk}) \\
&&-\nu_{i+k_n}^{n,gh}\nu_{i}^{n,lm}(C_{i+k_n}^{n,jk}-C_{i}^{n,jk})
+\nu_{i+k_n}^{n,gh}(C_{i+k_n}^{n,jk}-C_{i}^{n,jk})(C_{i+k_n}^{n,lm}-C_{i}^{n,lm})
\\
&&
-\nu_i^{n,gh}\nu_{i+k_n}^{n,jk}\nu_{i+k_n}^{n,lm}-\nu_i^{n,gh}\nu_{i}^{n,jk}%
\nu_{i}^{n,lm}
+\nu_i^{n,gh}\nu_{i}^{n,lm}\nu_{i+k_n}^{n,jk}+\nu_i^{n,gh}\nu_{i}^{n,lm}%
\nu_{i+k_n}^{n,jk} \\
&&-\nu_i^{n,gh}\nu_{i+k_n}^{n,jk}(C_{i+k_n}^{n,lm}-C_{i}^{n,lm})+%
\nu_i^{n,gh}\nu_{i}^{n,jk}(C_{i+k_n}^{n,lm}-C_{i}^{n,lm})-\nu_i^{n,gh}%
\nu_{i+k_n}^{n,lm}(C_{i+k_n}^{n,jk}-C_{i}^{n,jk}) \\
&&+\nu_i^{n,gh}\nu_{i}^{n,lm}(C_{i+k_n}^{n,jk}-C_{i}^{n,jk})
-\nu_i^{n,gh}(C_{i+k_n}^{n,jk}-C_{i}^{n,jk})(C_{i+k_n}^{n,lm}-C_{i}^{n,lm})
\\
&& +\nu_{i+k_n}^{n,jk}\nu_{i+k_n}^{n,lm}(C_{i+k_n}^{n,gh}-C_{i}^{n,gh})
+\nu_{i}^{n,jk}\nu_{i}^{n,lm}(C_{i+k_n}^{n,gh}-C_{i}^{n,gh})
-\nu_{i}^{n,lm}\nu_{i+k_n}^{n,jk}(C_{i+k_n}^{n,gh}-C_{i}^{n,gh}) \\
&&
-\nu_{i}^{n,lm}\nu_{i+k_n}^{n,jk}(C_{i+k_n}^{n,gh}-C_{i}^{n,gh})+%
\nu_{i+k_n}^{n,jk}(C_{i+k_n}^{n,lm}-C_{i}^{n,lm})(C_{i+k_n}^{n,gh}-C_{i}^{n,gh})
\\
&&-%
\nu_{i}^{n,jk}(C_{i+k_n}^{n,lm}-C_{i}^{n,lm})(C_{i+k_n}^{n,gh}-C_{i}^{n,gh})
+\nu_{i+k_n}^{n,lm}(C_{i+k_n}^{n,jk}-C_{i}^{n,jk})(C_{i+k_n}^{n,gh}-C_{i}^{n,gh})
\\
&&
-\nu_{i}^{n,lm}(C_{i+k_n}^{n,jk}-C_{i}^{n,jk})(C_{i+k_n}^{n,gh}-C_{i}^{n,gh}) +(C_{i+k_n}^{n,jk}-C_{i}^{n,jk})(C_{i+k_n}^{n,lm}-C_{i}^{n,lm})(C_{i+k_n}^{n,gh}-C_{i}^{n,gh}).
\end{eqnarray*}
From (\ref{eqn:BGD}), notice that $\nu_i^{n}$ is $\mathcal{F}_{i+k_n-1}^n$%
-measurable and satisfies $\|\mathbb{E}(\nu_i^{n}|\mathcal{F}_{i-1}^n)\|\leq
K\Delta_n^{1/2}$.\newline
The law of iterated expectations and existing bounds imply
\begin{eqnarray}
|\mathbb{E}(\nu_{i}^{n,lm}\nu_{i+k_n}^{n,jk}|\mathcal{F}_{i-1}^n)| & \leq &
K\Delta_n^{3/4},  \notag \\
|\mathbb{E}(\nu_{i}^{n,lm}\nu_{i}^{n,gh}\nu_{i+k_n}^{n,jk}|\mathcal{F}_{i-1}^n)|
& \leq & K\Delta_n,  \notag \\
|\mathbb{E}(\nu_{i}^{n,lm}(C_{i+k_n}^{n,gh}-C_{i}^{n,gh})\nu_{i+k_n}^{n,jk}|%
\mathcal{F}_{i-1}^n)| & \leq & K\Delta_n,  \notag \\
|\mathbb{E}(\nu_{i+k_n}^{n,lm}(C_{i+k_n}^{n,jk}-C_{i}^{n,jk})|\mathcal{F}%
_{i-1}^n)| & \leq & K\Delta_n^{3/4},  \notag \\
|\mathbb{E}%
((C_{i+k_n}^{n,jk}-C_{i}^{n,jk})(C_{i+k_n}^{n,lm}-C_{i}^{n,lm})(C_{i+k_n}^{n,gh}-C_{i}^{n,gh})|%
\mathcal{F}_{i-1}^n)| & \leq & K\Delta_n.  \label{eqn:firstbounds}
\end{eqnarray}
It can also be readily verified that
\begin{align*}
&|\mathbb{E}(\nu_{i+k_n}^{n,gh}\nu_{i+k_n}^{n,ab}|\mathcal{F}_{i+k_n-1}^n)-%
\frac{1}{k_n}%
(C_{i+k_n}^{n,ga}C_{i+k_n}^{n,hb}+C_{i+k_n}^{n,gb}C_{i+k_n}^{n,ha})-\frac{%
k_n\Delta_n}{3}\overline{C}_{i+k_n}^{n,gh,ab}| \\
&\leq K\sqrt{\Delta_n}(\Delta_n^{1/8}+\eta_{i+k_n,k_n}^n).
\end{align*}
Hence, for $\varphi_i^{n,gh}
\in\{\nu_i^{n,gh},C_{i+k_n}^{n,gh}-C_{i}^{n,gh}\}$, which satisfies $\mathbb{%
E}(|\varphi_i^{n,gh}|^q\Big| \mathcal{F}_{i-1}^n) \leq K\Delta_n^{q/4}$ and $%
\mathbb{E}(\varphi_i^{n,gh}|\mathcal{F}_{i-1}^n)\leq K\Delta_n^{1/2}$. One can
show that
\begin{align*}
&|\mathbb{E}(\varphi_i^{n,gh}\nu_{i+k_n}^{n,jk}\nu_{i+k_n}^{n,lm}|\mathcal{F}%
_{i-1}^n)-\mathbb{E}\Big(\varphi_i^{n,gh}\Big[\frac{1}{k_n}%
(C_{i+k_n}^{n,jl}C_{i+k_n}^{n,km}+C_{i+k_n}^{n,jm}C_{i+k_n}^{n,kl})-\frac{%
k_n\Delta_n}{3}\overline{C}_{i+k_n}^{n,jk,lm}\Big]|\mathcal{F}_{i-1}^n\Big)| \\
&\leq K\Delta_n^{3/4}(\Delta_n^{1/4}+\eta_{i,2k_n}^n).
\end{align*}
Next, by combining the successive conditioning together with existing
bounds, we have
\begin{eqnarray*}
|\mathbb{E}(\varphi_i^{n,gh}\overline{C}_{i+k_n}^{n,jk,lm})| & \leq &
K\Delta_n^{1/4}(\Delta_n^{1/4}+\eta_{i,k_n}^n) \\
|\mathbb{E}(\varphi_i^{n,gh}C_{i+k_n}^{n,jl}C_{i+k_n}^{n,km})| & \leq &
K\Delta_n^{1/2},
\end{eqnarray*}
which together imply
\begin{equation}
|\mathbb{E}(\varphi_i^{n,gh}\nu_{i+k_n}^{n,jk}\nu_{i+k_n}^{n,lm}|\mathcal{F}%
_{i-1}^n)|\leq K \Delta_n^{3/4}(\Delta_n^{1/4}+\eta_{i,2k_n}^n).
\label{eqn:secondbounds}
\end{equation}
It is easy to see that equations (\ref{eqn:R1}), (\ref{eqn:firstbounds}) and
(\ref{eqn:secondbounds}) and the inequality $\eta_{i,k_n}^n \leq
\eta_{i,2k_n}^n $ together yield equations (\ref{eqn:R4}) and (\ref{eqn:R5}%
).

\subsection{Proof of Lemma \protect\ref{lemma:convresults}}

Equation (\ref{eqn:convergence1}) can be proved easily using the bounds of $%
\rho (u,v)_{i}^{n,gh}$ in equation (E.60).
To show equations (\ref%
{eqn:convergence2}), (\ref{eqn:convergence3}) and (\ref{eqn:convergence4}),
we set
\begin{equation*}
\overline{\overline{A11}}(H,gh,u;G,ab,v)=\lambda (u,v)_{0}^{n}\sum_{i\in
L^{\prime }\left( n,T\right) }(\partial _{gh}H\partial _{ab}G)(C_{i-1})\zeta
(u)_{i}^{n,gh}\zeta (v)_{i}^{n,ab}.
\end{equation*}%
Then,
\begin{equation*}
\frac{1}{\Delta _{n}^{1/4}}\Big(\overline{\overline{A11}}(H,gh,u;G,ab,v)-%
\overline{A11}(H,gh,u;G,ab,v)\Big)\overset{\mathbb{P}}{\Rightarrow }0.
\end{equation*}%
The above result is proved following similar steps as for equation (\ref%
{eqn:approx1}) in case $w=1$ by replacing $\Theta (u,v)_{0}^{(C),i,n}$ by $%
\lambda (u,v)_{0}^{n}((\partial _{gh}H\partial _{ab}G)(C_{i-1})-(\partial
_{gh}H\partial _{ab}G)(C_{i-2k_{n}}))$, which has the same bounds as the
former. Next, decompose $\overline{\overline{A11}}$ as follows,
\begin{align*}
\overline{\overline{A11}}(H,gh,u;G,ab,v)& =\lambda (u,v)_{0}^{n}\Bigg[%
\sum_{i\in L^{\prime }\left( n,T\right) }(\partial _{gh}H\partial
_{ab}G)(C_{i-1})V_{i-1}^{n}\\
& +\sum_{i\in L^{\prime }\left( n,T\right) }(\partial _{gh}H\partial
_{ab}G)(C_{i-1})\Big(\mathbb{E}(\zeta (u)_{i}^{n,gh}\zeta (v)_{i}^{n,ab}|%
\mathcal{F}_{i-1}^{n})-V_{i-1}^{n}\Big) \\
& +\sum_{i\in L^{\prime }\left( n,T\right) }(\partial _{gh}H\partial
_{ab}G)(C_{i-1})\Big(\zeta (u)_{i}^{n,gh}\zeta (v)_{i}^{n,ab}-\mathbb{E}%
(\zeta (u)_{i}^{n,gh}\zeta (v)_{i}^{n,ab}|\mathcal{F}_{i-1}^{n})\Big)\Bigg].
\end{align*}%
We follow the proof of equation (\ref{eqn:approx2}) for $w=1$, and we
replace $\Theta (u,v)_{0}^{(C),i,n}$ by $\lambda (u,v)_{0}^{n}(\partial
_{gh}H\partial _{ab}G)(C_{i-1})$, which satisfies only the condition $%
|\lambda (u,v)_{0}^{n}(\partial _{gh}H\partial _{ab}G)(C_{i-1})|\leq
\widetilde{\lambda }_{u,v}^{n}$. This calculation shows that the last two
terms in the above decomposition vanish at a rate faster
than $\Delta _{n}^{1/4}$. Therefore,
\begin{equation*}
\frac{1}{\Delta _{n}^{1/4}}\Bigg(\overline{\overline{A11}}%
(H,gh,u;G,ab,v)-\lambda (u,v)_{0}^{n}\Big(\sum_{i\in L^{\prime }\left(
n,T\right) }(\partial _{gh}H\partial _{ab}G)(C_{i-1})V_{i-1}^{n}\Big)\Bigg)%
\Rightarrow 0.
\end{equation*}%
As a consequence, for $(u,v)=(1,2)$ and $(2,1)$,
\begin{equation*}
\frac{1}{\Delta _{n}^{1/4}}\overline{\overline{A11}}(H,gh,u;G,ab,v)%
\Rightarrow 0.
\end{equation*}%
The results follow from the following observation,
\begin{eqnarray*}
&&\frac{1}{\Delta _{n}^{1/4}}\Bigg(\lambda (u,v)_{0}^{n}\Big(%
\sum_{g,h,a,b=1}^{d}\sum_{i\in L^{\prime }\left( n,T\right) }(\partial
_{gh}H\partial _{ab}G)(C_{i-1})V_{i-1}^{n}(u,v)\Big) \\
&&\hspace{10mm}-\frac{3}{\theta ^{2}}\int_{0}^{T}(\partial _{gh}H\partial
_{ab}G)(C_{t})(C_{t}^{ga}C_{t}^{hb}+C_{t}^{gb}C_{t}^{ha})dt\Bigg)\Rightarrow
0,~~\hspace{2mm}\text{for }(u,v)=(2,2), \\
&&\frac{1}{\Delta _{n}^{1/4}}\Bigg(\sum_{g,h,a,b=1}^{d}\lambda (u,v)_{0}^{n}%
\Big(\sum_{i\in L^{\prime }\left( n,T\right) }(\partial _{gh}H\partial
_{ab}G)(C_{i-1})V_{i-1}^{n}(u,v)\Big)-[H(C),G(C)]_{T}\Bigg)\Rightarrow 0, \\
&&\hspace{10mm}\text{for }(u,v)=(1,1).
\end{eqnarray*}

\swpNormalsize

\captionsetup{font=small,justification=justified}

\FloatBarrier


\section{Numerical Implementation}
\label{sec:appendix_numerical}

We now discuss some details for the numerical implementation of our
estimators. Section \ref{sec:estimation_special_case} explains how the main
quantities of interest can be expressed in terms of $\left[ H(C),G(C)\right]
_{T}$, where $C$ is the spot variance matrix of all $d$ assets. However, in
practice many quantities of interest involve only a much smaller subset of
assets, which greatly reduces the computational burden.

For example, suppose we want to calculate the variance of the IdioVol for a
single stock, where R-FM is the CAPM, and IdioVol-FM has one volatility
factor -- the market volatility. Then, we only need to consider two assets,
the stock and the market, e.g., SPY, so $d_{S}=d_{F}=1$ and $d=2$. Denote
the relevant spot variance-covariance matrix by%
\begin{equation*}
C=\left( 
\begin{array}{cc}
C_{11} & C_{12} \\ 
C_{21} & C_{22}%
\end{array}%
\right) ,
\end{equation*}%
where $C_{22}=C_{F}$ is the spot variance of the market, and $C_{11}$ is the
spot variance of the individual stock. The quantity of interest is 
\begin{equation*}
\left[ H(C),H(C)\right] _{T}=\left[ C_{Z1},C_{Z1}\right] _{T},
\end{equation*}%
where $C_{Z1,t}=C_{11}-C_{12}C_{22}^{-1}C_{21}$. The estimators in equations
(\ref{eqn:estAN}) and (\ref{eqn:estLIN}) involve the first derivatives $%
\partial _{ab}H\left( C\right) $ for $a,b=1,...,d$, which are%
\begin{equation*}
\partial _{ab}H\left( C\right) \equiv \frac{\partial H\left( C\right) }{%
\partial C_{ab}}=\frac{\partial C_{Zj}}{\partial C_{ab}}=\frac{\partial
\left( C_{11}-C_{12}C_{22}^{-1}C_{21}\right) }{\partial C_{ab}}=\left\{ 
\begin{array}{cc}
C_{12}C_{22}^{-2}C_{21} & \text{if }\left( a,b\right) =\left( 2,2\right)  \\ 
-C_{22}^{-1}C_{21} & \text{if }\left( a,b\right) =\left( 1,2\right)  \\ 
-C_{12}C_{22}^{-1} & \text{if }\left( a,b\right) =\left( 2,1\right)  \\ 
1 & \text{if }\left( a,b\right) =\left( 1,1\right) 
\end{array}%
\right. 
\end{equation*}

If we are interested in the stock's IdioVol $\gamma _{Z}$, by equation
(\ref{eqn:ID_gamma}) we also need the volatility factor $\Pi _{t}=G\left(
C_{t}\right) =C_{22,t}$, and $\left[ \Pi ,C_{Z1}\right] _{T}^{c}$. The
derivatives are $\partial _{ab}G\left( C\right) \equiv \partial G\left(
C\right) \left/ \partial C_{ab}\right. =1\left\{ a=b=2\right\} $.

\clearpage

\section{Additional Figures}

\label{sec:additional_figures} 
\begin{figure}[!h]
\centering%
\includegraphics[trim = 45mm 0mm 50mm
0mm,scale=0.7]{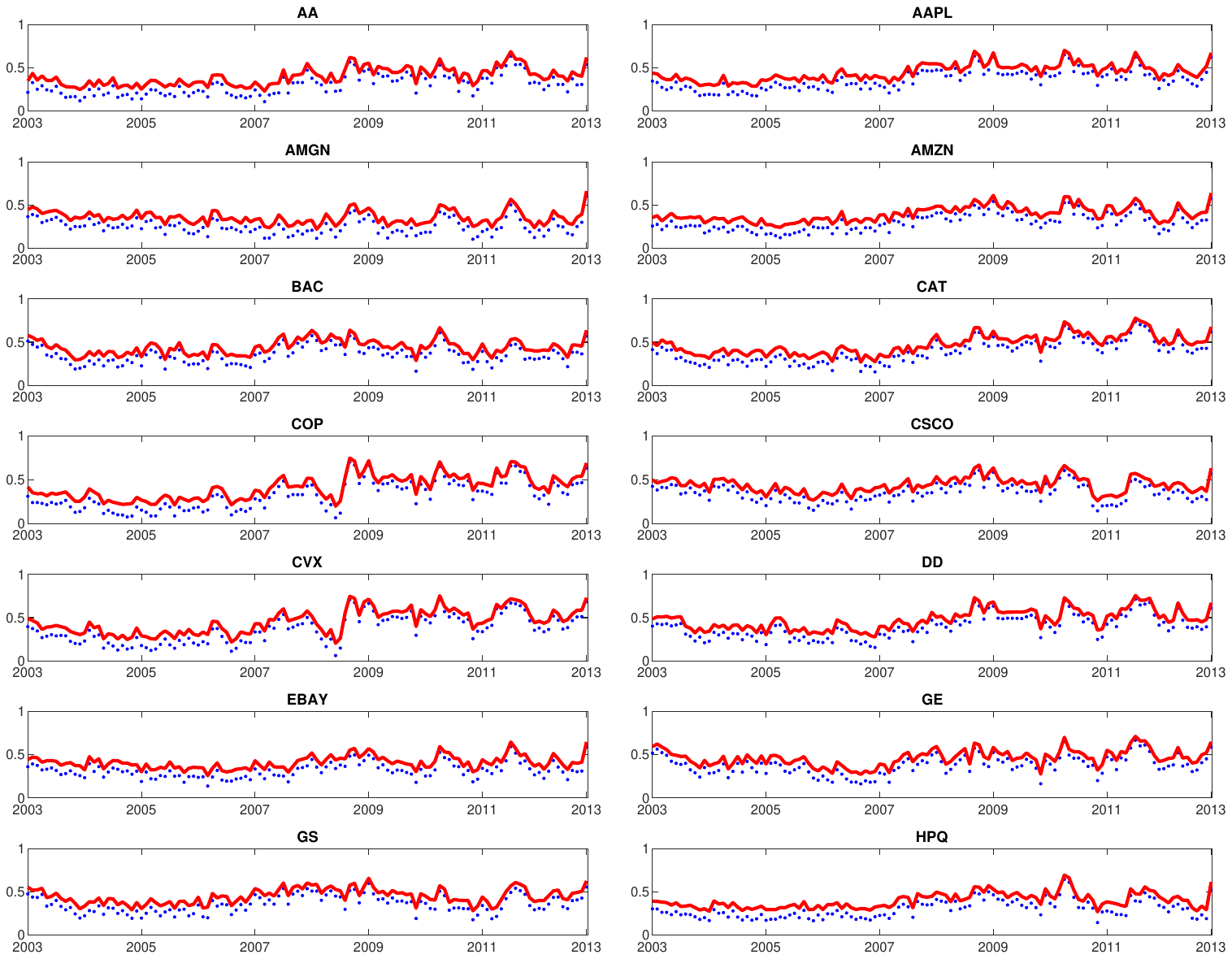}
\caption{Monthly $R^2$ of two Return Factor Models ($\protect\widehat{R}%
^2_{Yj}$): the CAPM (the blue dotted line) and the Fama-French three factor
model (the red solid line). Stocks are represented by tickers (see Table
\protect\ref{tbl:stocks} for full stock names). }
\label{fig:R2_a}
\end{figure}
\clearpage

\begin{figure}[!h]
\centering
\includegraphics[trim = 45mm 0mm 50mm
0mm,scale=0.7]{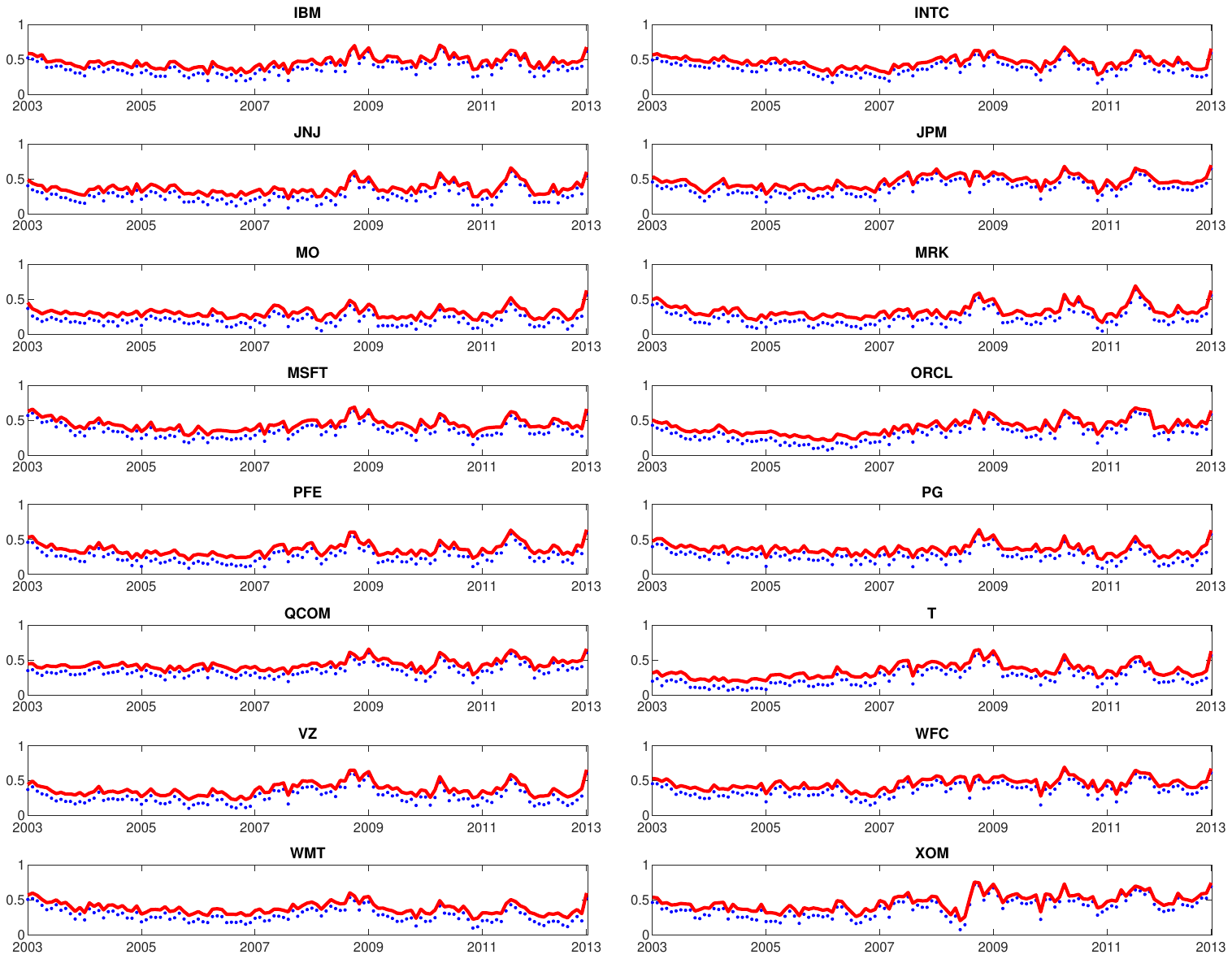}
\caption{Monthly $R^2$ of two Return Factor Models ($\protect\widehat{R}%
^2_{Yj}$): the CAPM (the blue dotted line) and the Fama-French three factor
model (the red solid line). Stocks are represented by tickers (see Table
\protect\ref{tbl:stocks} for full stock names). }
\label{fig:R2_b}
\end{figure}

\clearpage

\end{appendix}%

\end{document}